\documentclass[useAMS,usenatbib]{mn2e}

\usepackage{epsfig}
\usepackage{graphicx}
\usepackage{times}
\usepackage{natbib}
\usepackage{setspace}
\newif\ifAMStwofonts
\AMStwofontstrue

\renewcommand{\vec}[1]{\bmath{#1}}
\newcommand{\be}{\begin{equation}}
\newcommand{\ee}{\end{equation}}
\newcommand{\ba}{\begin{eqnarray}}
\newcommand{\ea}{\end{eqnarray}}
\newcommand{\brr}{\begin{array}}
\newcommand{\err}{\end{array}}
\newcommand{\bc}{\begin{center}}
\newcommand{\ec}{\end{center}}

\newcommand{\kms}{km s$^{-1}$}

\newcommand{\surf}{M$_\odot$ pc$^{-2}$}

\newcommand{\msunyr}{M$_\odot$ yr$^{-1}$}

\newcommand{\mincir}{\raise
  -2.truept\hbox{\rlap{\hbox{$\sim$}}\raise5.truept \hbox{$<$}\ }}
\newcommand{\magcir}{\raise
  -2.truept\hbox{\rlap{\hbox{$\sim$}}\raise5.truept \hbox{$>$}\ }}
\newcommand{\siml}{\raise
  -2.truept\hbox{\rlap{\hbox{$\sim$}}\raise5.truept \hbox{$<$}\ }}
\newcommand{\simg}{\raise
  -2.truept\hbox{\rlap{\hbox{$\sim$}}\raise5.truept \hbox{$>$}\ }}

\title{A sub-resolution multiphase interstellar medium model of
  star formation and SNe energy feedback }

\author[Murante et al.] {Giuseppe
  Murante$^1$, Pierluigi Monaco$^{2,3}$, Martina Giovalli$^{4,1,5}$, 
Stefano Borgani$^{3, 2, 6}$
\newauthor{\& Antonaldo Diaferio$^{4,5}$} \\
$^1$ INAF, Osservatorio Astronomico di Torino, Strada Osservatorio 20, I-10025
Pino Torinese (Italy) (murante@oato.inaf.it)\\ 
$^2$ INAF, Osservatorio Astronomico di Trieste, Via Tiepolo 11,
I-34131 Trieste (Italy) \\
$^3$ Dipartimento di Astronomia dell'Universit\`a di Trieste, via
Tiepolo 11, I- 34131 Trieste, Italy (borgani, monaco@oats.inaf.it)\\ 
$^4$ Dipartimento di Fisica Generale ``Amedeo Avogadro'', Universit\`a degli Studi di Torino, Torino (Italy) (giovalli@oato.inaf.it, diaferio@ph.unito.it)\\ 
$^5$ INFN, Istituto Nazionale di Fisica Nuclare, Torino (Italy) \\
$^6$ INFN, Istituto Nazionale di Fisica Nuclare, Trieste (Italy) \\ 
}

\begin{document}

\maketitle

\label{firstpage}

\begin{abstract}
We present a new multi-phase sub-resolution model for star formation
and feedback in SPH numerical simulations of galaxy formation.  Our
model, called MUPPI (MUlti-Phase Particle Integrator), describes each
gas particle as a multi-phase system, with cold and hot gas phases,
coexisting in pressure equilibrium, and
a stellar component.
Cooling of the hot tenuous gas phase feeds the cold gas phase.  We
compute the cold gas molecular fraction using the phenomenological
relation of Blitz \& Rosolowsky between this fraction
and the external disk pressure, that we identify with the SPH
pressure. Stars are formed out of molecular gas with a
given efficiency, which scales with the dynamical time of the cold
phase. 
Our prescription for star formation is not based on imposing the
Schmidt-Kennicutt relation, which is instead naturally produced by
MUPPI. 
Energy from supernova explosions is deposited partly
into the hot phase of the gas particles, and partly to that of
neighboring particles.  Mass and energy flows among the different
phases of each particle are
described by a set of ordinary differential equations which we
explicitly integrate for each gas particle, instead of relying on
equilibrium 
solutions. This system of equations also includes the response of the
multi-phase structure to energy changes associated to the
thermodynamics of the gas. Our model has an intrinsically runaway
behavior: energy from supernovae increases gas 
pressure which increases in turn the star formation rate through the molecular
fraction. This runaway is stabilized in simulations by the hydrodynamic
response of the gas: when it receives enough energy, it expands
thereby decreasing its pressure.

We apply our model to two isolated disk galaxy simulations and two
spherical cooling flows. MUPPI is able to reproduce the
Schmidt-Kennicutt relation for disc galaxies.  It also reproduces the
basic properties of the inter-stellar medium in disc galaxies, the
surface densities of cold and molecular gas, of stars and of star
formation rate, the vertical velocity dispersion of cold clouds and
the flows connected to the galactic fountains. Quite remarkably, MUPPI
also provides efficient stellar feedback without the need to include a
scheme of kinetic energy feedback.
\end{abstract}


\section{Introduction}
\label{section:introduction}

Numerical simulations are a powerful tool to study structure formation
and evolution in a cosmological context, and their use has become
standard in the last decades.  The gravitational evolution of the {\it
  concordance} $\Lambda$ Cold Dark Matter ($\Lambda$CDM) model is now
studied and understood in detail through N-body simulations covering
large dynamic ranges. However, a direct comparison of the results of
numerical simulations with observations clearly requires the treatment
of baryonic physics.  Moreover, while $\Lambda$CDM cosmogony is quite
successful in reproducing the observed large-scale properties of the
Universe \citep[][]{Springel_etal06}, on galactic scales a number of
issues, like ab-initio formation of disk galaxies and abundance and
properties of small ``satellite'' galaxies \citep[see, e.g.,][for a
  review]{Mayer08} , are still debated.

Including astrophysical processes in simulations such as radiative gas
cooling, star formation and energy feedback from Supernovae (SNe)
\citep[see e.g.][for a review]{dolag2008}, is a hard task for several
reasons. The physics of star formation is complex and currently not
understood in detail; moreover, the dynamical range needed to
simultaneously resolve the formation of cosmic structures and the
formation of stars is huge, since the former process happens on Mpc
scales and the latter on sub-pc scales.

Several authors included simple schemes to transform the cold dense
gas (depicted as a single fluid) into stars
(e.g. \citealp{1978MNRAS.183..341W}, \citealp{1992ApJ...393...22C},
\citealp{1993MNRAS.265..271N}, \citealp{Katz96}). For instance,
\citealp{Katz96} based their recipe of star formation simply on an
overdensity criterion, thus ignoring the multi-phase nature of the
Inter-Stellar Medium (ISM). However, in the ISM of observed galaxies,
the gas is not single-phase, being instead characterized by a wide
range of density and temperature states, as illustrated, e.g., by
\cite{Cox05}.  As shown in this review, such a multi-phase medium
results from the interplay of processes such as gravity,
hydrodynamics, turbulence, heating by SNe and stellar winds, magnetic
fields, cosmic rays, chemical enrichment and dust formation.  
The
problem of star formation can thus be dealt with algorithms which
explicitly model, to some level of approximation, the multi-phase
structure of the ISM. 
 A number of recipes to tackle such a problem has been proposed
(\citealp{1997MNRAS.284..235Y}, \citealp{ThacCouch00},
\citealp{SpringHern03}, \citealp{2003MNRAS.345..561M},
\citealp{2006MNRAS.371.1125S}, \citealp{Stinson06}, \citealp{Booth07},
\citealp{2008MNRAS.387.1431D}). While not all of them make an
  attempt to model the multi-phase behavior of single gas particles,
 the aim of all these prescriptions is to reproduce some basic
observations like the Schmidt-Kennicutt (SK) relation \citep{1998ApJ...498..541K} while
regulating the final amount of stars with energetic feedback from Type
II and, sometimes, Type Ia SNe. Still, only a few of these
prescriptions include some realistic (though necessarily simplified)
modeling of a multi-phase, turbulent ISM.

Early attempts to introduce stellar feedback into simulations
(\citealp{1987ApJ...322..585B}, \citealp{1992ApJ...393...22C},
\citealp{Katz92}, \citealp{1993MNRAS.265..271N}) already showed that
if SN energy is deposited as thermal energy onto the star-forming gas
particle, it is quickly dissipated through radiative cooling before it
has any relevant effects. This is due to the fact that the
characteristic timescale of radiative cooling at the typical density
of star-forming regions is far shorter then the free-fall
gravitational timescale \citep{Katz92}.

Several solutions have been proposed to increase the efficiency of
feedback in regulating star formation. For instance,
\cite{SpringHern03} introduced a sub-resolution description of a
two-phase ISM. In their model, the amount of gas in the cold and hot
phases is regulated by the thermodynamical properties of the
star-forming gas particle: the two phases are always considered to be
in thermal pressure equilibrium.  Star formation is proportional to the amount
of gas in the cold phase. Thermal energy of Type II SN participates in the
determination of equilibrium between the phases; since the particle
interacts hydrodynamically using its average temperature, the
efficiency of such a feedback is not high enough to suppress star
formation to observed level.  For this reason, \cite{SpringHern03}
also included a phenomenological prescription of kinetic feedback, in
which star-forming gas particles are stochastically selected to
receive a velocity ``kick'', with probability proportional to their
star formation rate (SFR hereafter), and hydrodynamically decouple for
a given time from the surrounding gas.

\cite{2006MNRAS.371.1125S} tried to overcome the overestimation of
the cooling rate by decoupling gas phases using separate thermodynamic
variables i.e. hot, diffuse gas particles do not ``see'' cold, dense
gas particles as neighbors. \cite{Booth07} took a different
approach and decoupled the cold molecular phase from the hot phase by
describing the former with sticky particles and treating their
aggregations with a sub-resolution kinetic coagulation model.
\cite{2008MNRAS.387.1431D} implemented a variation of
\cite{SpringHern03} prescription in which winds are produced locally
by neighboring star-forming particles and are not hydrodynamically
decoupled. These authors showed that coupled winds generate a large
bipolar outflow in a dwarf-like galaxy and a galactic fountain in a
massive galaxy, while the \cite{SpringHern03} decoupled wind produces
isotropic outflows in both cases.

A different solution consists in simply suppressing radiative cooling
of gas particles which have just been heated by SNe (typically for 30
Myr, the observed lifespan of a SN blast: \citealp{Ger97},
\citealp{ThacCouch00}, \citealp{2003ApJ...596...47S}).
Apart from providing an artificial method for reducing cooling {\it
  and} star formation rates, this scheme is strongly
resolution-dependent \citep[e.g.,][]{ThacCouch00}.
\cite{Stinson06} found a numerical implementation able to partly
remove the dependence on resolution. This was used by \cite{Gov07} in
a cosmological simulation of a disk galaxy, and was demonstrated to be
very efficient in reducing the angular momentum loss of gas by pushing
it in the outskirts of DM haloes while they merge.

Very recently, direct simulations of the ISM in extended galaxies
have become computationally feasible. For instance, \cite{Pelupessy06,
  Robertson08, Tasker08} simulated isolated galaxies
with a force resolution of tens of pc, explicitly following the formation of
molecular clouds,
while \cite{Ceverino09} simulated the formation and evolution of a
disk galaxy in a cosmological context (but only up to redshift $z
\sim 3$) with comparable resolution.
Such first works show that it is possible to simulate galaxies with
reasonable ISM properties; however, applying these codes to a
cosmological box having sizes of $\ga10$ Mpc 
will be
unfeasible for several years to come. This confirms the necessity of
developing realistic sub-resolution models of star formation and feedback,
which are able to capture in a physically motivated way the complex
nature of star formation and energy feedback.

In this paper, we present a novel algorithm of this kind, MUPPI
(\textbf{MU}lti-\textbf{P}hase \textbf{P}article \textbf{I}ntegrator),
implemented in the GADGET-2 Tree-SPH code \citep{GADGET2}.  It is
based on a modified version of the analytic model of star formation
and feedback in a two-phase ISM developed by \cite{Monaco04a} (M04
hereafter). The main characteristic of this algorithm is that it
performs, for each multi-phase (MP) gas particle, the integration of a
system of ordinary differential equations that regulate mass and
energy flows among the ISM phases.  This integration is performed
within the SPH time-step, and allows to follow all the transients of
the system, instead of assuming an equilibrium solution for the ISM,
which in fact turns out to never be in equilibrium.  The model is
heavily based on the assumption of a correlation between the ratio of
molecular to neutral hydrogen and gas pressure. Such relation has been
observed to hold in local galaxies by \cite{Blitz04,Blitz06}, and is
used as a phenomenological prescription to bypass the need of a
detailed description of molecular cloud formation. The interaction
between MUPPI and the SPH part of GADGET-2 allows each particle to
immediately respond to the injected energy; because most thermal
energy is contained in the diluted hot phase, the cooling time of
MP particles is relatively long, thus allowing an efficient
injection of thermal energy.

The organization of the paper is as follows.
Section~\ref{section:model} provides a detailed description of MUPPI:
the scientific rationale behind it, the system of equations, the
entrance and exit conditions, the stochastic star formation algorithm,
the interaction with SPH, the redistribution of energy to neighboring
particles. Section~\ref{section:results} presents the suite of
numerical tests performed to assess the validity of the code. We
describe here the results of simulations carried out for (i) an
isolated Milky Way-like (MW) galaxy, (ii) an isolated Dwarf-like (DW)
galaxy, (iii) two analogous isolated, non-rotating haloes with gas
initially in hydrostatic equilibrium. 
We finally report on a study
of resolution dependence and exploration of model parameter space.  We
summarize our main results and conclusions in
Section~\ref{section:conclusions}.

\section{The model}
\label{section:model}

A successful sub-resolution model for hydrodynamical simulations of galaxy
formation should find the best compromise between simplicity of
implementation and complexity of evolution, aiming at describing as
closely as possible the behavior of star-forming regions.

Our starting point is the M04 analytic model of star formation and
feedback.  In that paper the ISM is described as a two-phase medium in
thermal pressure equilibrium. The formation of molecular clouds is due
to the balance between kinetic aggregations and gravitational collapse
of super-Jeans clouds.  SNe exploding in molecular clouds give raise
to super-bubbles that sweep the ISM.  The energy given to the ISM by
the blast is explicitely computed under the simple assumption that
this propagates into the most pervasive and diluted hot phase
\citep{Ostriker88}.  Different regimes of feedback are found,
according to whether the super-bubbles are able (or not) to blow out
of the system before they are pressure-confined by the hot phase.

The M04 model was constructed to shed light on the way energy from SNe
is given to the ISM, and to quantify the efficiency of this process as
a function of fundamental environmental parameters like gas surface
density or galaxy vertical scale-length.  However, we consider the M04
model unsuitable for a direct implementation in an SPH code, for several
reasons.  First, the system on which the M04 model is constructed
consists of a region of ISM which contains many cold and star-forming
clouds.  The mass of a molecular cloud can reach $\sim10^7$ M$_\odot$
in the Milky Way, while particle masses in simulations of isolated
galaxies can range from $10^4$ to $10^6$ M$_\odot$.  Clearly, a single
star-forming cloud will be sampled by many particles in realistic
cases, so that the starting assumption of M04 should be revised.

Our approach here is to consider each SPH gas particle as a potential
  parcel of the multi-phase ISM, when its density and temperature allows
  it. Whenever this is the case, the gas particle will samples the ISM
  for a period of time, during which it represents both part of a
  giant molecular cloud and part of the tenuous hot gas. The sampled
  molecular cloud is assumed to be destroyed after some time, after
  which the gas particle becomes again single phase. It can become
  again multi-phase, in case it is allowed by its physical conditions,
  thus sampling another portion of the ISM. Gas particles are
  therefore allowed to have several distinct multi-phase stages.

Second, in M04 the system is open, receiving mass from and ejecting
mass (and energy) to an external reservoir, called ``Halo''. In a
typical simulation infall and outflow of gas in/from the galaxy will
be resolved. For simplicity of implementation, we prefer to neglect
any inward or outward mass flow in our subgrid model, only taking into
account external energy flows.

Third, M04 explicitly computes the energy received by the hot phase of
the ISM from expanding super-bubbles generated by multiple SNe
explosions in star-forming clouds.  Our choice here is to parameterize
such process by assuming that the hot phase receives a constant
fraction of the produced SN energy,
using M04 and \cite{Monaco04b} to suggest fiducial values for the parameters.
This approach has the advantage of giving a much simpler and
transparent grasp to the distribution of energy.

Fourth, an important feature of the M04 model is to predict different
behavior for systems with different geometries: depending on the cold
gas surface density, in a ``thin'' disk system super-bubbles will
manage to blow out, thus dispersing most energy in the vertical
direction, while in a ``thick'' system most energy will be deposited
into the local ISM, thus leading to much higher pressure.  In a hydro
simulation the geometry of the system is resolved, at least within the
force resolution of the code.  The question is then whether the
vertical scale-length of the system may be guessed starting from a
particle's local (in the SPH sense) neighborhood, and the two
different regimes inserted in the sub-resolution model.  We choose to avoid
this guess and adopt a different strategy: all particles behave like in
the ``thin system'' case of M04, i.e. they inject energy to
neighboring particles along the ``least resistance path'' defined by
(minus) the gradient of the SPH density.  In a truly thin disk this
energy will be directed outwards, while in a thick or very
dense system the energy will be trapped within the star-forming
region.  In other words, the ``thin'' or ``thick'' regimes are
resolved by the simulation instead of being assumed as part of the
sub-resolution model.

A further fundamental difference is related to the modeling of
star-forming (molecular) cloud masses, which is done in M04 using a
kinetic aggregation model.  However, different authors
\citep{Blitz04,Blitz06,Leroy09} have shown that in local
  galaxies the ratio between the surface densities of molecular and
cold gas scales almost linearly with the so-called ``external
pressure'', defined as the total pressure required at the midplane to
support a hydrostatic disc.  This behavior is not obtained in the M04
kinetic approach, where pressure determines the size of clouds and
higher pressure implies smaller clouds and then smaller cross section.
As a result, a higher pressure results in a smaller fraction of cold
mass in molecular clouds, which is clearly at variance with
observations.  These observations are naturally explained in a context
in which pressure is driven by turbulence and molecular cooling in the
turbulent ISM is properly taken into account \citep{Robertson08}.  M04
assumed that the formation of ``molecular'' (star-forming) clouds is
triggered by gravitational collapse through a Jeans-like criterion,
and this may well be a naive description of the dynamics of the ISM.
We then adopt the strategy of using the observed relation of
\cite{Blitz06} to construct a phenomenological model that computes the
fraction of molecular gas, thus bypassing all the difficulties
connected with the poorly understood formation of molecular clouds.
We use the SPH pressure of the particle in place of the external
pressure.  Clearly the use of a relation found for local galaxies
  and not tested at high redshift should be considered as an {\it
    anstatz} to the true solution. However, star formation in high
  redshift galaxies is known to take place in more gas-rich, compact,
  pressurized and molecular-dominated environments than in local
  galaxies. Therefore, an evolution of this relation with redshift
  should have a minor impact.

We have implemented our model in a version of the GADGET-2 code which
adopts an explicitly entropy-conserving SPH formulation, and includes
radiative cooling computed for a primordial plasma with vanishing
metalicity \citep{SpringHern03}.

\begin{figure}
\centerline{
\includegraphics[width=\linewidth]{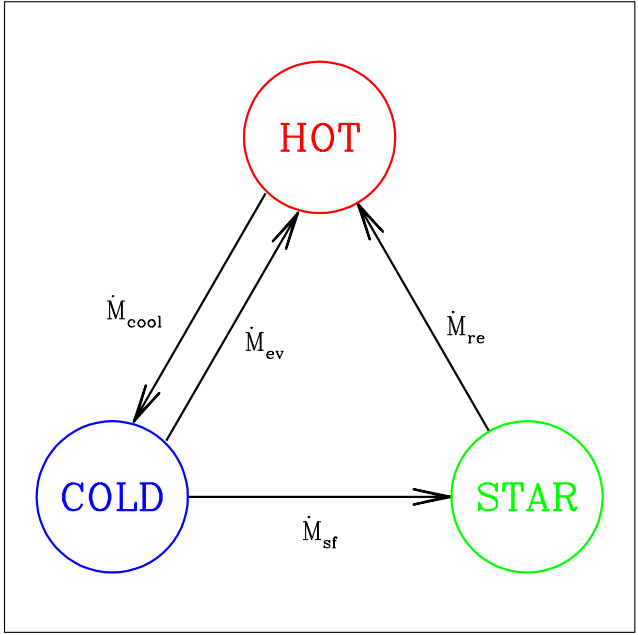}}
\caption{Schematic illustration of the mass flows between the
  different phases.}
\label{fig:mflows}
\end{figure}

\subsection{The sub-resolution model}
\label{section:sub-resolution}

We assume that a MP SPH particle is made up of two gas
phases, namely cold and hot gas, and a stellar component. The two gas
phases are assumed to be in thermal pressure equilibrium: 

\be n_{\rm h}\cdot T_{\rm h} = n_{\rm c} \cdot T_{\rm c} \ee 

\noindent
Here and in the following $n$ denotes particle number density,
e.g. $n_{\rm h} =
M_{\rm h}/(\mu_{\rm h} m_p)$, where $m_p$ is the proton mass, $\mu_{\rm h} = 4 / (5
f_{\rm HI} + 3) \simeq 0.6$ and $\mu_{\rm c} = 4/ (3 f_{\rm HI} + 1) \simeq 1.2$ are
the molecular weights, $f_{\rm HI}=0.76$ being the fraction of neutral
hydrogen. To most effectively single out the behavior of MUPPI, we
decided to develop and test the code neglecting any chemical
evolution. This means that the molecular weights of the hot and cold
phases are constant and the cooling time is computed for a gas with
zero metalicity.

The temperature of the cold phase is set to $T_{\rm c}=1000$ K; this is in
line with \cite{SpringHern03}.  This temperature regulates the
duration of each multi-phase stage and the star formation rate: the
lower $T_{\rm c}$ is, the shorter the multi-phase stage and the higher the
star formation rate. The value we chose gives multi-phase stages
lasting some tens of million years and a reasonable SFR (see below,
Section~\ref{section:parameters}). If $F_{\rm h}$ is the fraction of gas
mass in the hot phase, then its filling factor $f_{\rm h}$ is:

\be 
f_{\rm h} = \frac{1}{ 1 +  \frac{F_{\rm c}}{F_{\rm h}}\cdot \frac{\mu_{\rm h}}{\mu_{\rm c}} \cdot \frac{T_{\rm c}}{T_{\rm h}}}
\ee

\noindent
with $f_{\rm c}=1-f_{\rm h}$. Mass is exchanged among the components as in Figure~\ref{fig:mflows}.
A fraction $f_{\rm mol}$ of the cold gas will be in molecular form, and another
fraction $f_{\star}$, per
dynamical time $t_{\rm dyn}$ of the molecular cold phase,
will be consumed into stars. The dynamical time is:

\be
t_{\rm dyn} = \sqrt{\frac{3\pi}{32G\rho_{\rm c}}} \simeq 5.15\cdot
10^{7} (\mu_{\rm c} n_{\rm c})^{-1/2} \ {\rm yr} \label{eq:tdyn}
\ee

\noindent This gives rise to a star formation rate of: 

\be
\dot{M}_{\rm sf} = f_{\star} \cdot \frac{f_{\rm mol}\cdot M_{\rm c}}{t_{\rm dyn}}
\label{eq:sfr}
\ee

\noindent
As mentioned above, we compute the fraction of molecular gas, $f_{\rm
  mol}$, using the results of \cite{Blitz06}, that can be recast as
follows:

\be
f_{\rm mol} = \frac{1}{1 + P_0/P}
\label{eq:fcoll}
\ee 

\noindent
where P is the pressure. Note that here for simplicity we adopt a linear
scaling with pressure, in place of the $0.92\pm0.07$ exponent found in
\cite{Blitz06}. Following this paper, the pressure $P_0$ at which half of the
cold gas is molecular is set to $P_0/k_B = 35000$ K cm$^{-3}$, with $k_B$
being the Boltzmann constant. As mentioned above, we use the SPH pressure as
an estimate of the external one.

A fraction $f_{\rm re}$ of the mass involved in star formation is
promptly restored to the hot phase through the death of massive stars:

\be 
\dot{M}_{\rm re} = f_{\rm re} \cdot \dot{M}_{\rm sf}  
\ee

\noindent
We use $f_{\rm re}=0.2$, roughly consistent with a Salpeter stellar
Initial Mass Function (IMF).  Since we neglect metal production, the
restored material will have the same composition as the original one.
Radiative cooling creates a cooling flow which is modeled as follows:

\be
\dot{M}_{\rm cool} =  \frac{M_{\rm h}}{t_{\rm cool}}
\ee

\noindent
where $t_{\rm cool}$ is the radiative cooling time derived from the
GADGET-2 cooling function which uses the tabulated cooling rates given
by \citet{Katz96}.  Following M04 and \cite{Monaco04b}, the
evaporation rate is not connected to thermal conduction within the
expanding super-bubbles, as in \cite{McKee77} and in
\cite{SpringHern03}, but with the destruction of the molecular cloud
by the action of massive stars. \cite{Monaco04b} estimates that
$\sim10$ per cent of the cloud is evaporated, the rest being
snow-ploughed back to the cold phase by the first SNe that explode in
the cloud.  We then model the evaporation flow as:

\be
\dot{M}_{\rm ev} = f_{\rm ev} \cdot \dot{M}_{\rm sf}   
\ee

\noindent where $f_{\rm ev}=0.1$ in our runs, as suggested by \cite{Monaco04b}.

Calling $M_\star$, $M_{\rm c}$ and $M_{\rm h}$ the mass in stars and in the cold
and hot gas phases, the resulting system for the mass flows is:

\be
\dot{M}_{\star} = \dot{M}_{\rm sf} - \dot{M}_{\rm re} 
\label{eq:sf}
\ee

\be
\dot{M}_{\rm c} = \dot{M}_{\rm cool} - \dot{M}_{\rm sf} - \dot{M}_{\rm ev}
\label{eq:mc}
\ee

\be
\dot{M}_{\rm h} = -\dot{M}_{\rm cool} + \dot{M}_{\rm re} + \dot{M}_{\rm ev}
\label{eq:mh}
\ee

Energy flows are associated to mass flows. We follow the evolution of
the thermal energy of the hot phase, $E_{\rm h}$, that is connected to its
temperature as follows:

\be
T_{\rm h} = \frac{E_{\rm h}}{M_{\rm h}}\frac{(\gamma-1)\mu_{\rm h} m_p}{k_B}
\ee

\noindent 
where we assume $\gamma=5/3$ for the adiabatic index of a mono-atomic
gas.  Energy is lost by cooling and acquired from SN
explosions. Moreover, energy is exchanged with other particles through
the hydro energy term $\dot{E}_{\rm hydro}$, provided by the SPH part
of the code, which gives the energy variation due to adiabatic
contraction/expansion, shocks etc.  The cooling term is simply:

\be
\dot{E}_{\rm cool} =  \frac{E_{\rm h}}{t_{\rm cool}}
\ee

\noindent
while we assume that a small fraction $f_{\rm fb,local}=0.02$ of the SN energy is
directly injected into the local ISM:

\be
\dot{E}_{\rm heat,local} = E_{\rm SN} \cdot f_{\rm fb,local} \cdot \frac{\dot{M}_{\rm sf}}{M_{\rm\star,SN}} \,.
\label{eq:esn}
\ee

\noindent
Here $E_{\rm SN}$ is the energy of one single SN and $M_{\star\rm,SN}$
the mass of stars formed for each SN. We assume for this parameter a
value of 120 M$_\odot$, roughly compatible with a Salpeter IMF, once
we assume a limiting mass of 8 M$_\odot$ for the stars exploding as
type-II SNe. The exact value of $f_{\rm fb,local}$ has a small influence
on the behavior of our model, as long as the energy injected in the
local ISM is 
not larger
than that distributed to neighbors.
We note that in our model, when a
particle is multi-phase, radiative cooling only involves its hot
phase.

The resulting equation for the evolution of the hot
phase thermal energy is:

\be
\dot{E}_{\rm h} = \dot{E}_{\rm heat,local} - \dot{E}_{\rm cool} + \dot{E}_{\rm hydro} \qquad \qquad
\label{eq:eh}
\ee

The term $\dot{E}_{\rm hydro}$ also includes the energy contributed by SN
  explosions within neighboring MP particles.  This energy is computed
  distributing a fraction $f_{\rm fb,out}$ of SN energy to neighboring
  particles. This point will be further discussed in
  Section~\ref{section:redistrib}.

In this version of the code we do not attempt to model the evolution
of the kinetic energy of the cold phase, which would give a kinetic
pressure term. We tested some simple implementations and noticed that
they added to the complexity (and number of parameters) of the
formulation without giving any obvious improvement, so we prefer here
to keep the simpler formulation.

\subsection{Entrance and exit conditions from multi-phase and the SPH/MUPPI interface}
\label{section:entrance}

A particle enters the multi-phase regime whenever its density is higher
than the threshold value $n_{\rm thr}=0.01$ cm$^{-3}$ and its temperature is
below $T_{\rm thr}=5\times10^4$ K.  Since pressure determines the molecular
fraction (equation~\ref{eq:fcoll}), and thus the cut in the SK relation, we
use a very low value of the density threshold such that most particles in our
tests are always in multi-phase regime. Moreover, a temperature threshold is
imposed to prevent hot dense gas particles to spuriously become multi-phase.
As soon as a particle becomes multi-phase, it is then initialized with all of its
mass in the hot phase, although the temperature is relatively low at the
beginning.

At this point, all the gas mass is in the hot phase, and maintains
  its SPH density. Its temperature is obtained from the SPH
  entropy. During the first stage of the multi-phase evolution, the hot gas
  cools and deposits mass onto the cold phase. If pressure is
  sufficiently high, a fraction of such a cold gas is treated as
  molecular and gives rise to star formation. Then, SN feedback heats
  the remaining hot gas and also evaporates some of the cold gas,
  moving it to the hot phase, thus contrasting the cooling flow.

The two gas phases and the stellar component are clearly constrained to stay
within the same volume. However, in a realistic case these two phases should
not respond in the same way to pressure forces, because the hot phase tends to
flow away while the cold phase is only partially dragged.  In other words,
describing a star-forming, multi-phase ISM as a single particle cannot be
valid over a long time.  For this reason, we decide to use a single
time-scale to regulate both star formation rate of a multiphase particle and
the duration of the multi-phase stage.  

We use the dynamical time of the cold phase (equation~\ref{eq:tdyn}), but
  we keep it fixed when, for the first time since the start of the multi-phase
  stage,
  the mass fraction in the cold phase exceeds 95 per cent of the total
  mass. Cooling of the hot phase is very fast for many MP particles and the
  condition of 95 per cent of cooled mass is usually met within the first
  MUPPI integration. On the other hand, particles that enter in the multi-phase stage
  with temperature $T\la10^4$ K, i.e. when all hydrogen is neutral and no
  effective coolant is present, have to wait until the temperature rises (by
  compression of by feedback from other star-forming particles) to start
  depositing mass into the cold phase.

This dynamical time $t_{\rm dyn}$ thus regulates both star formation and the
exit from the multi-phase stage.  We adopt M04 suggestion fixing $t_{\rm clock} = 2
t_{\rm dyn}$; this is the typical time after which the molecular cloud that
gives rise to star formation is destroyed.  In this way, the particle exits
from the multi-phase stage after the time $t_{\rm clock}$ has elapsed since the
"freezing" of the dynamical time discussed above.  Moreover, the particle also
exits the multi-phase stage whenever its SPH density drops below $n_{\rm out}
= (2/3)n_{\rm thr}$. Usually, a multi-phase stage lasts for tens or up to hundreds of
simulation time-steps.  We evaluate our exit conditions before the multi-phase
integration takes place. Therefore, when a particle exits its multi-phase stage, it
maintain its SPH entropy as given by the SPH calculation.  

Within each SPH time-step, MUPPI is implemented in the GADGET-2 flow as
follows: {\em (i)} SPH densities $\rho$ of gas particles are evaluated in
the standard way, along with gravitational and hydrodynamical forces; {\em
  (ii)} MUPPI integration is performed for single-phase particles that match
the entrance criterion and for multi-phase particles that did not match the
exit criterion at the previous time-step; {\em (iii)} stochastic
star formation on multi-phase particles is performed; {\em (iv)} SN energy is
distributed to neighboring particles.

More in detail, MUPPI integration in the above step {\em (ii)}
proceeds as follows.  For each particle, the increase in entropy $dS$
resulting from the computation of hydrodynamic forces, is translated
into an energy increment $dE_{\rm hydro}= E_{\rm new} - E_{\rm
    ave,old}$. Note that such an energy increment includes the effect
  of an adiabatic expansion or compression, if the density of the
  particle has changed since the previous time-step.  We use the {\it
    new} SPH density to convert entropy to energy at the beginning of
  our MUPPI integration, so the difference with the previous value of
  the particle average energy $E_{\rm ave,old}$ includes the $PdV$
  work. The energy $E_{\rm new}$ also includes contribution due to SNe
  feedback coming from neighboring particles (see
  Section~\ref{section:redistrib}).

Then, the energy flux $\dot{E}_{\rm hydro}$ is calculated, by dividing the
energy increment by the duration of the SPH time-step.  This energy is
gradually injected into (or extracted from) the hot phase during the
integration (equation~\ref{eq:eh}). Since $\dot{E}_{\rm hydro}$ includes
  the effect of the entropy change $\Delta S$ due to hydrodynamics, $\Delta S$
  itself is set to zero. 

The evolution of properties of the ISM represented by the MP gas
  particle is then calculated by integrating equations \ref{eq:sf},
  \ref{eq:mc}, \ref{eq:mh} and \ref{eq:eh} for the duration of the SPH
  time-step. The average density of a MP particle never changes during
  a multi-phase integration, while the cold and hot number densities,
  $n_c$ and $n_h$, evolve according to our system of ordinary differential equations.

At the end of the integration, the updated entropy of the particle is
re-assigned based on the final mass-averaged thermal energy, $E_{\rm ave,new}$,
and on the density, $\rho$, of the same particle:

\be
 S_{\rm new} = (\gamma - 1) \frac{E_{\rm ave,new}}{\rho_{\rm ave,new}^{\gamma-1}}\,.
\ee

\noindent
As a consequence of the pressure equilibrium hypothesis, the
mass-averaged values used to compute the final energy and entropy
provide a particle pressure which is equal to that of the hot and cold
MUPPI phases.  This allows the particle to respond hydrodynamically to
the energy injection and pressurization caused by stellar feedback.
At the same time, cooling of the MP particles is performed by
MUPPI on the basis of the hot phase density. In this way the injected
energy is not quickly radiated away since the hot phase density is
much lower than the average density.  The energy $E_{\rm
    ave,new}$ will be used at the beginning of the next multi-phase integration
  to estimate the term $dE_{\rm hydro}$.

At relatively low pressure, when the molecular fraction and thus the
star formation rate is low, it can happen that the cooling caused by
an expansion (and driven by $\dot{E}_{\rm hydro}$ in the energy
equation) leads to a catastrophic loss of hot phase thermal energy.
In this case the energy input from SNe is too low to substain a
multi-phase ISM, so we simply force the particle to exit the
multi-phase regime.

Finally, it can happen that a gas particle is supposed to spawn a star
  (see Section~\ref{section:spawn} for details), but its mass in stars and
  cold gas is less than the mass of the star to be spawned.  In this rather
  unlikely event we force the gas particle to exit the multi-phase stage. It is the
  only case in which the exit happens after an multi-phase integration; the SPH entropy
  is set to $S_{\rm new}$.

\subsection{Stochastic star formation}
\label{section:spawn}

Star formation is implemented with the stochastic algorithm of
\cite{SpringHern03}.  In an SPH time-step, MUPPI transforms a mass
$\Delta M_\star$ of gas into stars\footnote{We compute this quantity
  {\it after restoration}, so the spawned stars are an ``old''
  population.}.
If $M_p=M_{\rm h}+M_{\rm c}+M_\star$ is the total particle mass, then a star
particle of mass $M_{p\star}$ is spawned by the gas particle with
probability:

\be 
P=\frac{M_p}{M_{p\star}}  \left[1 - \exp\left(-\frac{\Delta M_\star}{M_p}\right)\right]
\ee

\noindent
The mass of the spawned star is set as a fraction $1/N$ of the initial
mass of gas particles, so as to have $N$ generations of stars per gas
particle. In the following we use $N=4$ as a reasonable compromise
between the needs of providing a continuous description of star
formation and of preventing profileration of low-mass star
particles. The mass of the spawned star particle is taken from the
mass $M_\star$ of the stellar component of the MP particle
and, if this is insufficient, from the cold phase.  If the spawning
consumes also all of the cold phase, then the particle exits the
multi-phase regime.

Whenever a particle exits the multi-phase regime, its accumulated
stellar component, that has not been used to spawn a star, is
nulled. This is consistent with stochastic star formation: the
sub-resolution model determines the probability of spawning a star, but this
probability is connected with the amount of stars formed in an SPH
time-step and not with the stellar mass accumulated in a MP
gas particle.

\subsection{Redistribution of energy to neighboring particles}
\label{section:redistrib}

As already discussed in Section 2.2, only a small fraction, $f_{\rm
  fb,local}$, of the energy produced by SNe during a MUPPI integration is
deposited in the local hot phase. The rest of this energy is made
available to be distributed to neighboring particles along the
least-resistance path, as described in the following.

The total thermal energy flowing out of a MP particle 
is a fraction $f_{\rm fb,out}$ of the total budget:

\be 
\Delta E_{\rm heat,o} = E_{\rm SN} \cdot f_{\rm fb,out}\cdot \frac{\Delta
  M_\star}{M_{\rm\star,SN}}\,. 
\label{eq:en}
\ee 

\noindent
Here $\Delta M_\star$ is the mass in stars formed within the timestep,
computed {\it before restoration} (massive stars are the ones that
give rise to stellar feedback), while, as in eq.~\ref{eq:esn}, $E_{\rm SN}$ is the energy
released by one supernova and $M_{\star,SN}$ the mass of formed
stars per SN.

\begin{figure}
\centerline{
\includegraphics[width=\linewidth]{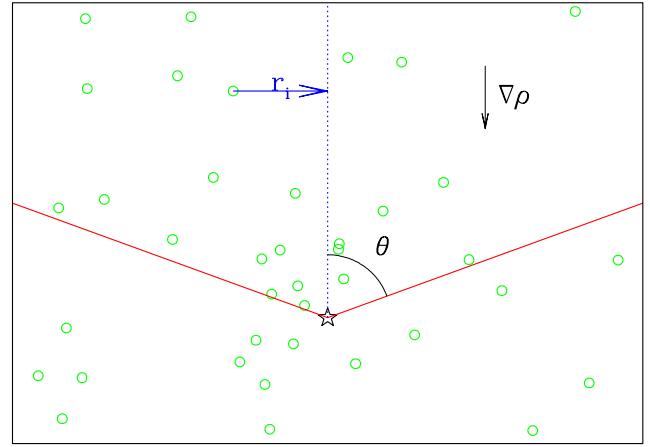}}
\caption{Schematic illustration of the energy redistribution mechanism.
The star indicates the particle that emits energy over a cone with
aperture angle $\theta$ and aligned with minus the density gradient.
$r_i$ is the distance of the $i$-th particle from the cone axis.}
\label{fig:ered}
\end{figure}

Consistently with the SPH formalism, we define neighboring particles as those
lying within the sphere of radius given by the MP particle smoothing
length $h$. For each SN explosion event, the maximum number of particles that
can receive the SN energy is fixed by the number of SPH
neighbors \footnote{Since particle mass varies in the current version of the
  code, the smoothing length is defined as the radius of a sphere including
  the kernel-weighted mass corresponding to 32 times the initial mass of a gas
  particle.  Since we do not allow the smoothing kernel to be less than $1/2$
  of the gravitational softening, MP particles turn out to have
  typically more than 32 neighbors.}.  Among the SPH neighbors, we select
those lying within the semi-cone with vertex at the position of the
MP particle, axis aligned with (minus) the direction of the local
density gradient $-\vec{\nabla}\rho$ and aperture $\theta= 60^\circ$ (see
Fig.~\ref{fig:ered}).

We then distribute the energy $\Delta E_{\rm heat,o}$ to all
neighbors, by weighting the amount of energy assigned to each
particle lying
within the cone according to its distance $r_i$ from the
axis.  
Accordingly, the SN energy fraction received by a neighbor $i$ is: 

\be
\Delta E_i = \frac{m_i \cdot W(r_i,h_{\star})K \Delta
E_{\rm heat,o}}{\rho_i}\,. \label{ered_eq} 
\ee 

\noindent
Here, $h_{\star}$ is the SPH smoothing length of the particle which
distribute energy and $K$ a normalization constant to guarantee that
$\sum_i \Delta E_i = \Delta E_{\rm heat,o}$.

The redistributed energy is recast in terms of entropy.
If a receiving particle is multi-phase,
the redistributed energy enters the $\dot{E}_{\rm hydro}$ term of the next
time-step. Otherwise, it will be added 
to the entropy variation due to hydrodynamics.  Note that we use the
same SPH kernel $W$ used for the hydrodynamical calculations, so that
particles farther from the axis of the cone receive an energy fraction
which is significantly lower than the ones lying closer to it. 
The influence of energy distribution on simulated galaxy properties will be 
discussed in Section~\ref{section:parameters}.

We choose this scheme of energy re-distribution to mimic the ejection
of energy by blowing-out super-bubbles along the least resistance
path. As mentioned above, we do not attempt to model a transfer of hot
gas between particles. If powerful enough, this energy ejection will
drive a gas outflow, which will then be resolved in our simulations,
and not treated as a sub-resolution event.

With this scheme, each particle keeps a small part of its SN energy
and gives another part to its neighbors.  Thus, as already mentioned
in Section~\ref{section:model}, the feedback blow-out and
pressure-confinement regimes of the M04 model may take place
depending on the actual spatial distribution of particles. In the
following we present numerical tests showing the effect of energy
redistribution in cases characterized by very different geometries,
like a thin disc or a spherical cooling flow.

While the computational cost of the integration of our ordinary
  differential equations system
  is small, of the order of 5 per cent of the total cost of a typical
  simulation, the cost of redistribution between different processors
  in a parallel run is higher. We need two communication rounds, one
  for calculating the normalization constant $K$, appearing in
  equation \ref{ered_eq}, for each particle and another one to
  redistribute energy. Each round has a CPU cost comparable to that of
  the SPH communication round. The total overhead is of about 25 per
  cent of the total CPU cost in a typical simulation. 


\section{Results}
\label{section:results}
The implementation of MUPPI in the GADGET-2 code has been extensively
tested by running it on a suite of initial conditions and using different
combinations of parameters. The basic properties of the structures
that we simulated are described in Tables~\ref{table:runs} and ~\ref{table:ic}. They are:
{\em (i)} an isolated Milky Way-like galaxy (MW); {\em (ii)} an
isolated low-surface brightness dwarf galaxy (DW); {\em (iii)} two isolated spherical,
non-rotating halos, with masses similar to MW and DW, where gas,
initially sitting in hydrostatic equilibrium within the gravitational
potential well, generates a cooling flow (CFMW and CFDW,
respectively).
Initial conditions for the simulations are
described below in Section~\ref{section:ic}.

\begin{table*}
  \caption{Basic characteristics of the different runs. Column 1: simulation
    name; Column 2: mass of the DM particles; Column 3:
    mass of the gas particles; Column 4: mass of the star (bulge and
    disk) particles in
    the ICs (only present in DW and MW galaxies);
    Column 5: number of DM
    particles within the virial radius; Column 6: number of gas particles within the
    virial radius; Column 7: number of star particles within the virial
    radius; Column 8: Plummer-equivalent softening length for
    gravitational forces.
    Masses are expressed in units of M$_\odot$ and  
    softening lengths in units of kpc.   
  }
\begin{tabular}{c c c c c c c c}
\hline\hline Simulation & $M_{\rm DM}$ & $M_{\rm gas}$& $M_{\rm star}$ & $
N_{\rm DM}$ & $N_{\rm gas}$ & $N_{\rm star}$ & $\epsilon_{\rm Pl}$ \\
\hline
MW & $3.6 \cdot 10^6$ & $7.6 \cdot 10^4$ & $1.4 \cdot 10^6$ & 300000 & 50000 & 50000 & 0.71 \\
DW & $8.3 \cdot 10^5$ & $4.0 \cdot 10^4$ & $1.6 \cdot 10^5$ & 300000 & 50000 & 50000 & 0.43 \\
CDMW & $3.6 \cdot 10^6$ & $4.7 \cdot 10^5$ &  0 & 300000 & 50000 & 0 & 0.71 \\
CFDW & $8.3 \cdot 10^5$ & $2.0 \cdot 10^5$ & 0 & 300000 & 50000 & 0 & 0.43 \\
\hline
\end{tabular}
\label{table:runs}
\end{table*}

\begin{table*}
  \caption{Parameters of MW and DW isolated galaxies. Column 1: simulation
    name; Column 2: DM halo virial mass; Column 3: spin parameter;
    Column 4: disk mass (stars and
    gas); Column 5: bulge mass; Column 6: DM halo virial radius;
    Column 7: disk radial scale length; 
    Column 8: bulge radius; Column 9: gas fraction inside the disk. Masses
    and radii are expressed in units of M$_\odot$ and of kpc, 
    respectively. We estimate virial quantities at a (formal)
    overdensity of 200 times the critical density of the Universe, while
    the spin
    parameters are computed as in \citet{Bullock01}. }
\begin{tabular}{c c c c c c c c c}
\hline\hline Simulation & $M_{\rm halo}$ & $\lambda$ &  $M_{\rm disk}$
& $M_{\rm bulge}$ & $r_{\rm halo}$ &
$r_{\rm disk}$ & $r_{\rm bulge}$ & $f_{\rm gas}$  \\
\hline
MW & $9.4 \cdot 10^{11}$ & 0.04 & $3.7 \cdot 10^{10}$ & $9.4 \cdot 10^9$ & $197$ & $2.9$ &
$0.58$ & $0.1$ \\
DW & $1.6 \cdot 10^{11}$ & 0.04 & $1.0 \cdot 10^{10}$    &       $0$  & $80$  & $3.5$ & $0$ &
$0.2$ \\
\hline
\end{tabular}
\label{table:ic}
\end{table*}

We first present results from the MW simulation, which is carried
out using the reference choices for the MUPPI parameters, as reported
in Table~\ref{table:params}, that will be justified in
Section~\ref{section:parameters}. In Section~\ref{section:ism} we 
describe the evolution of two MP particles residing in
different regions of the disc, so as to show the behavior of the
sub-resolution variables as a function of time. In
Section~\ref{section:mw} we address the global properties of the MW
galaxy and in Section~\ref{section:dw} those of the DW galaxy.
In Section~\ref{section:cf} we use the CFMW and CFDW simulations
to demonstrate that the effect of feedback depends on the geometry of
the system. In Section~\ref{section:parameters} we discuss the
sensitivity of the final results on the choice of the MUPPI
parameters, and the procedure adopted to fix such parameters. Finally,
we show  the resolution tests for
the MW in Section~\ref{section:resolution}.

\begin{table*}
\caption{Standard values for MUPPI parameters.}
\begin{center}
\begin{tabular}{|cccccccccccccc|}
 $f_\star$ & $P_{0}$ (K cm$^{-3}$) & $T_{\rm c}$ (K) & $f_{\rm fb,out}$ & $f_{\rm fb,local}$ &  $f_{\rm ev}$ & $\theta$ ($^\circ$)  &$f_{\rm re}$
& $\frac{E_{\rm SN}\ ({\rm erg})}{M_{\star,\rm SN}\ ({\rm M}_\odot)}$ & $n_{\rm thr}$ (cm$^{-3}$)& $T_{\rm thr}$ (K) & $\frac{n_{\rm out}}{n_{\rm thr}}$ & $\frac{t_{\rm clock}}{t_{\rm dyn}} $\\ \hline
0.02 & 35000 & 1000 & 0.3 & 0.02 & 0.1 & 60 & 0.2 & $10^{51}/120$ & 0.01 & 50000 & 2/3 & 2 
\end{tabular}
\end{center}
\label{table:params}
\end{table*}

\subsection{Initial Conditions (ICs)}
\label{section:ic}

\subsubsection{Isolated galaxy models (MW, DW)} 
IC for these simulations have been generated following the procedure
described in \cite{GADGET2} and were kindly provided by S. Callegari
and L. Mayer. They are near-equilibrium distributions of particles
consisting of a rotationally supported disc of gas and stars\footnote
{These star (collisionless) particles are not related to the newly
  formed stars that are generated by the code during the
  evolution. They are more massive than the latters.}
and a
dark matter halo. For the MW only, also a stellar bulge component is
included. Bulge and halo components are modeled as spheres with
\cite{Hernquist90} profiles, while the gaseous and stellar discs are
modeled with exponential surface density profiles. The values of the
relevant parameters describing the MW and DW galaxies
are reported in Table~\ref{table:ic}. To make sure that we start from
a relaxed and stable configuration, we first evolve the two galaxy
models for 10 dynamical times with non-radiative hydrodynamics. 
We then use the configurations
evolved after 4 dynamical times as initial conditions for our MUPPI
simulations.
To perform a resolution study of the model, 
for the MW galaxy we used higher and lower resolution ICs, as
discussed later in Section~\ref{section:resolution}.

\subsubsection{Isolated halos (CFMW, CFDW)}
The procedure to generate the initial conditions for the isolated
non-rotating halos is described in detail by \cite{Viola08}. We used
DM halos with an NFW \citep{Navarro96} profile, with hot gas in
hydrostatic equilibrium within the halo potential well. Gas thermal
energy is fixed, by following the prescription by \cite{Komatsu01},
from the requirement that gas and dark matter density profiles have
the same logarithmic slope at the virial radius.  As for the MW and DW
models, we evolved the systems without cooling and star formation for
10 dynamical times, so as to let them relax. The resulting
configurations are then taken as initial conditions for our
simulations. DM mass, gas mass and virial radius for these haloes
are the same as for the corresponding MW and DW haloes, as reported in
Table~\ref{table:ic}. We set the NFW concentration for MW and DW haloes
  to $C_{\rm \Delta=100}=13$ and $20$ respectively. 

\begin{figure}
\centerline{
\includegraphics[width=0.5\linewidth,angle=-90]{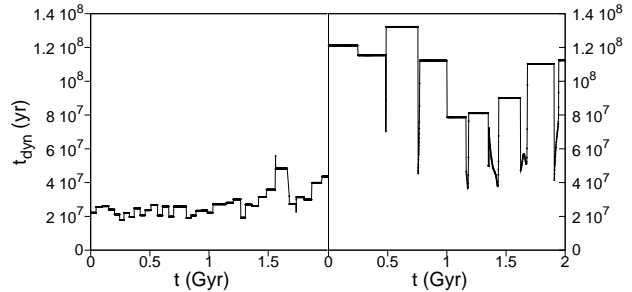}}
\caption{Evolution of the dynamical time of the cold phase, kept
  frozen after 95 per cent of mass has cooled, for two gas particles
  selected (see text) at distances $\sim2$ kpc and $\sim8$ kpc from
  the center of the isolated MW simulation (left and right panel,
  respectively).  The time span of the figures is 2 Gyr after $\sim1.3$ Gyr.  
}
\label{fig:tdyn}
\end{figure}

\begin{figure*}
\centerline{
\includegraphics[width=0.8\linewidth,angle=-90]{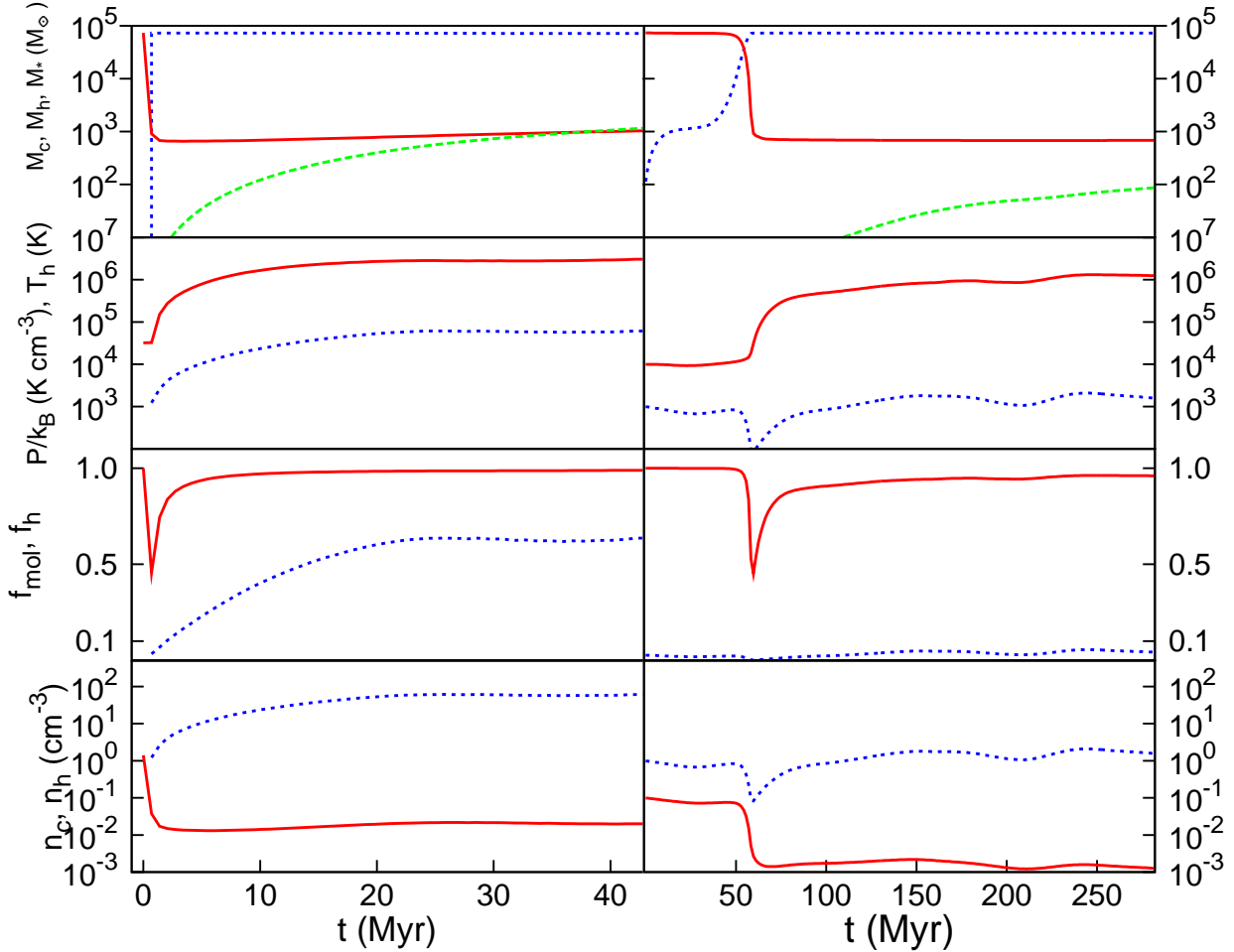}}
\caption{Evolution of ISM variables for the same two MW particles
  shown in Fig.~\protect\ref{fig:tdyn} during one single multi-phase
  stage. Left panels: bulge particle, right panels: disk particle.
  From top to bottom: hot, cold and stellar mass gas ($M_{\rm h}$, red
  continuous line, $M_{\rm c}$, blue short-dashed line, $M_\star$,
  green long-dashed line); pressure $P/k_B$ (blue dashed line) and hot
  phase temperature $T_{\rm h}$ (red continuous line); molecular
  fraction $f_{\rm mol}$ (blue dashed line) and hot phase filling
  factor $f_{\rm h}$ (red continuous line); particle number densities
  of cold and hot phases ($n_{\rm c}$: blue dashed line; $n_{\rm h}$:
  red continuous line).  The time is measured since the entrance of
  the particles in the multi-phase regime.}
\label{fig:ism}
\end{figure*}

\subsection{Evolution of multi-phase particles}
\label{section:ism}

A close look to the system of equations~\ref{eq:sf}-\ref{eq:mh} and
\ref{eq:eh} reveals that this system, if integrated in isolation
(constant average density and $\dot{E}_{\rm hydro}=0$), leads to a
runaway of the molecular fraction. In fact, stellar feedback increases
gas pressure, and this leads to higher molecular fraction which
further increases feedback. This runaway stops when the molecular
fraction saturates to unity.  This instability is interrupted when the
particle can exchange energy with neighbors through the hydro term
$\dot{E}_{\rm hydro}$.  In this case the particle starts expanding
when its pressure is higher than the one of its neighbors. This gives
origin to a negative $\dot{E}_{\rm hydro}$ term which cools the
particle.  For this reason the existence of a runaway does not imply
that all particles reach very high pressure, irrespective of their
environment.  This fact highlights the need of an explicit integration
of the system of equations, because assuming an equilibrium solution
would imply on the one hand some more strong assumptions on the
system (for instance, a simplified star formation law which is
independent of pressure), on the other hand the transient which leads
to a quasi-equilibrium solution through the pressurization of the
particle would be missed.

To illustrate the behavior of the ISM as described by our sub-resolution
model, we show the evolution of two MP gas particles. We use the MW
initial condition, with MUPPI parameters set to the fiducial values given in
Table~\ref{table:params}.  We choose two particles, whose initial positions
are located at $\sim 2$ kpc (bulge particle) and $\sim 8$ kpc (disk particle)
from the galaxy center, respectively.  Figure~\ref{fig:tdyn} shows how the
dynamical time of each particle, computed within each multi-phase stage,
changes during 2 Gyr of evolution, starting from $\sim1.3$ Gyr so as to
avoid the initial transient during which all the gas in the MW simulation
simultaneously enters in the multi-phase condition.  As explained in
Section~\ref{section:entrance}, the dynamical time is kept frozen after 95 per
cent of cold mass is accumulated.  We note that the dynamical times scatter
around values of $2\sim10^7$ yr and $\sim\times10^8$ yr, showing that the
bulge particle has a higher evolution rate.  We also note that in some cases
the dynamical time of the disc particle reaches a stable value only after some
tens or hundreds Myr.

Figure~\ref{fig:ism} shows for the same two particles (from top to bottom
panels) and for one specific multi-phase stage, the evolution of mass of the
three components ($M_{\rm c}$, $M_{\rm h}$ and $M_\star$), pressure $P/k_B$
and temperature of the hot phase $T_{\rm h}$, hot phase filling factor $f_{\rm
  h}$ and fraction of molecular gas in the cold phase $f_{\rm mol}$, particle
number densities $n_{\rm c}$ and $n_{\rm h}$ of the two gas phases.  Their
evolution is followed from their entrance in multi-phase to their exit, which
takes place after two dynamical times of the cold phase.  These two particles
have been chosen so as not to spawn a star particle before or during this
cycle. For the second particle, a cycle has been chosen in which deposition of
cold phase, and thus the computation of the cold phase dynamical time, does
not start immediately after entrance in multi-phase.

As for the bulge particle, although initial conditions are set with all mass
in the hot phase, most mass cools already during the first SPH time-step. The
disc particle instead enters the multi-phase regime at a relatively low
temperature and density. Therefore, due to the absence of any molecular
cooling in our simulations, it has to wait until it is heated (by compression
or by feedback) to high enough temperature to develop a significant cooling
flow and deposit mass to the cold gas phase. In this second case, the drop in
pressure visible after the onset of cooling is due to the decrease of the mean
particle temperature caused by the accumulation of cold phase.  Then, for both
particles energy from SN explosions pressurizes the gas, without causing a
pressure runaway, thanks to the hydrodynamical response of the particles to
this energy input. As a result, pressure increases by more than one order of
magnitude for the ``bulge'' particle, while the ``disc'' particle only
recovers from the drop in pressure due to cooling. This difference in pressure
reflects in very different molecular fractions. For both particles the hot
phase temperature settles in the range $10^6$--$10^7$ K with its filling
factor remaining high.  Number densities reach values of $<10^{-2}-10^{-3}$
for the hot phase in both cases, while the cold phase number density scales
with ISM pressure, being the temperature of the cold phase forced to be
constant at the value $T_{\rm c}=10^3$K.

After the first transient, the system settles into a quasi-equilibrium
state where cold gas is slowly turned into stars and ISM properties
evolve quite slowly.  At the end of the multi-phase stage, stellar
mass amounts only to $\sim1$ and $\sim0.1$ per cent of the total mass,
so the probability of spawning a star is low. The multi-phase stage
lasts for $\sim60$ and $\sim250$ SPH time-steps, in the two
cases. After exiting the multi-phase stage, both particles are still
in high density regions, so their multi-phase variables are reset and
MUPPI integration starts again. Other particles are pushed out of the
disc, reach densities below the threshold value for multi-phase,
become single-phase and later fall back in a galactic fountain. This
happens roughly in about 3 per cent of the cases.

\subsection{Global properties of the MW simulation}
\label{section:mw}

\begin{figure*}
\centerline{
\includegraphics[width=0.4\linewidth]{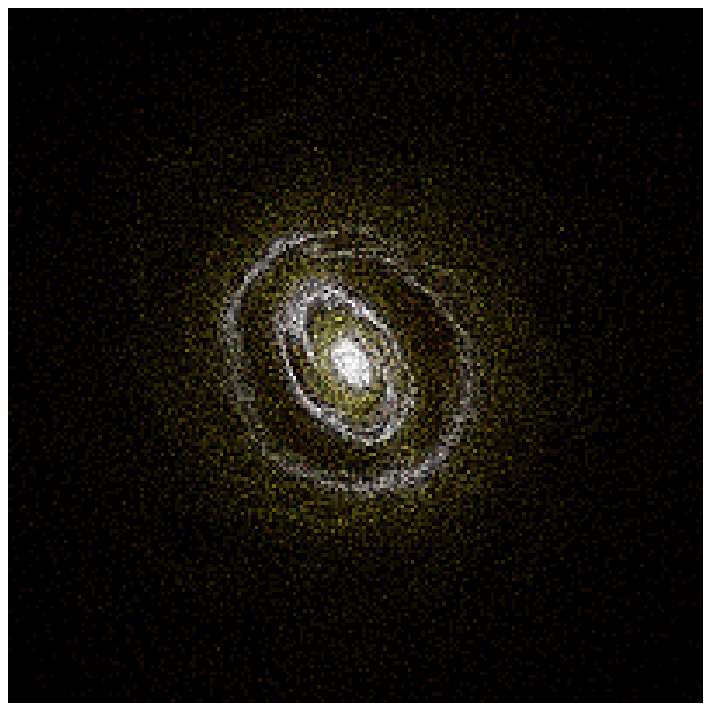}
\includegraphics[width=0.4\linewidth]{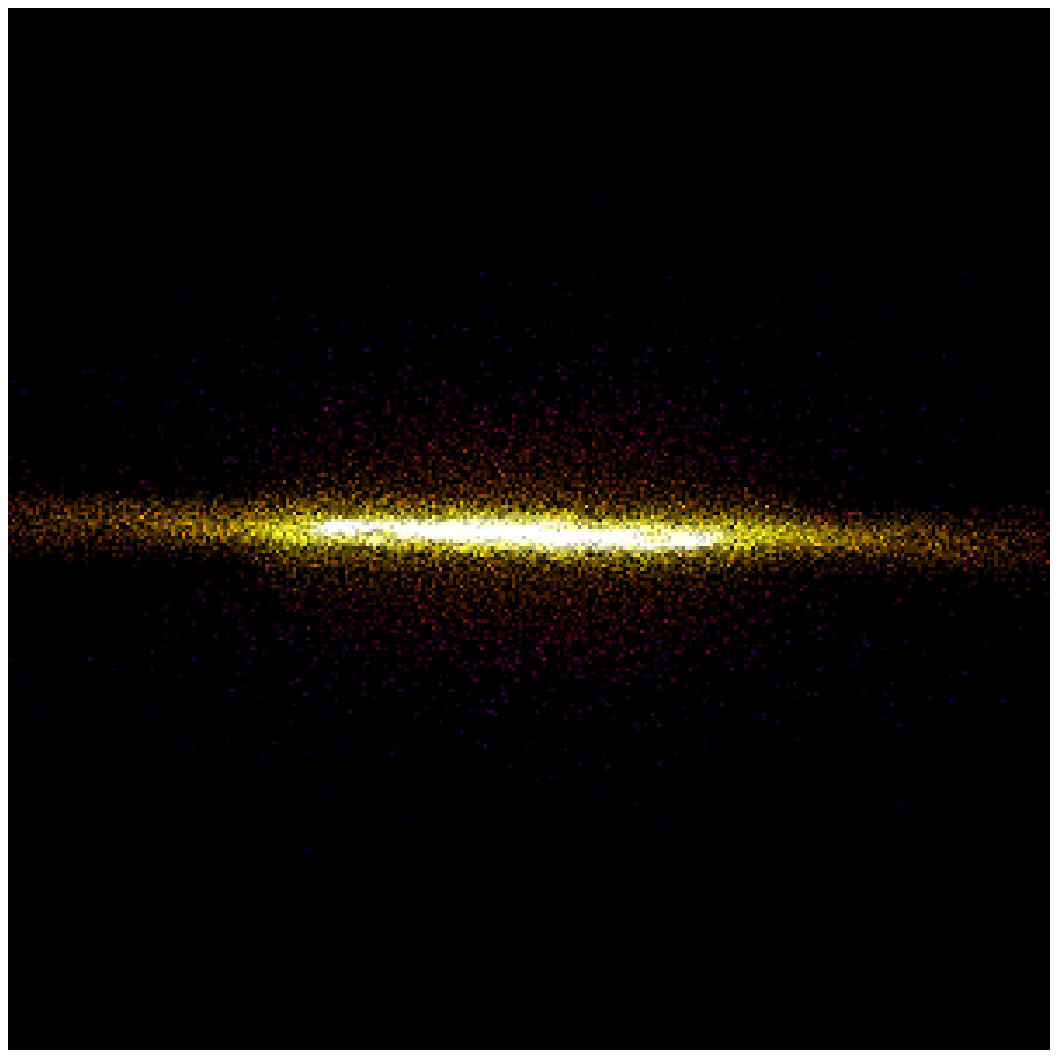}
}
\centerline{
\includegraphics[width=0.4\linewidth]{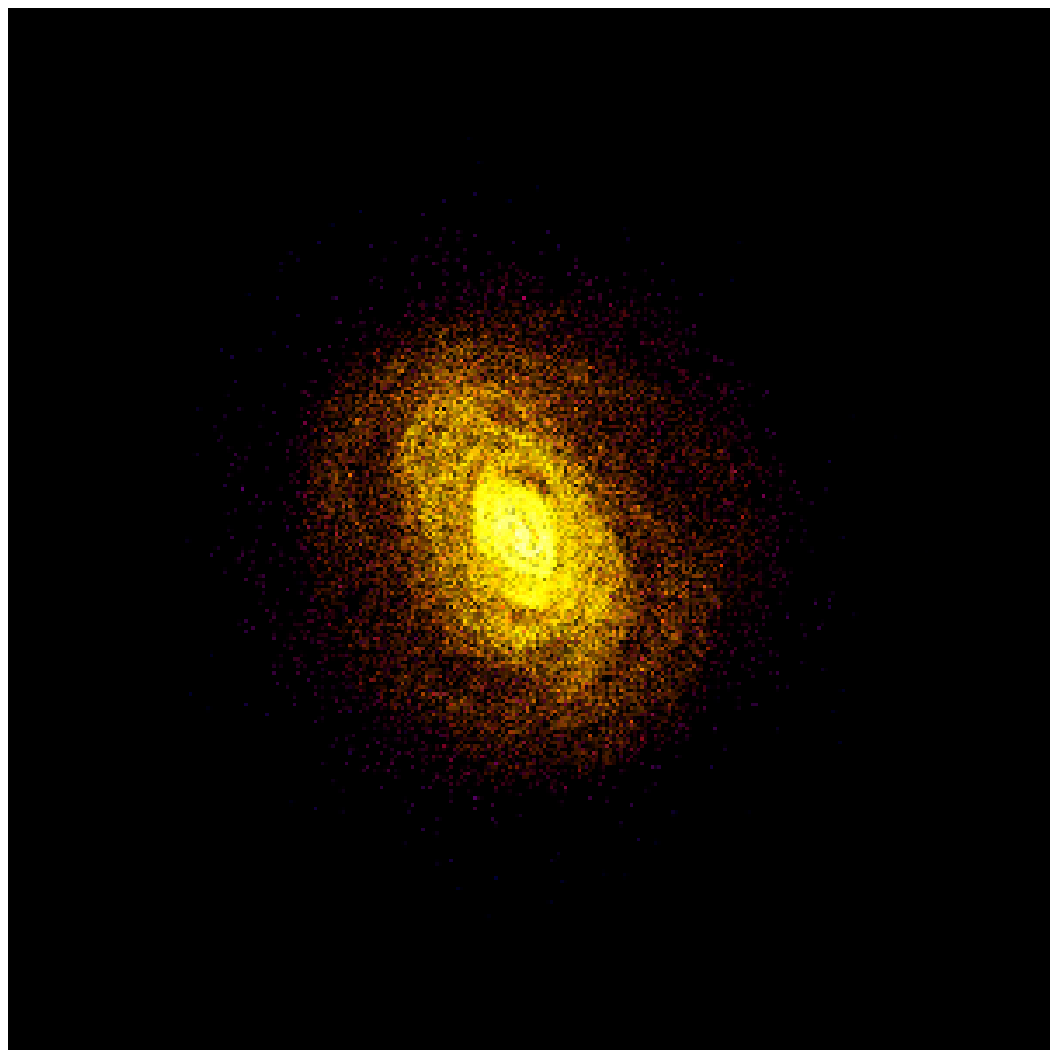}
\includegraphics[width=0.4\linewidth]{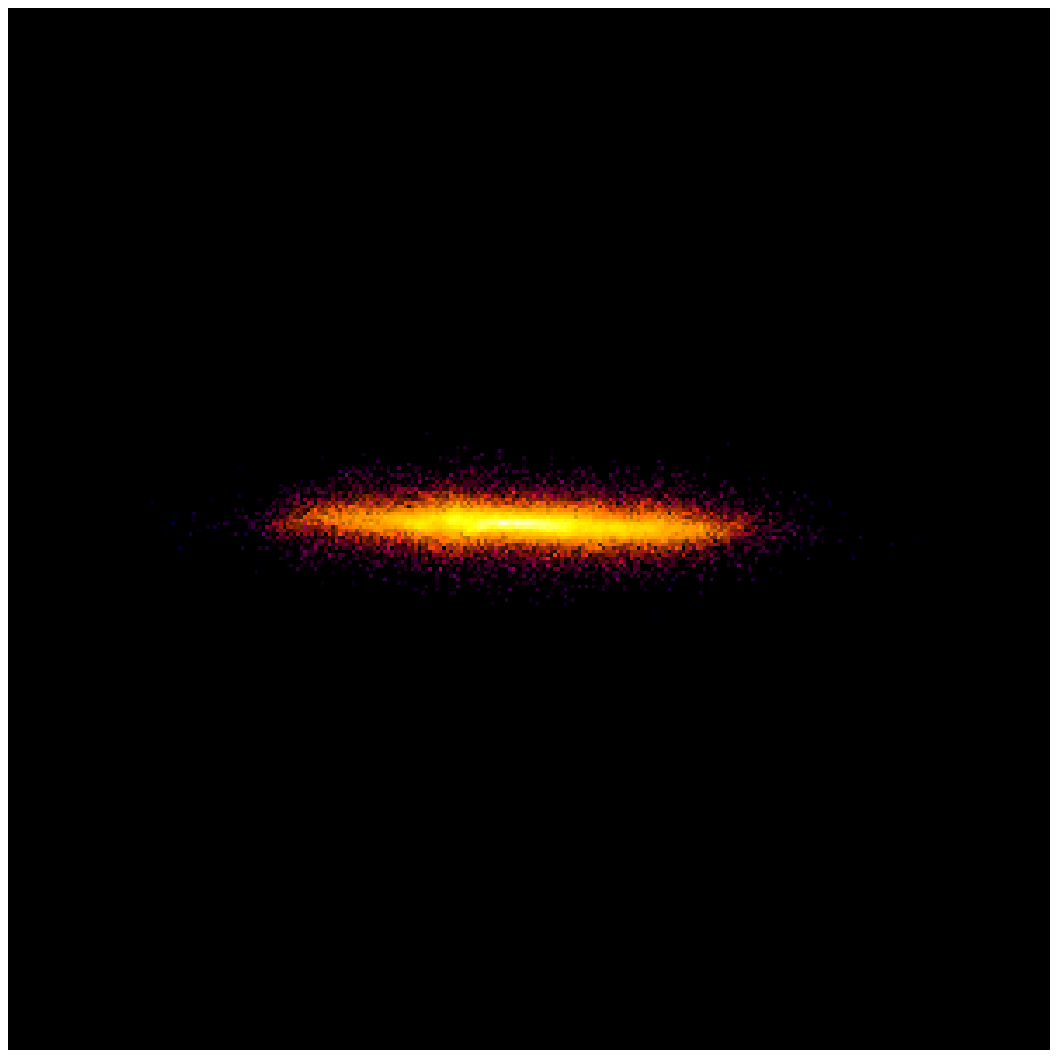}
}
\centerline{
\includegraphics[width=0.4\linewidth]{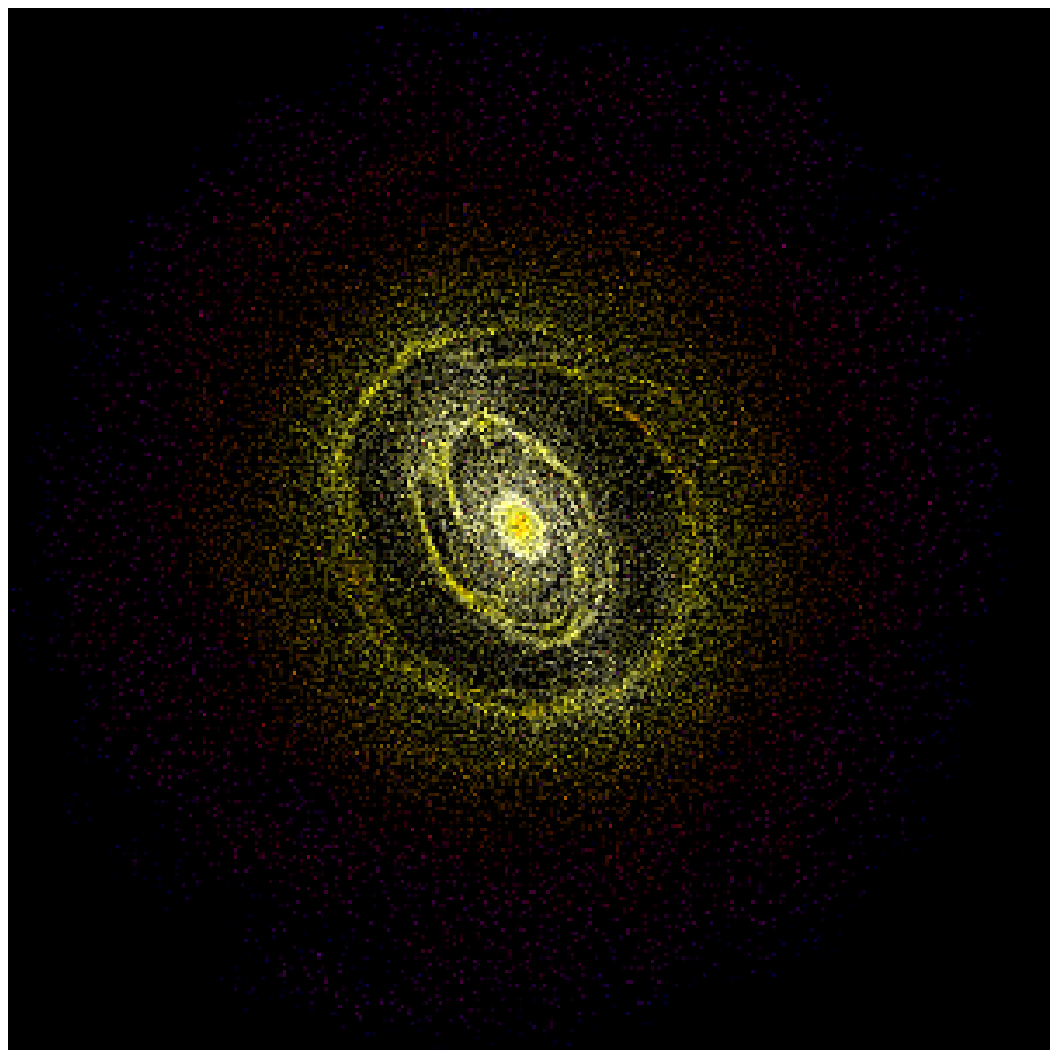}
\includegraphics[width=0.4\linewidth]{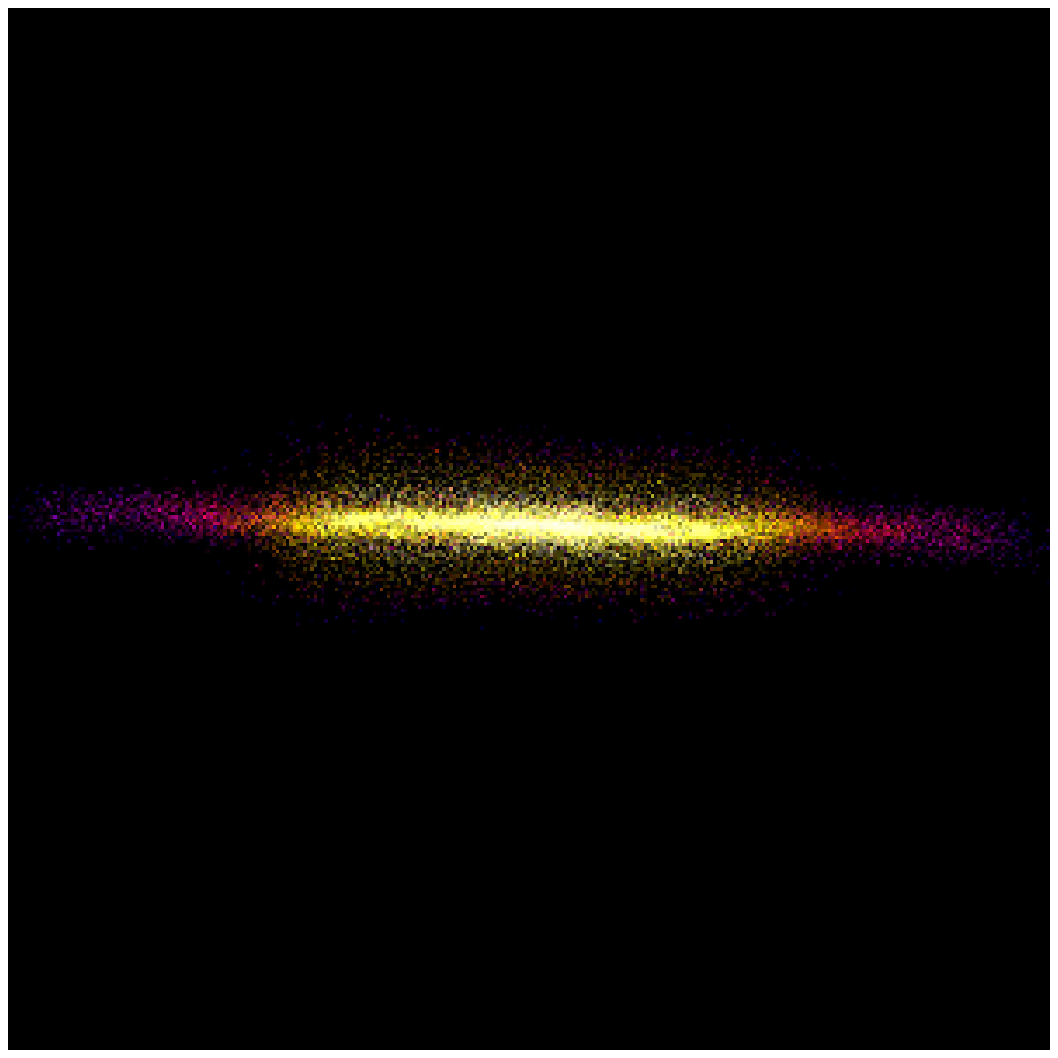}
}
\caption{ 
Distribution of gas particles (upper panels), star particles (mid panels) and
SFR (lower panels) for the MW simulation with standard parameters at the end
of the simulation. Particles are color coded by logarithm of their SPH density
(upper and middle panels), decreasing from white to yellow to red to
blue. Lower panels show again the distribution of gas particles, but the color
code is the logarithm of the star formation rate.  Left and right panels are
for the face-on and edge-on projections, respectively. The box size is 35
kpc.
}

\label{fig:mw}
\end{figure*}

The MW simulation has been evolved for 3 Gyr.  Figure~\ref{fig:mw} shows maps
of gas density (upper panels), stellar mass density (central panels) and SFR
(lower panels) of the MW at the end of the simulation, seen face-on and
edge-on (left and right panels, respectively). In this simulation, the star
formation and feedback scheme of MUPPI generates at the end of the simulation a
barred galaxy with a typical spiral-like pattern, consisting of a central
concentration, a star-forming circular ring at corotation, a second more
distant ring and an extended gaseous disc with low molecular fraction and star
formation.  We remark that such a pattern is less noticeable in the first Gyrs
of evolution. The edge-on view shows a characteristic of our scheme: feedback
leads to a modest heating of the gas disc.  We will discuss this point in more
detail below.

\begin{figure}
\centerline{
\includegraphics[width=\linewidth]{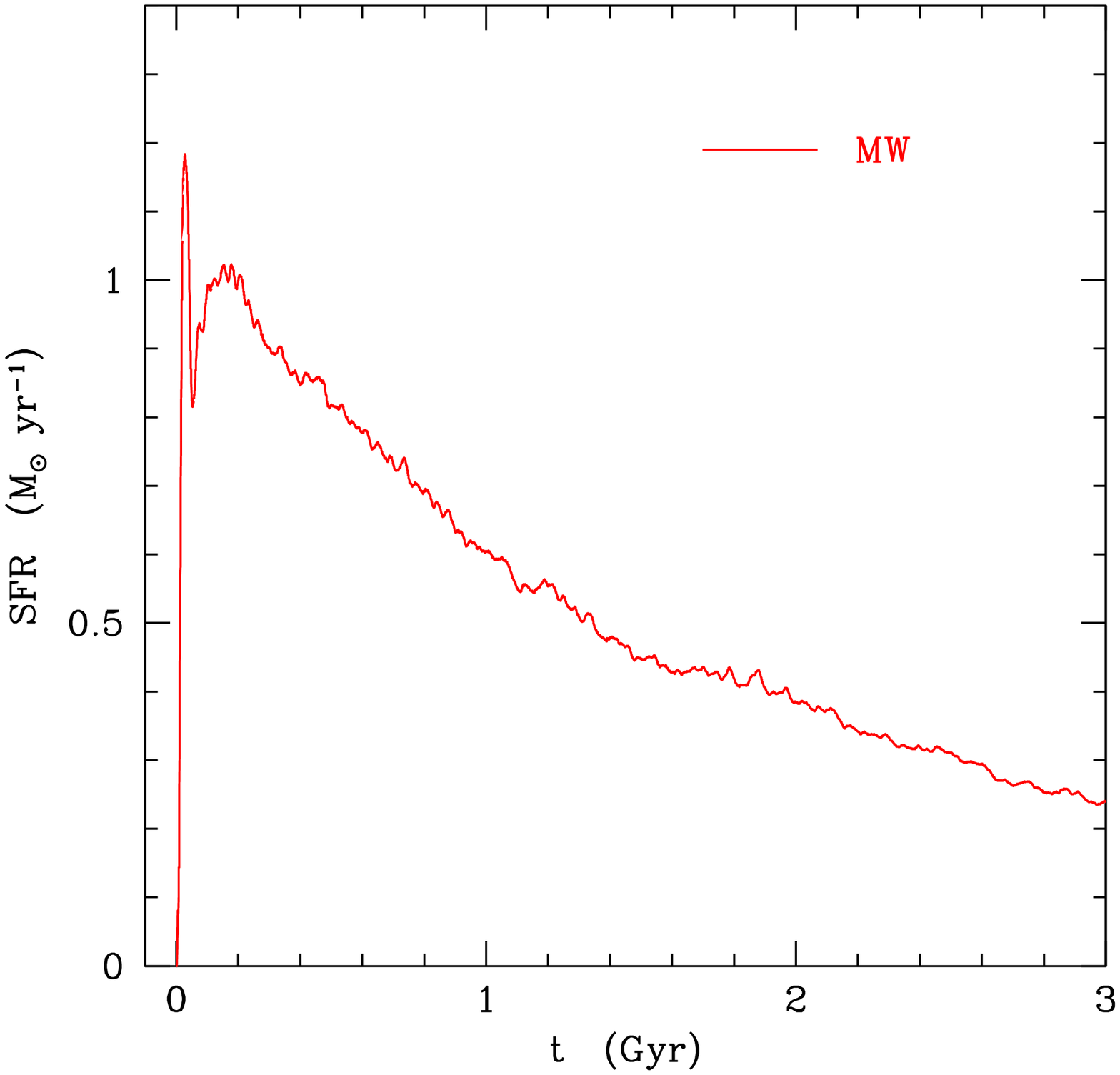}}
\caption{Star formation rate as a function of time for the MW
  simulation with standard parameters.}
\label{fig:sfr_mw}
\end{figure}

\begin{figure}
\centerline{
\includegraphics[width=\linewidth]{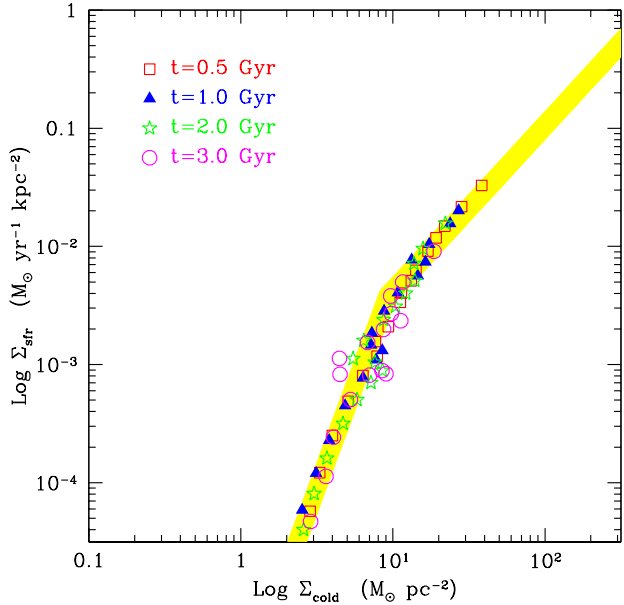}}
\caption{SK relation at four different times for
  the MW simulation with standard parameters, computed in concentric
  circles aligned with the galaxy disc. Red squares, blue triangles,
  green stars and magenta circles are for $t=0.5$, 1, 2 and 3 Gyr.
  The shaded yellow area gives
  a fit to the data of Kennicutt (1998) at high gas surface densities,
  with a break at 8 {\surf} to fit the data by Bigiel et al. (2008).}
\label{fig:ken_mw}
\end{figure}

\begin{figure}
\centerline{
\includegraphics[width=\linewidth]{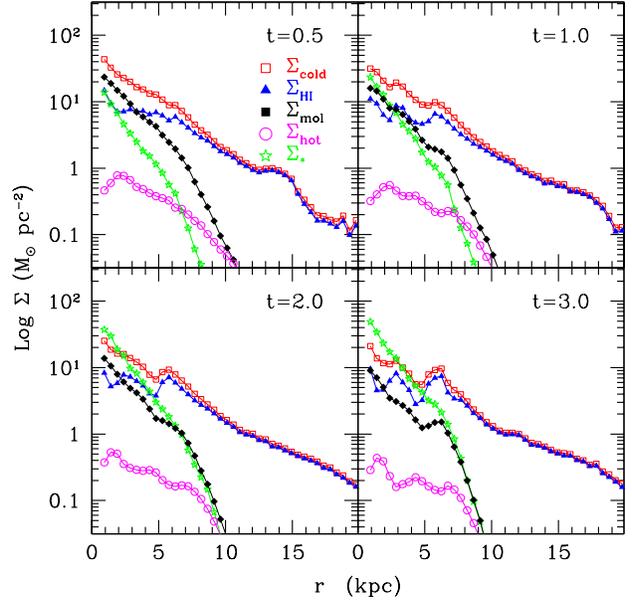}}
\caption{Surface density mass profiles of cold (red open squares),
  HI (blue triangles), molecular (black filled squares), hot gas
  (magenta circles) and (newly formed) stellar mass (green stars) for
  the MW simulation with standard parameters, at four different
  times as indicated in the panels.}
\label{fig:profili_mw}
\end{figure}

\begin{figure*}
\centerline{
\includegraphics[width=\linewidth]{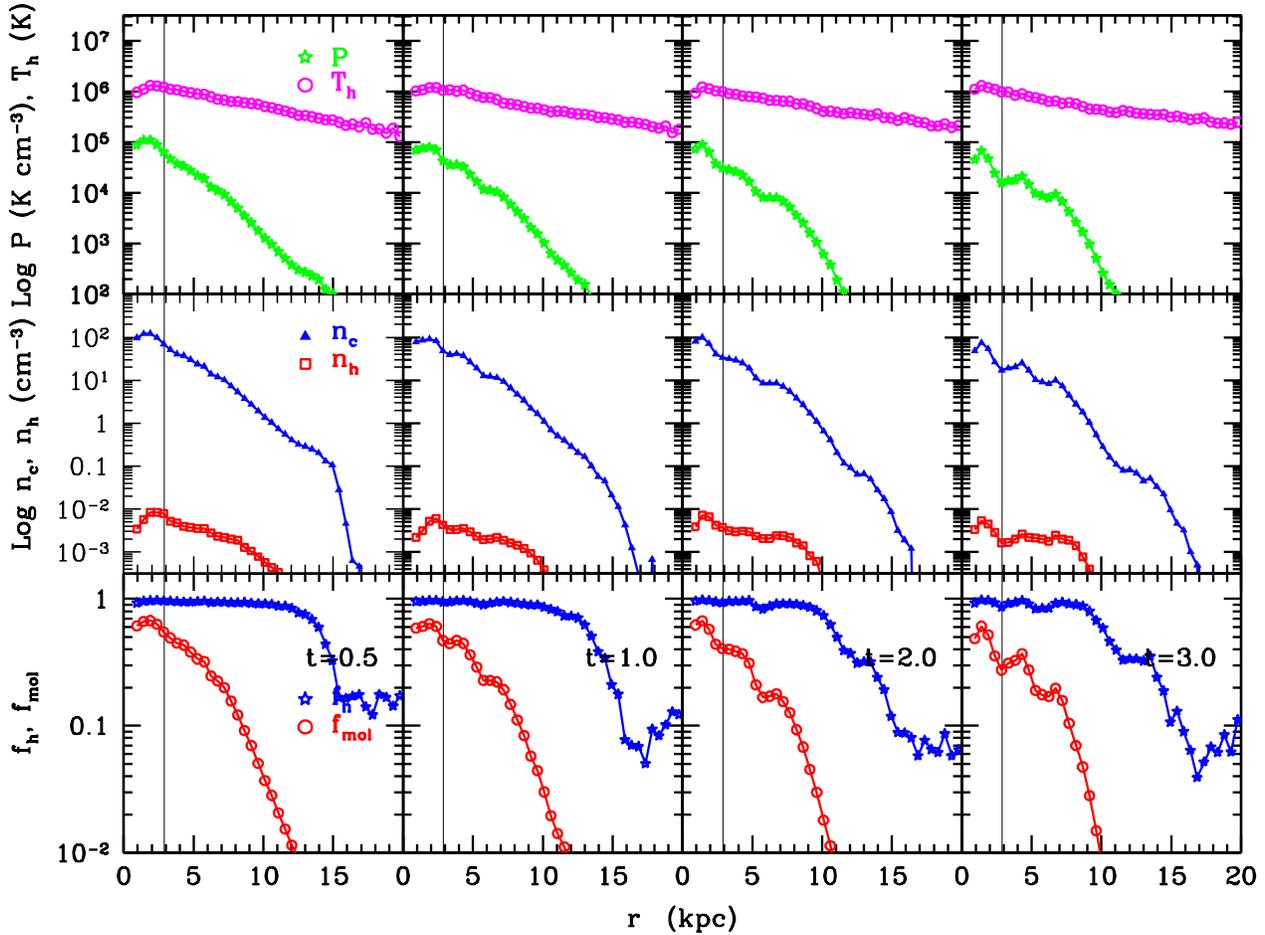}}
\vspace{-4.5truecm}
\caption{Average properties of the ISM as a function of radius for the
  MW simulation with standard parameters at four different times as
  indicated in the panels. Upper panels: pressure (green stars) and
  temperature (magenta circles) of the hot phase; central panels:
  number densities of the cold and hot phases (blue triangles and red
  squares, respectively); lower panels: fraction of hot and molecular
  gas (blue stars and red circles, respectively). The black vertical line
  marks the scale radius of our MW disk.} 
\label{fig:ism_mw}
\end{figure*}

Figure~\ref{fig:sfr_mw} shows the total star formation rate of the
galaxy. The first peak, reaching values of nearly 1.2 {\msunyr} and
lasting for a few tens of Myr, is a numerical transient due to the
fact that many particles are cold and satisfy the star formation
criterion already in the initial conditions. For this reason, they are
out of the equilibrium when MUPPI is switched on, so that they all
start forming stars at the same time, and their feedback tends to
quench star formation. The second, broader peak marks the
pressurization of the galaxy disk and the start of the
quasi-equilibrium star-forming phase.  The star formation rate then
declines on a time-scale of $\sim2$ Gyr, as expected for a disc which
obeys the SK relation \citep[see][]{Leroy08} and receives no mass
inflow from the outside.

Figure~\ref{fig:ken_mw} shows the comparison between the observed and the
predicted SK relation, i.e. the relation between surface density of star
formation rate, $\Sigma_{\rm sfr}$, and total surface gas density $\Sigma_{\rm
  cold}$, at four different times (0.5, 1, 2 and 3 Gyr). As we will discuss
extensively in Section~\ref{section:parameters}, this relation is actually
used to tune the parameters of our model. The interpretation of this relation
will be the subject of a forthcoming paper, while we focus here only on the
main numerical aspects of this comparison. The shaded area shows a double
power-law fit to observational data, with slope and normalization as proposed
by \cite{Kennicutt98} for surface densities higher than 8 {\surf}, and slope
3.5 for lower gas surface densities, roughly consistent with the break
reported by \cite{Bigiel08}.  The width of this area is compatible with the
error on the zero point quoted by \cite{Kennicutt98}, and should not be taken
as indicative of the total observational uncertainty, which is larger.
Observed gas surface densities, corresponding to $HI+H_2$, have been corrected
by a factor 1.31 to take into account the presence of helium.  Quite clearly,
the simulated relation is remarkably stable with time and agrees well with
the observational fit. The simulated SK relation at $t=3.0$ Gyr of evolution
shows a higher degree of scatter, caused by the development of the structure
in the gas distribution shown in Figure~\ref{fig:mw}.

Figure~\ref{fig:profili_mw} shows the mass surface densities of cold,
molecular, $HI$ (defined as $(1-f_{\rm mol})M_{\rm c}$) and hot gas, along
with the SFR surface density.  While $HI$ gas density flattens at $\sim8$
{\surf}, molecular and stellar surface densities have comparable scale
lengths. This is again in line with the observations of the THINGS group
\cite{Bigiel08,Leroy08}.  The hot gas density remains at $\sim1$ per cent of
that of cold gas (here and in the following we neglect ``hot'' gas in
particles that have not started to cool and pressurize). The pattern in the
gas distribution is here visible as bumps in the density profile, which
develop at later times ($t\ga1$ Gyr).

Figure~\ref{fig:ism_mw} shows the predicted (mass-weighted) average
values of ISM properties, namely pressure and hot phase temperature
(upper panels), number densities of cold and hot phases (central
panels), molecular fraction and hot phase filling factor (lower
panels), computed in cylindric volumes around the galaxy center of
mass, as a function of radius. All profiles are computed only over
MP particles. We note that pressure and hot phase temperature
both follow an exponential profile, but the latter quantity is
remarkably flatter. The same trend is noticeable also for the cold and
hot gas number densities.
The molecular fraction is high at the galaxy center but quickly
  decreases beyond $\sim$6 kpc, i.e. where the gas surface density
  drops below 10 {\surf}, while the hot phase filling factor is very
  high throughout the region where star formation is present.

\begin{figure}
\centerline{
\hspace{-1.3truecm}
\includegraphics[height=0.6\linewidth]{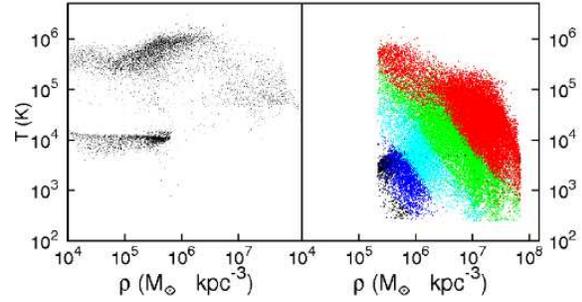}}
\caption{Phase diagram for particles in the MW simulation with
  standard parameters, after 2 Gyr.  Left and right panels show
  respectively single- and MP particles. The latters are
  color-coded according to their SFR as follows (all SFRs being in
  \msunyr). Black: $[0,10^{-8})$; blue: $[10^{-8},10^{-7})$; cyan:
      $[10^{-7},10^{-6})$; green: $[10^{-6},10^{-5})$; red:
          $>10^{-5}$.}
\label{fig:phase}
\end{figure}

Figure~\ref{fig:phase} shows the density--temperature phase diagram
for single-phase (left panel) and MP (right panel) gas
particles after 2 Gyr of evolution. Mass-averaged temperature and SPH
gas density are shown for the MP particles.  At low
densities, most single-phase particles lie in two sequences at
temperature $\sim10^4$ and $\sim5\times10^5$ K respectively. They
correspond to cold particles lying on the disc and to particles heated
by thermal feedback. Some single-phase particles reside above the density
threshold. These are particles that have exited the multi-phase stage
at high average temperature and can not re-enter it immediately after.
 The multi-phase sequence starts with a drop in
temperature, which corresponds to the initial accumulation of cold
mass lowering the average temperature. The bump at $T\sim10^5$ K
corresponds to the pressurized and star-forming particles with
$T_{\rm h}\sim10^6-10^7$ K and hot mass fraction of $\sim1$ per cent. Two
denser regions in the MP particle phase diagram, at densities
$\rho \sim 10^7$ and $\sim 3\cdot10^7$ M$_\odot$ kpc$^{-3}$
corresponds to regions of the galaxy where density is enhanced
by spiral arms.

\begin{figure}
\centerline{
\includegraphics[width=\linewidth]{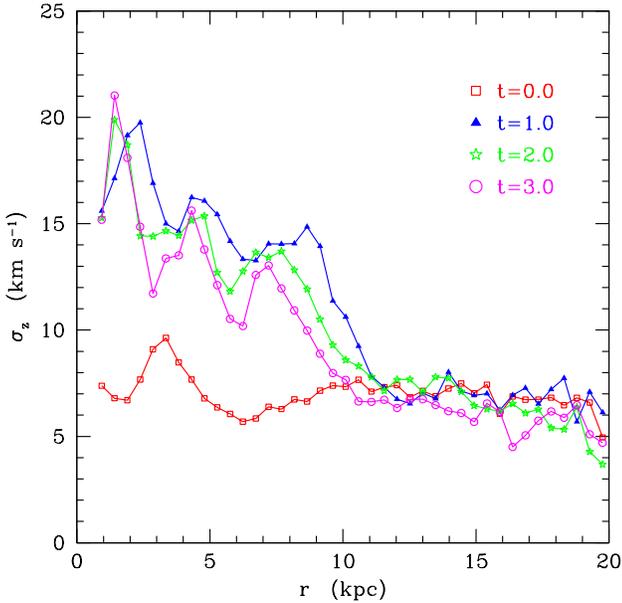}}
\caption{Profiles of r.m.s. vertical velocity as a function of radius
  for the MW simulation at four different times, including the initial
  conditions corresponding to $t=0$.}
\label{fig:vel_mw}
\end{figure}

Figure~\ref{fig:vel_mw} shows the profiles of the r.m.s. value of the
vertical velocity, $\sigma_z$, for gas particles in the initial
conditions and at three different times.  As already noticed from the
edge-on view of the gas disc in Fig.~\ref{fig:mw}, feedback leads to a
thickening of the disc, which is visible here as an increase of the
vertical velocity from $\sim7$ {\kms} to a peak value of $\sim20$
{\kms} in the inner kpc.  These values are in line with, though on the
high side of, the observational determination of \cite{Tamburro09}
based on THINGS 21-cm data.

\begin{figure}
\centerline{
\includegraphics[width=\linewidth,angle=-90]{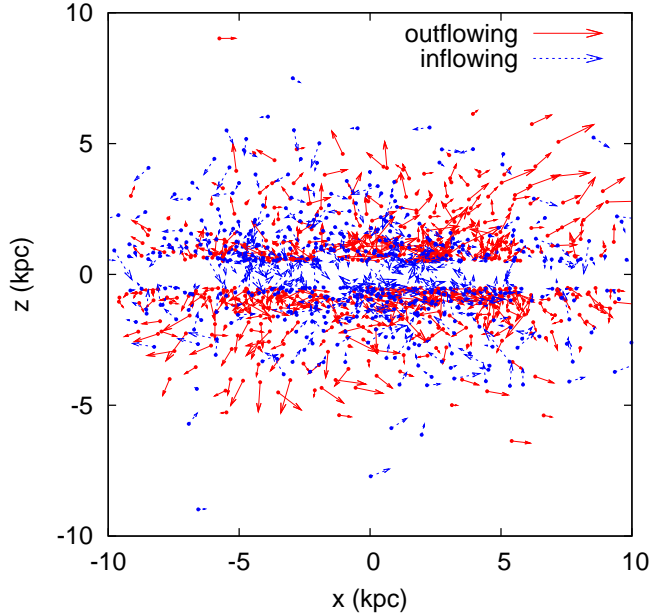}}
\caption{Particle velocities above and below the disc for the MW
  simulation after 1 Gyr. Gas
  particles with $|y|<0.5$ kpc are shown in the xz-plane; particles
  with $|z|<0.5$ kpc are removed for sake of clarity.  Red
  (continuous) and blue (dashed) particles are respectively outflowing
  and inflowing ones.  Vector lengths are scaled with velocity, 1 kpc
  corresponding to 50 km s$^{-1}$.}
\label{fig:vector_flows_mw}
\end{figure}

The edge-on images of Figure~\ref{fig:mw} show that the disc is
surrounded by a thick corona of gas particles. 
Figure~\ref{fig:vector_flows_mw} shows velocities for such corona
  particles at $t=1$ Gyr; only a slice with $|y|<0.5$ kpc is shown, and particles
  with $|z|<0.5$ kpc are removed for sake of clarity. Outflowing and
  inflowing particles are shown with different colors and line
  styles, vector lengths are scaled with velocity, 1 kpc corresponding
  to 50 km s$^{-1}$.  This figure shows that such particles are
  ejected from the disc with modest velocities, reaching values of
  about 50 \kms, in agreement with, e.g., \cite{Spitoni08}, 
and falling back in a fountain-like fashion.  We computed the
resulting outward and inward mass flows that cross the planes located
1 kpc above and below the disk midplane. We verified that they are
nearly equal at all radii and their values integrated along the radius
are similar to the average SFR.

From the results shown in this section, we conclude that MUPPI
provides a realistic description of the ISM of a Milky Way-like
galaxy. The hot phase has negligible mass but high filling factor,
and its properties are relatively stable with radius and time.  Cold
non-molecular ($HI$) gas surface density roughly flattens at $5-8$ {\surf},
while molecular gas dominates in the internal parts of the galaxy.
From the one hand, this is a consequence of an exponential pressure
profile, with a molecular fraction scaling with pressure as in
equation~\ref{eq:fcoll}. On the other hand, pressure results from
the interplay of an intrinsically unstable system, which would tend to
a pressure runaway up to saturation of the molecular fraction, and
the hydrodynamical feedback of this system on the surrounding gas.

\subsection{Global properties of the DW simulation}
\label{section:dw}

\begin{figure*}
\centerline{
\includegraphics[width=0.4\linewidth]{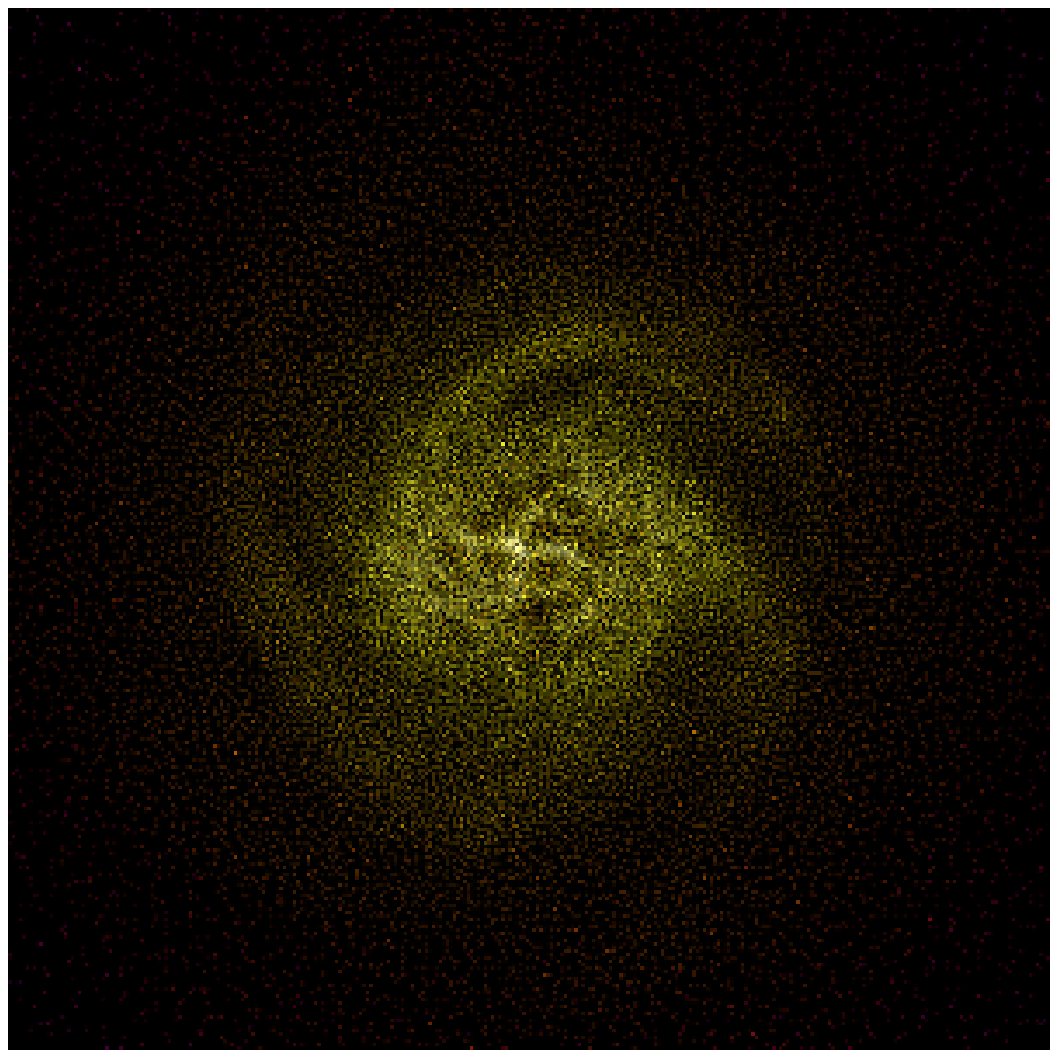}
\includegraphics[width=0.4\linewidth]{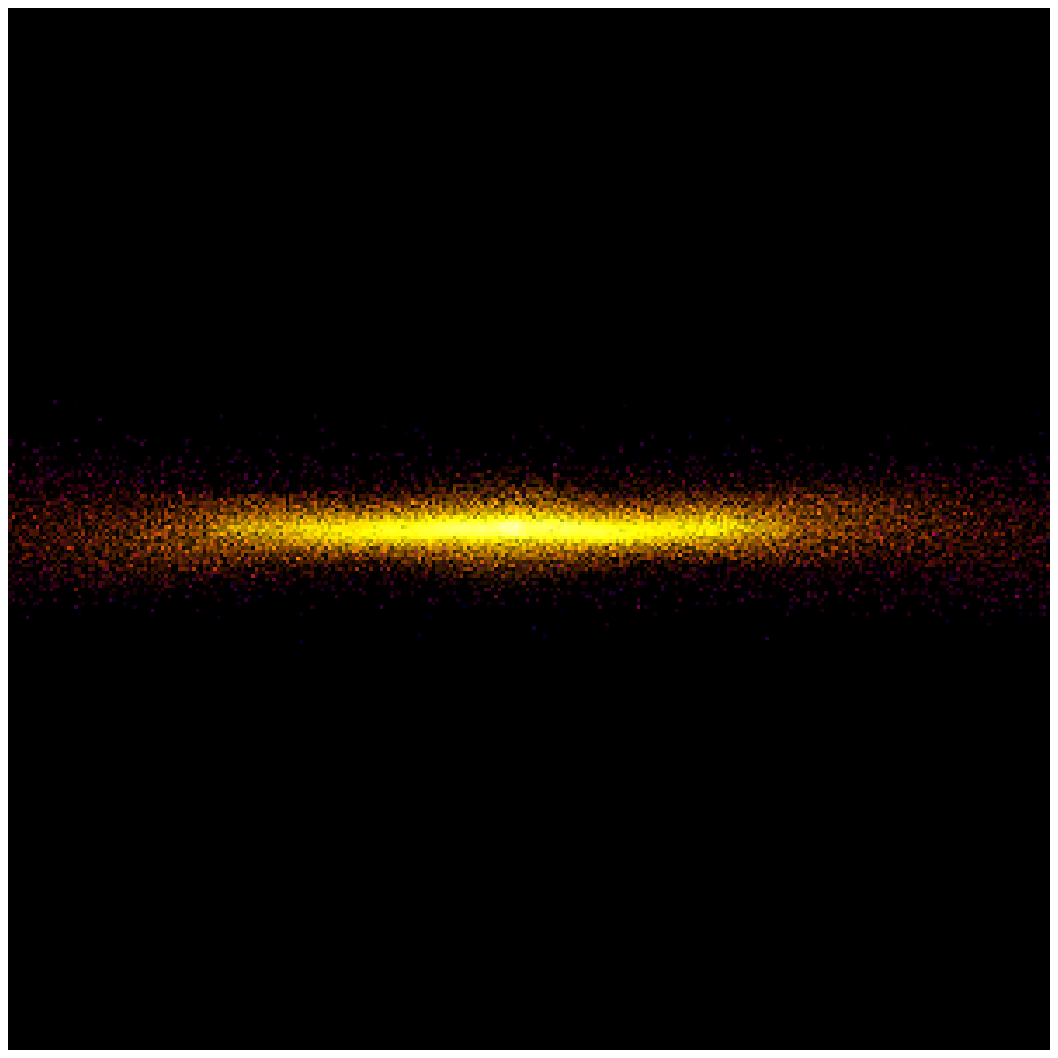}
}
\centerline{
\includegraphics[width=0.4\linewidth]{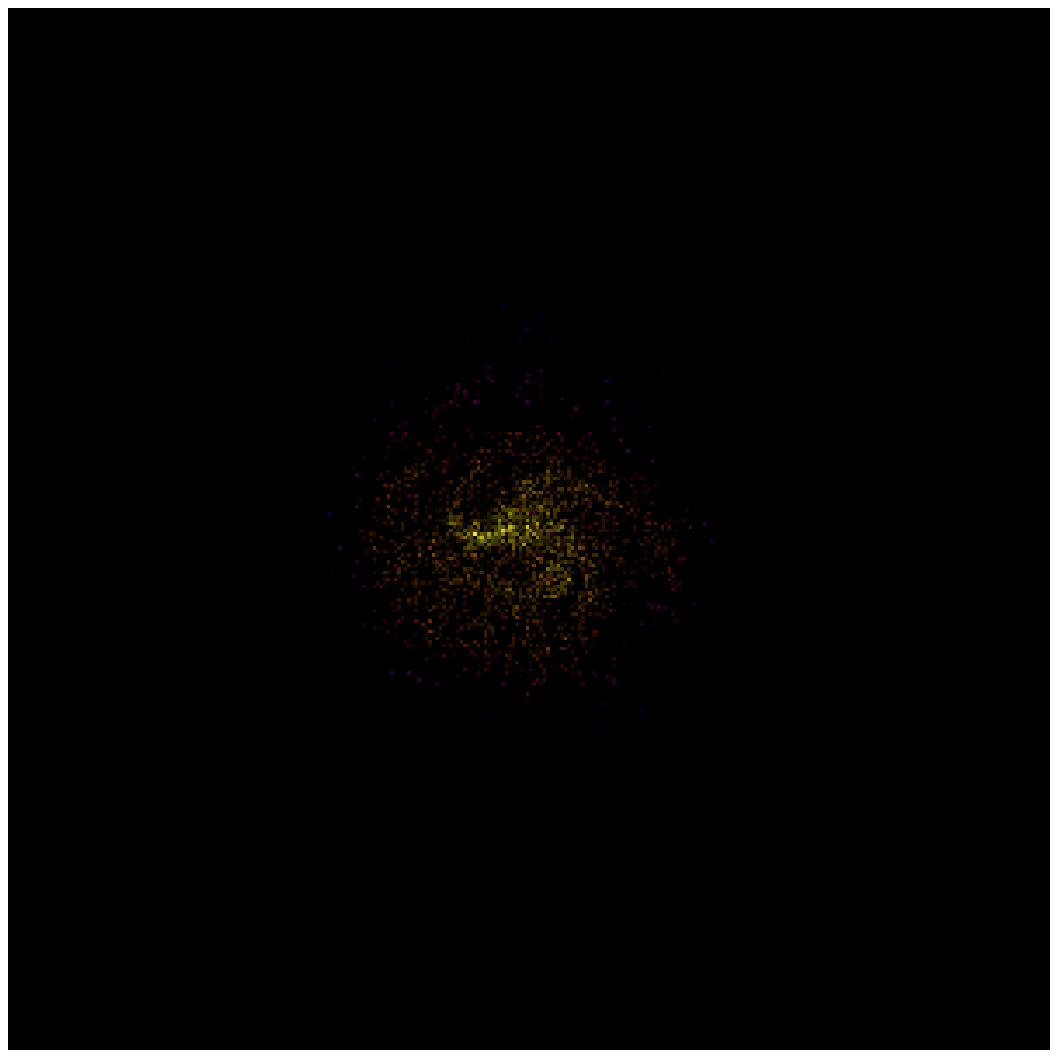}
\includegraphics[width=0.4\linewidth]{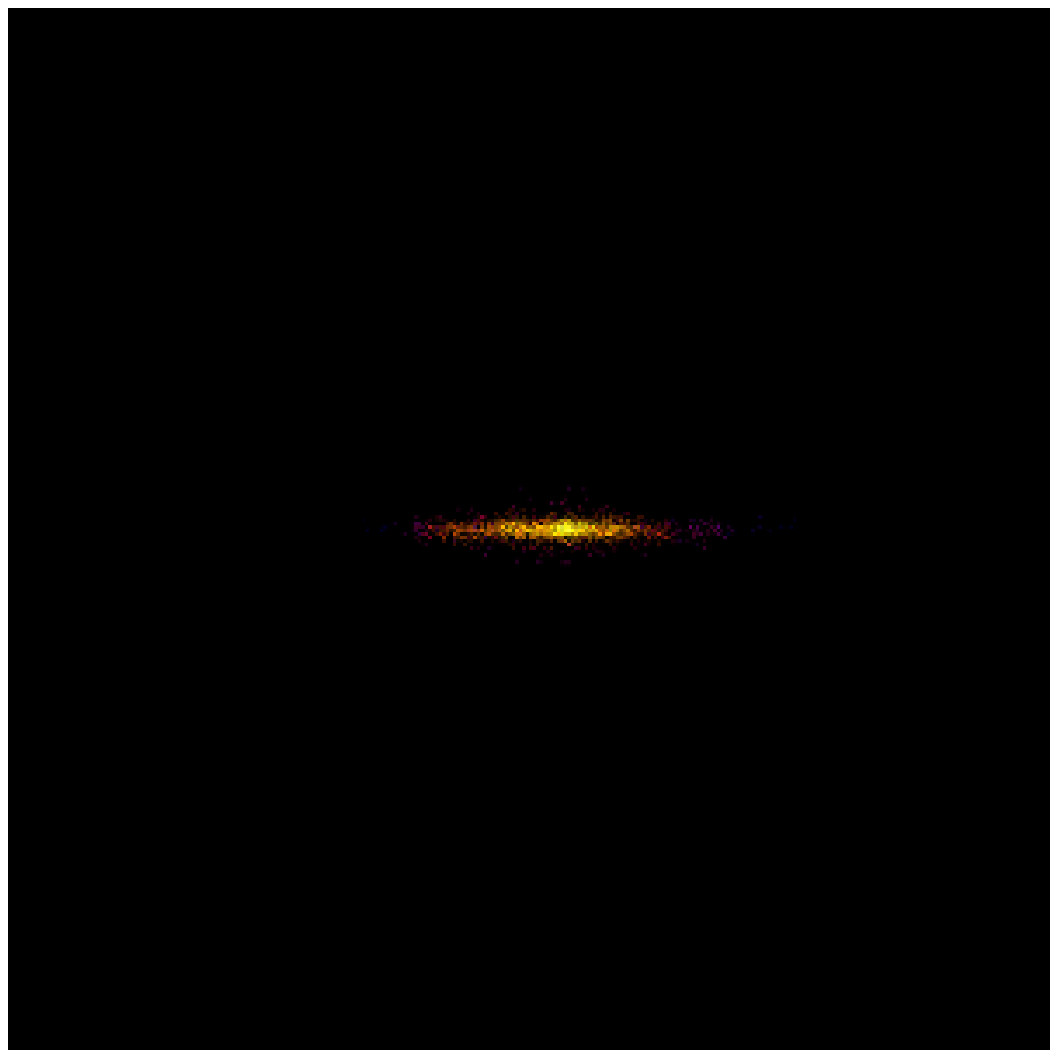}
}
\centerline{
\includegraphics[width=0.4\linewidth]{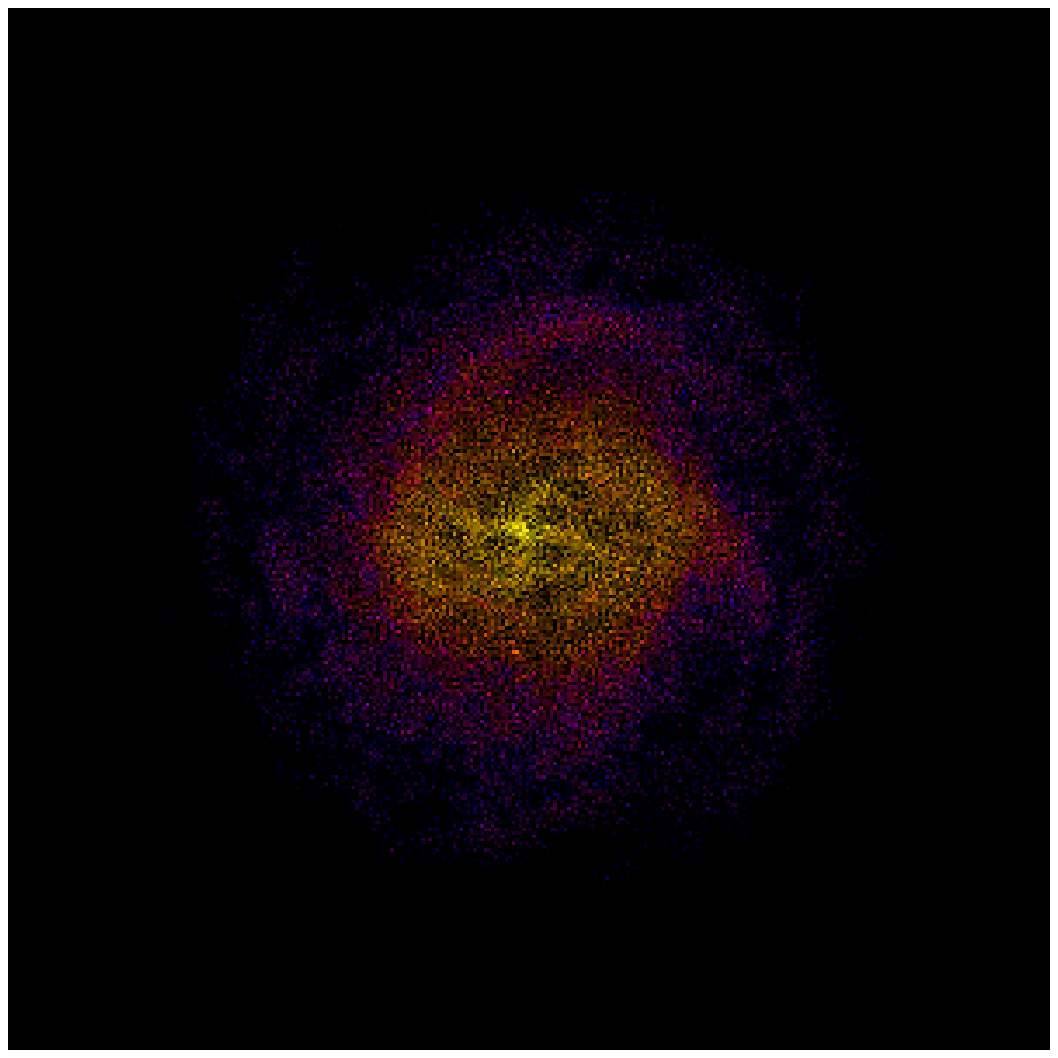}
\includegraphics[width=0.4\linewidth]{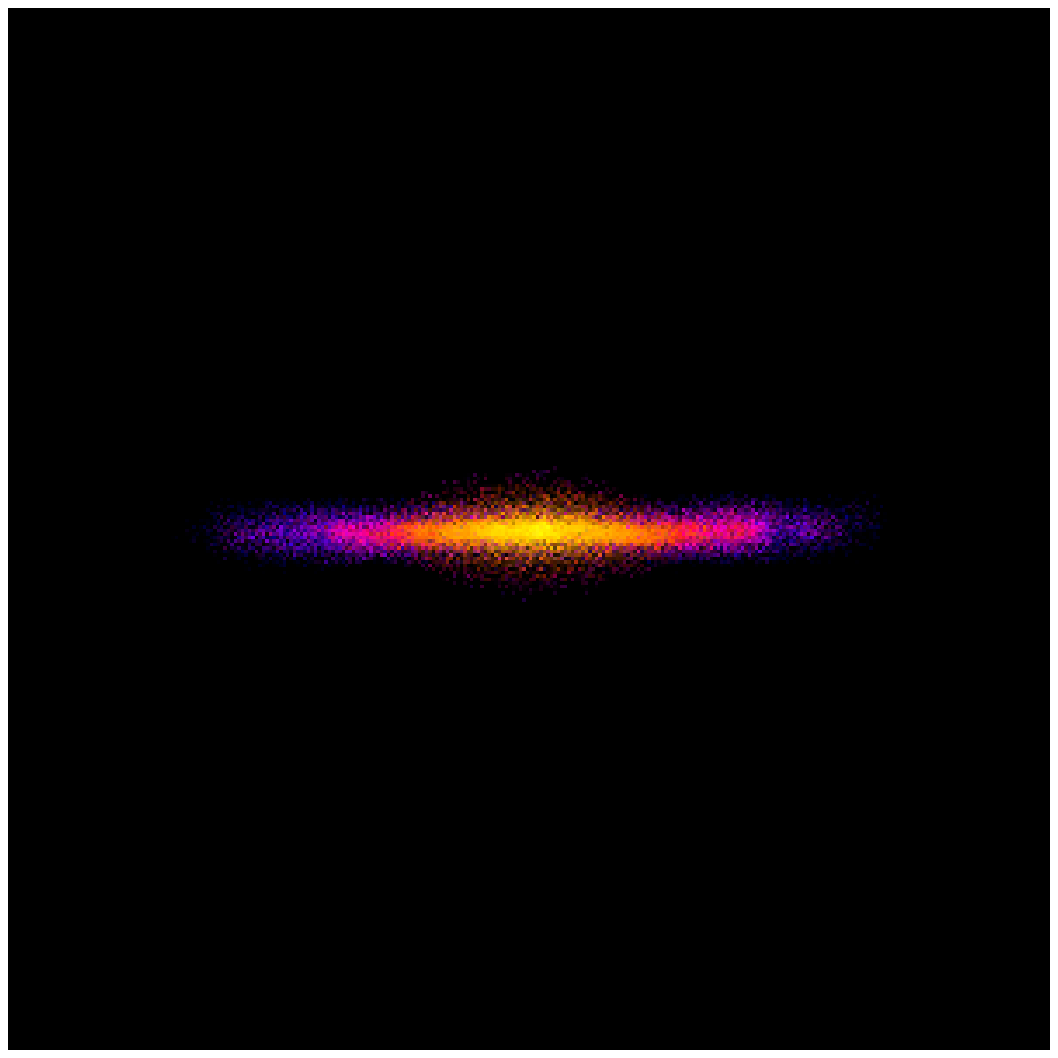}
}
\caption{ Distribution of gas particles (upper panels), star particles (mid
  panels) and SFR (lower panels) for the DW simulation with standard
  parameters after 3 Gyr.  The color code is the same as in
  Figure~\ref{fig:mw}, but we took lower maximum values for SFR and stellar
  density to enhance the color contrast. The box size is 35 kpc. }
\label{fig:dw}
\end{figure*}

\begin{figure}
\centerline{
\includegraphics[width=\linewidth]{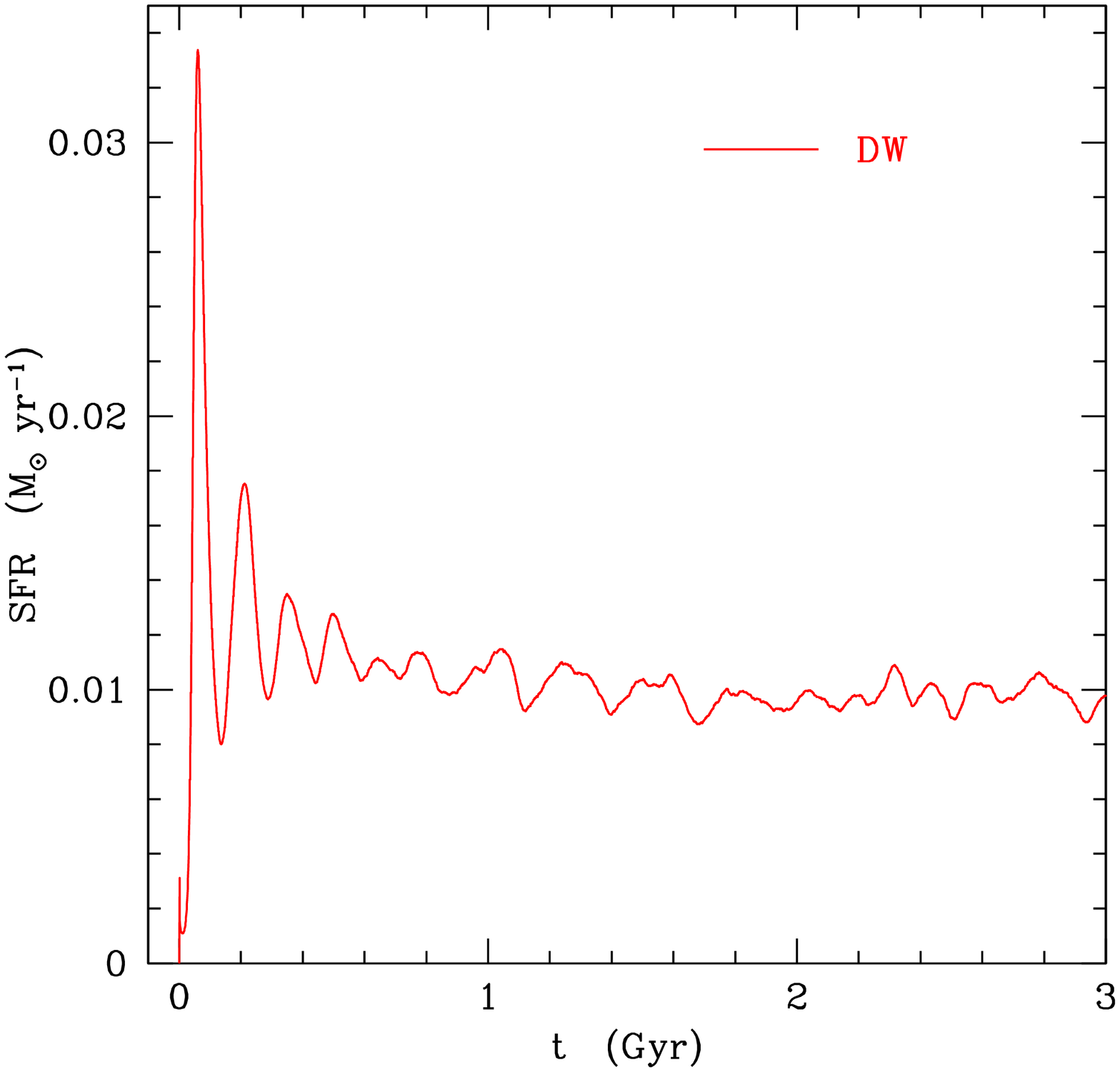}}
\caption{Star formation rate as a function of time for the DW
  simulation with standard parameters.}
\label{fig:sfr_dw}
\end{figure}

\begin{figure}
\centerline{
\includegraphics[width=\linewidth]{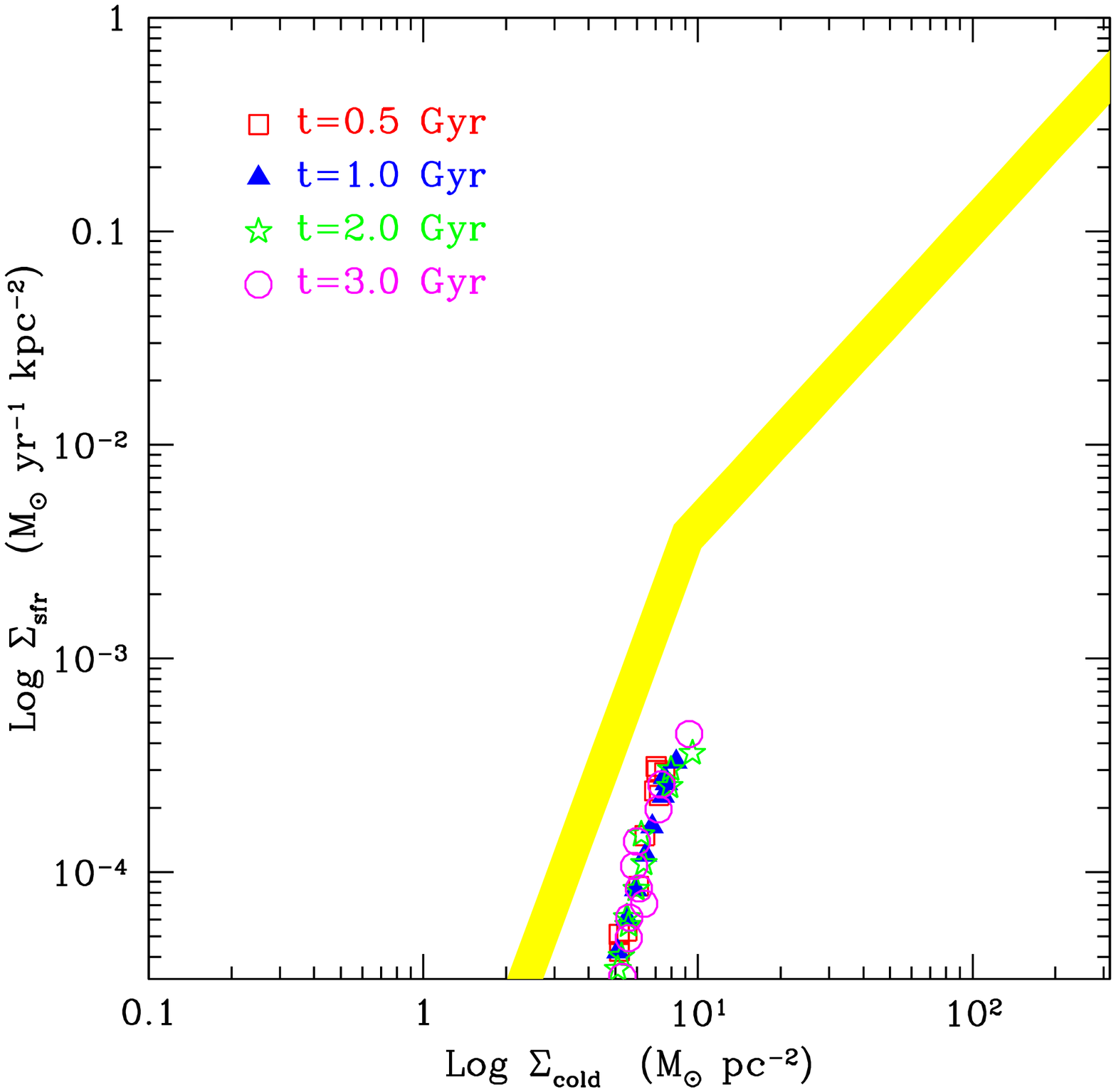}}
\caption{The same as Figure \protect\ref{fig:ken_mw}, but for the 
  simulation of the DW galaxy.}
\label{fig:ken_dw}
\end{figure}

\begin{figure}
\centerline{
\includegraphics[width=\linewidth]{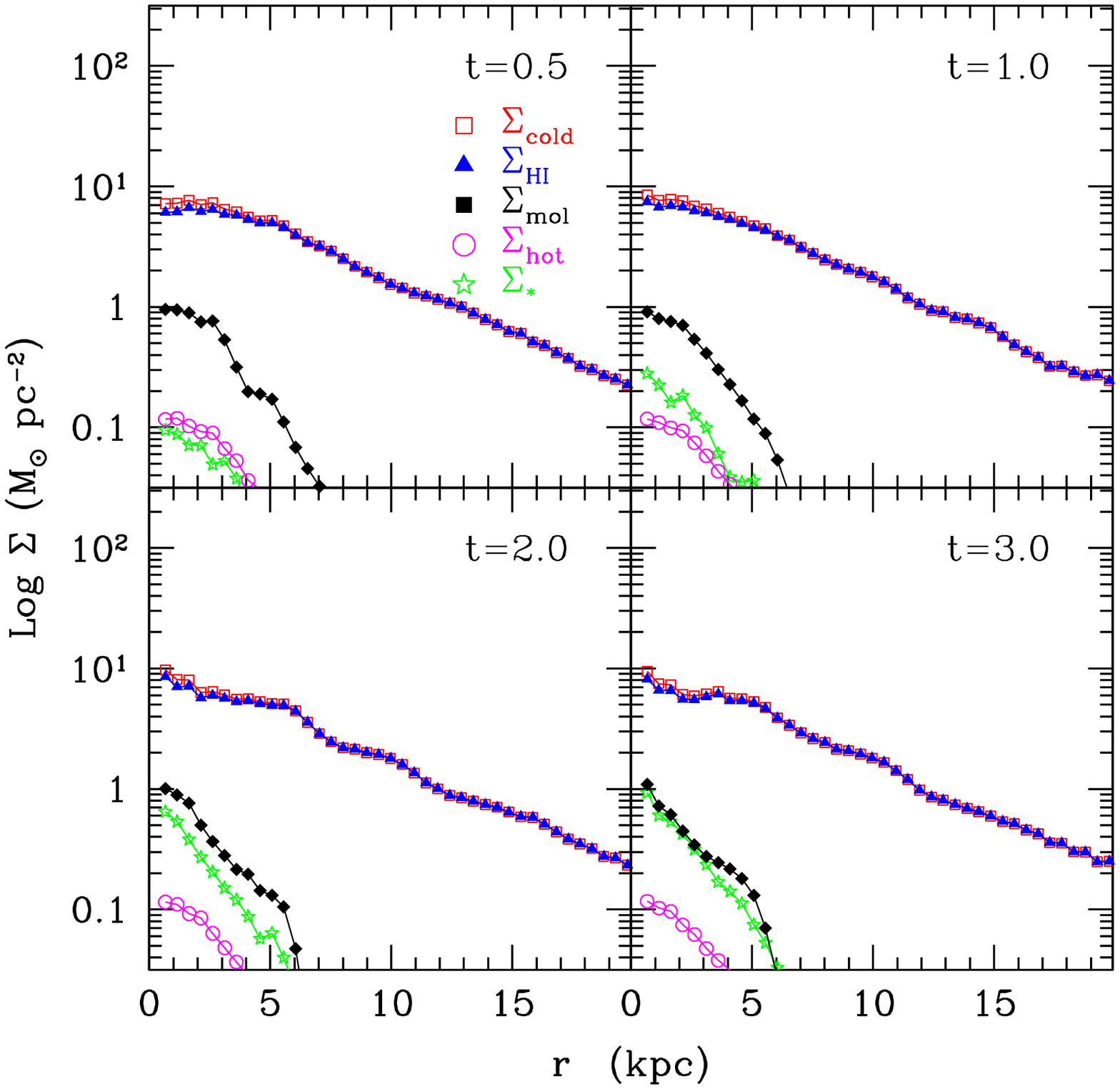}}
\caption{The same as Figure \protect\ref{fig:profili_mw} but for the
  simulation of the DW galaxy.}
\label{fig:profili_dw}
\end{figure}

\begin{figure*}
\centerline{
\includegraphics[width=\linewidth]{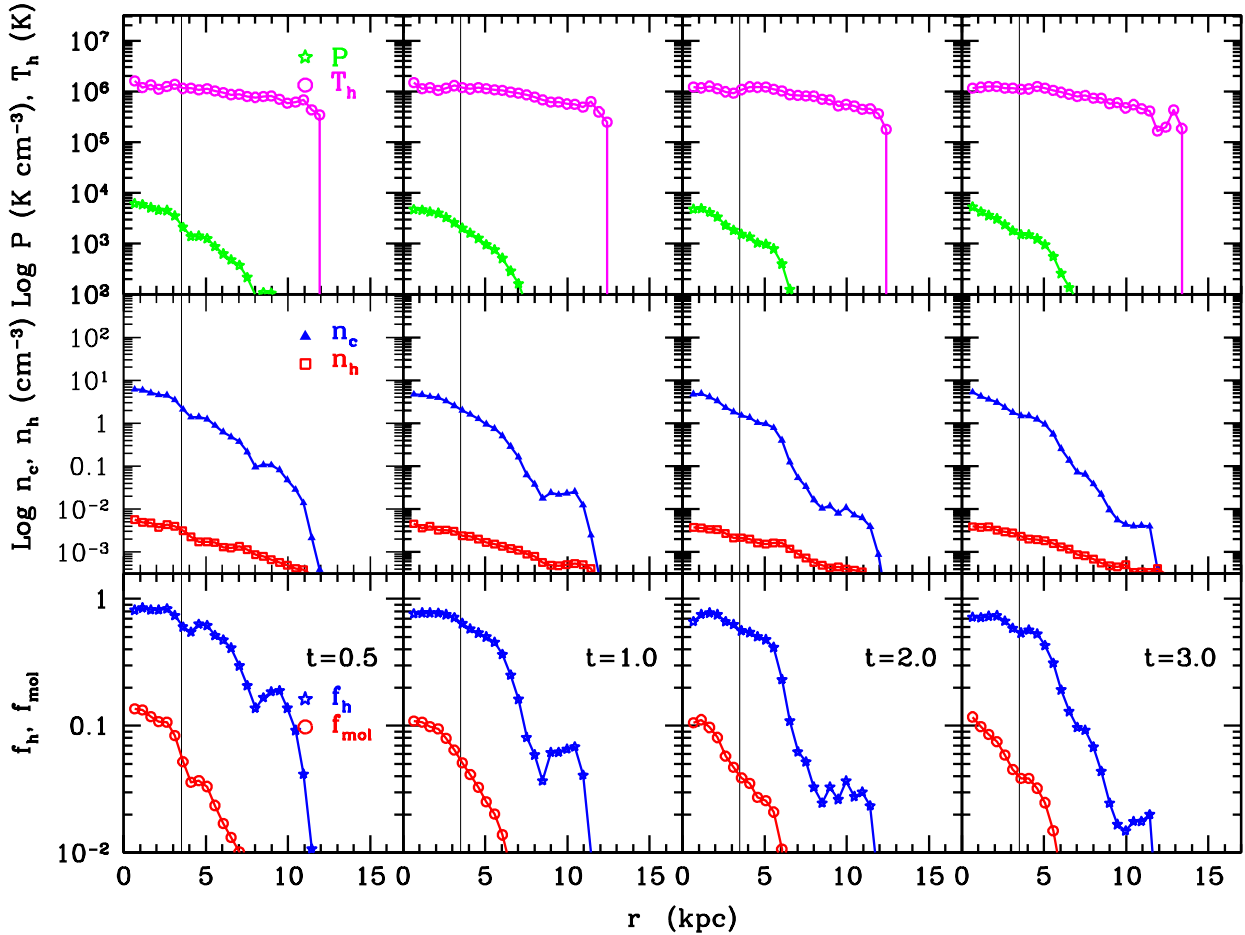}}
\vspace{-4.5truecm}
\caption{The same as in Figure \protect\ref{fig:ism_mw}, but for the
  simulation of the DW galaxy. Here, the black vertical line marks the
    scale radius of our DW disk.}
\label{fig:ism_dw}
\end{figure*}

\begin{figure}
\centerline{
\includegraphics[width=\linewidth]{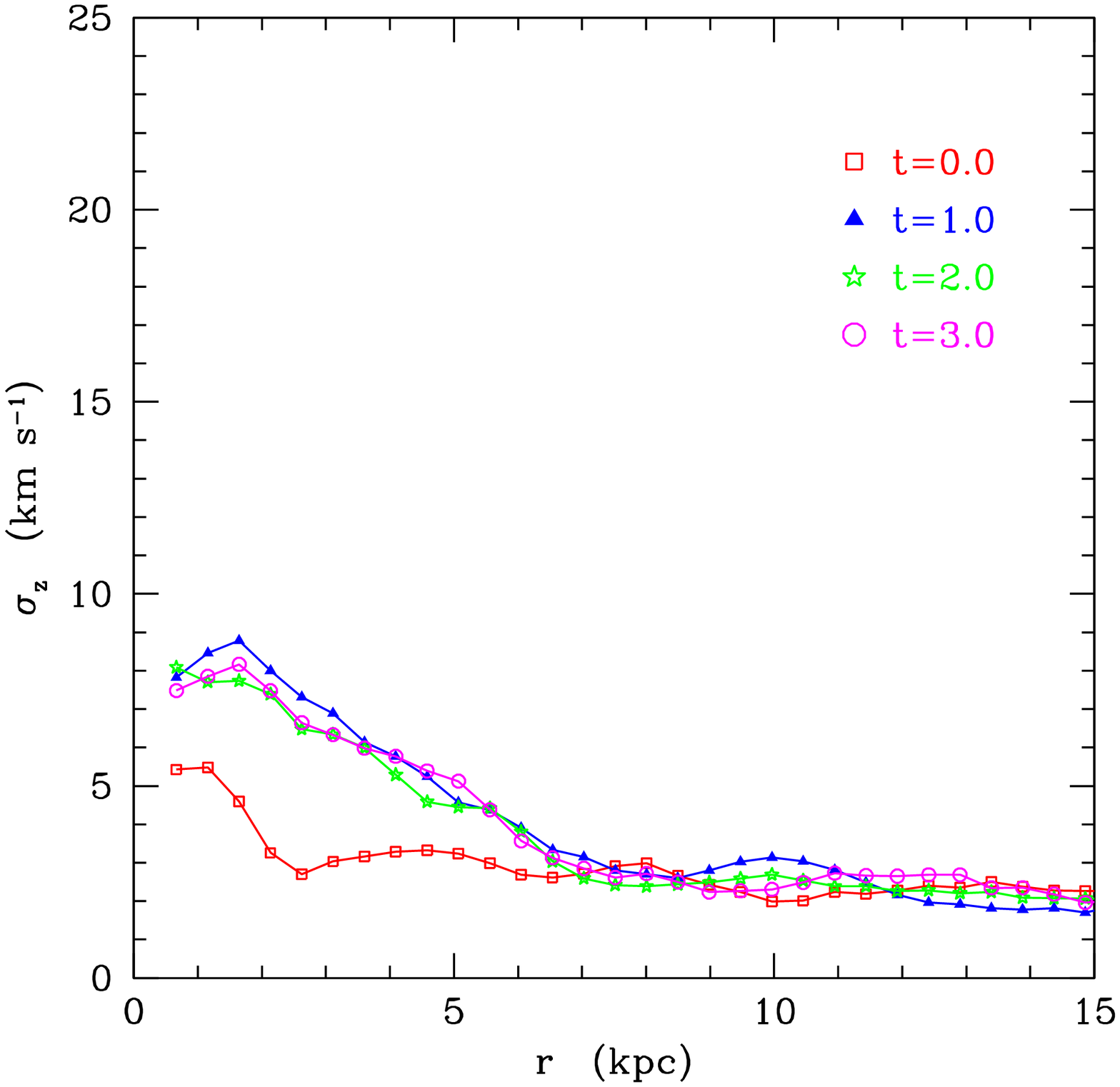}}
\caption{The same as in Figure \ref{fig:vel_mw}, but for the
  simulation of the DW galaxy.}
\label{fig:vel_dw}
\end{figure}

\begin{figure}
\centerline{
\includegraphics[width=\linewidth,angle=-90]{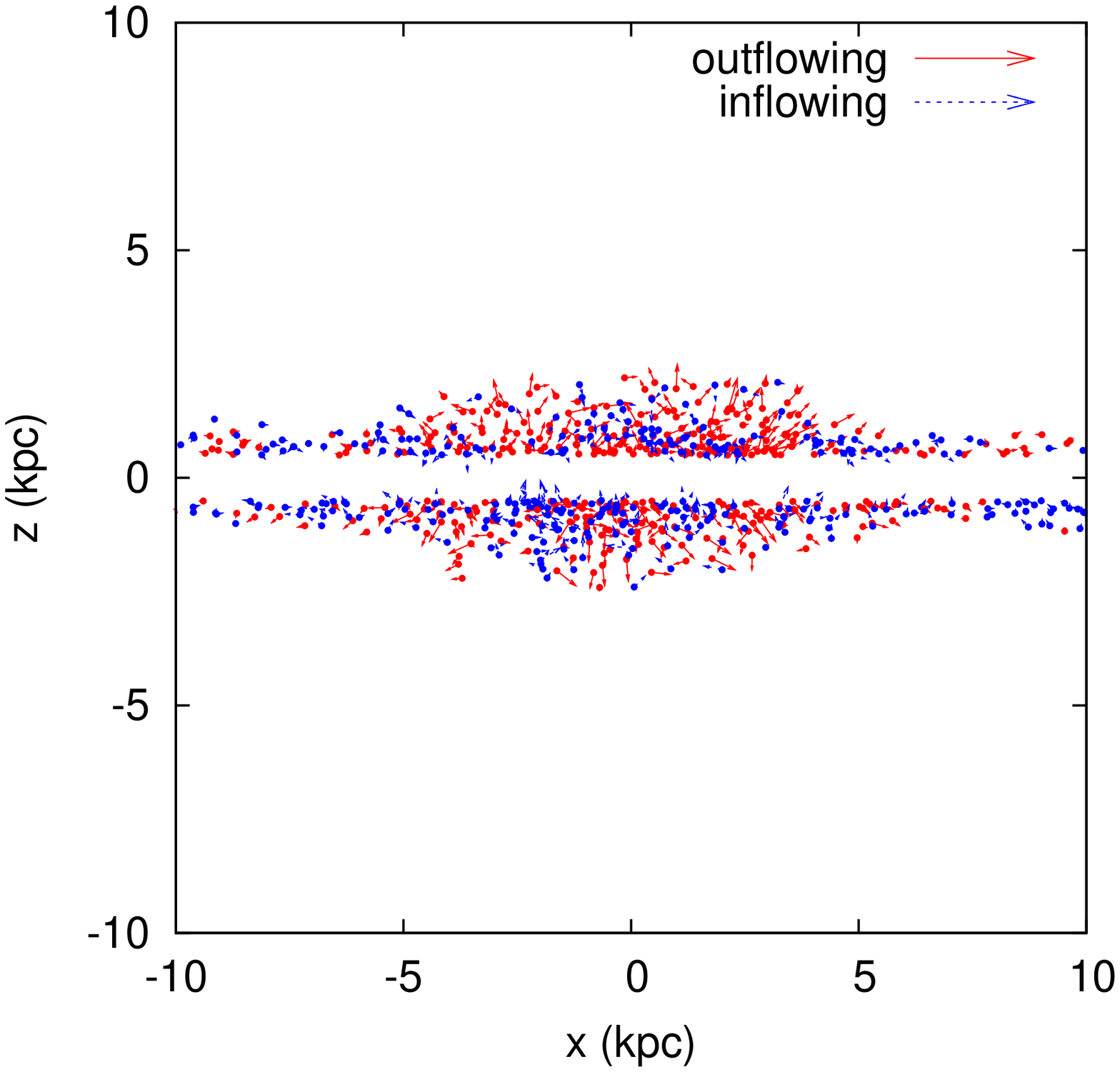}}
\caption{The same as in Figure \ref{fig:vector_flows_mw}, but for the
  simulation of the DW galaxy. }
\label{fig:vector_flows_dw}
\end{figure}

As for the simulation of the DW galaxy, Figure~\ref{fig:dw} shows gas, star
density and SFR maps at the end of the simulation, while
Figures~\ref{fig:sfr_dw}, \ref{fig:ken_dw}, \ref{fig:profili_dw},
\ref{fig:ism_dw}, \ref{fig:vel_dw} and \ref{fig:vector_flows_dw} show SFR, SK
relation, surface density profiles, ISM properties, vertical velocities and particle velocities above or below the disc. 
The gas surface density of
this galaxy is low, reaching values of $\sim10$ {\surf} only at the very
center. The molecular content of the galaxy is thus small, and the SFR
proceeds at the rate of $\sim$0.01 {\msunyr}, mildly declining with time.  The
transition to MUPPI dynamics causes oscillations of period $\sim100$ Myr, that
are slowly damped during the evolution. Unlike for the MW case, the SK
relation is now below the observational estimate, is steep and slightly
displaced with respect to that of the external regions of the MW, but
preserves the stability with evolution. Pressure and molecular fraction are
always low, so the galaxy is dominated by $HI$ gas. Hot phase temperature is
again the most stable quantity, it stays around $10^6$ K up to $\sim12$ kpc,
where the SFR drops to very low value; hot gas number and mass surface
densities are lower than for the MW case.  Vertical velocities are also lower
than the MW counterpart.  The spiral pattern is hardly visible in the stellar
component but more apparent in the gas component, where the flocculent
morphology in the center is reminiscent of the $HI$ holes seen in nearby dwarf
galaxies.  
Outflowing velocities are lower than for the MW case, but the
  extent of the corona generated by the fountain is a few kpc,
  thus indicating that the shallower potential well allows gas
  particles to be ejected to similar heights despite the fact that
  they receive a much lower energy input.

These differences with respect to the MW simulation are in line with the
observational evidence \citep{Bigiel08,Leroy08,Tamburro09} of dwarf galaxies
being morphologically irregular, $HI$-dominated, with very small values of
star formation and vertical velocity, and a steep SK relation roughly
coinciding (but slightly displaced) with respect to that of the external parts
of normal disc galaxies.

\subsection{Non--rotating halos}
\label{section:cf}

\begin{figure*}
\centerline{
\includegraphics[width=0.5\linewidth]{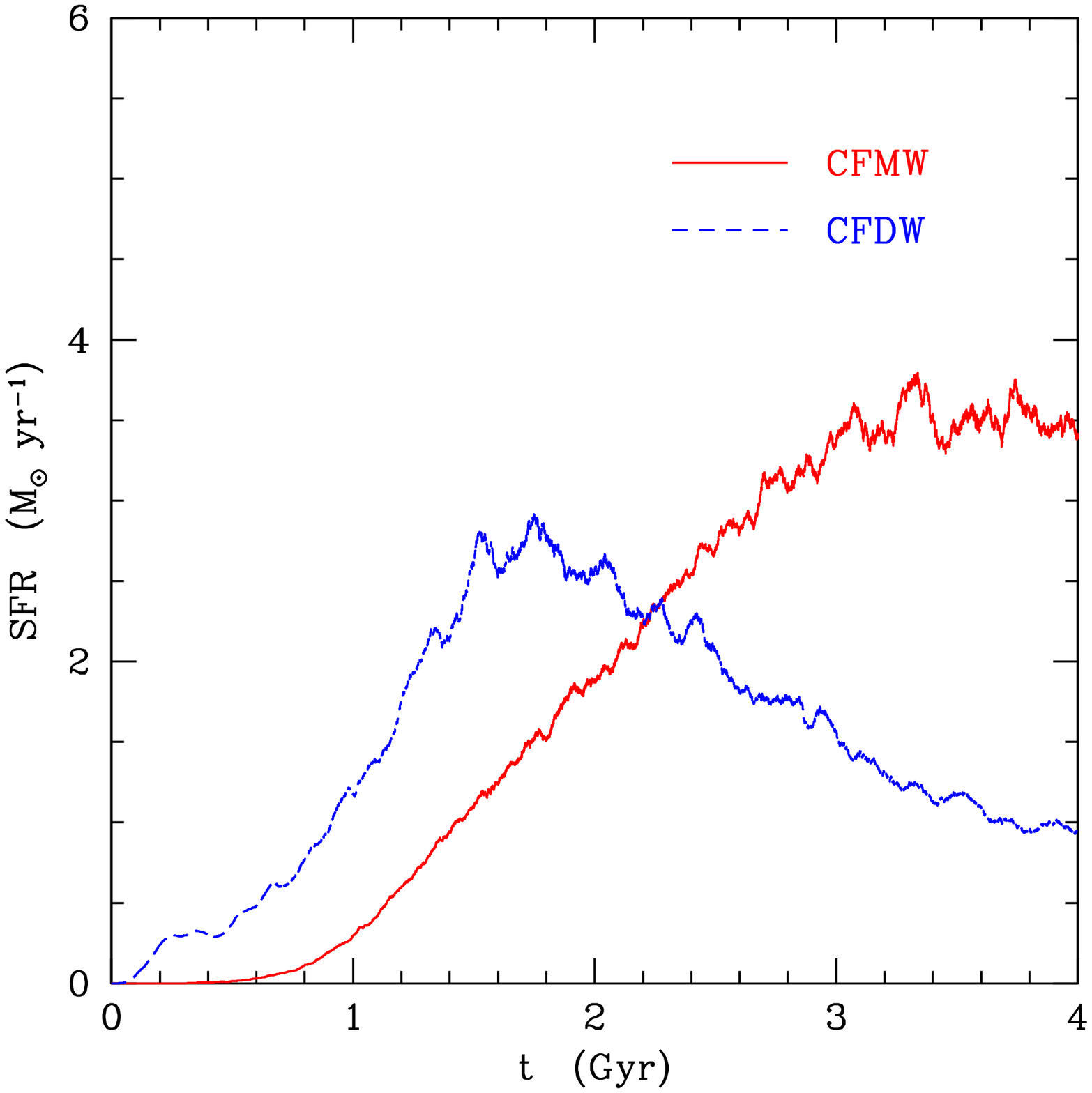}
\includegraphics[width=0.5\linewidth]{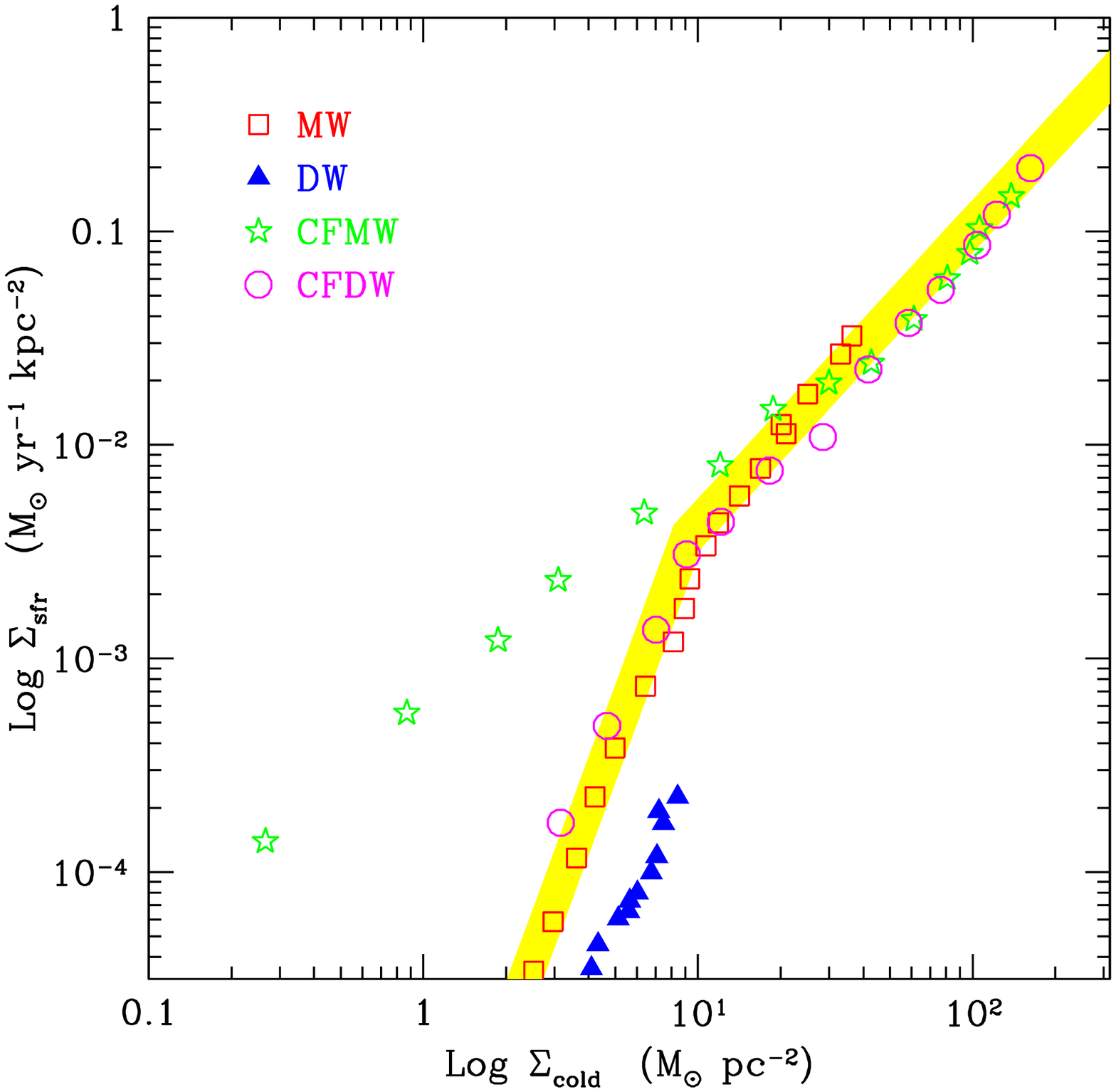}}
\caption{Evolution of the star formation rate (left panel) and
  SK relation after 2.78 Gyr of evolution (right
  panel) for the simulations of the Milky Way and dwarf non--rotating
  halos (CFMW and CFDW runs, respectively). Also shown for reference
  in the right panel are data points from the results of the SK relation for the
  rotating MW and DW simulations after 278 Myr.}
\label{fig:cf}
\end{figure*}

The main reason to test MUPPI on spherical, non-rotating, isolated
cooling flow halos is that this configuration leads to a galaxy with a
geometry which is completely different from that of a rotating disc.
In this case we can directly test whether the implemented scheme of
energy redistribution causes a different behavior of feedback in
different geometries.  In Figure~\ref{fig:cf} we show the SFR (left
panel) and the SK relation after 2.78 Gyr (right panel), for the Milky
Way (CFMW) and dwarf (CFDW); for reference we also report data points
from the SK relation for the rotating MW and DW simulated galaxies,
shown at a relatively early time (278 Myr) to probe the relation at
higher surface densities.  
The SFRs show an initial rise
due to the onset of the cooling flow, a peak and then, for the CFDW case,
a decrease, while the CFMW curve is flat down to 4 Gyr.
The difference in the onset of cooling is due to the different
virial temperature of gas in the initial conditions
and thus to the different cooling times of the halos. Moreover, the concentration of our
CFDW halo is higher than that of the CFMW. For these reasons, in the central region the
cooling time is shorter and star formation starts earlier.
The peak is mainly determined by feedback, which pressurize the multi-phase
gas and enhance its SFR.\footnote{
We checked that, running CFDW
at the same mass and force resolution as CFMW, the CFDW
SFR converges to the CFMW one after $\sim1.5$ Gyr, then peaks at $t\sim2$ Gyr
and declines thereafter.
}. The more massive halo peaks
at higher values as expected from its higher mass and deeper potential well.

We show the SK relation at the latest
output of our CFDW run.  The two cooling flows reach much higher surface
densities than the isolated discs.  The CFDW simulation is very similar to the
MW at low gas surface densities, but drops below the MW result at $\sim30$
{\surf}, while steepening at higher densities. The CFMW has a more peculiar
behavior: it stays well below the observed relation at high gas surface
density, then converging to the CFDW simulation (and below the MW one) at
higher densities.  The first point highlighted by this test is that the SK
relation is not trivially fixed by the model, instead it is obtained, although
after a suitable choice of model parameters, only in the case of a thin
rotating disc.  At high densities, the SK relation of cooling flows tends to
stay at the lower boundary allowed by the observed relations, as shown in
Figure~\ref{fig:cf}, and below the SK relation of the the MW simulation. As we
will discuss in the following, the latter tends to steepen at higher
resolution, while the two cooling flow simulations show a remarkable stability
against mass and force resolution.  This result can be explained as follows:
in a thin system, like the MW disc, energy is deposited preferentially along
the vertical direction. Therefore, particles that receive most energy are
those that stay just above or below the disk midplane.  This causes the
fountain-like flow described above and allows mid-plane particles to remain
cold and dense.  In the spherically-symmetric system this does not happen. In
this case, energy is injected much more effectively in the gas particles, so
that these are much more pressurized. At the same time, such particles expand
more and, compared with regions with similar gas surface density in the disk
midplane, they achieve lower densities.

The SK relation of the CFMW simulation at low surface densities is more
surprising. Due to the higher virial temperature and longer cooling time of
the hot gas initially present in this halo, single-phase particles flowing
toward the center are hot and thus not eligible to enter in the multi-phase
regime. As a consequence, MP particles outflowing from the central
star-forming region are surrounded by hot particles that pressurize them but,
being hot, cannot contribute to the cold gas surface density.  Indeed, if all
gas particles are (incorrectly) included in the calculation of the gas surface
densities, the other SK relations are moved to the right by a small amount,
while the SK relation of the CFMW run is shifted slightly below the observed
one at all surface densities. In a realistic case, this mechanism could be
relevant for weak star formation episodes taking place outside the main body
of large galaxies, like in winds or tidal tails embedded in a hot rarified
medium.

\subsection{Fixing the parameters}
\label{section:parameters}
We performed a number of test runs to control the behavior of our
model as we vary its parameters, and to determine our reference set of
parameters. Some of them influence the model in a quite predictable
way, so we will only briefly discuss them and concentrate in the
following on those that need a more detailed discussion. Unless otherwise
stated, we use the isolated MW initial conditions to carry out our
tests.

The full set of MUPPI parameters, shown in Table~\ref{table:params}
amounts to 8 values: $f_\star$, $P_{0}$, $T_{\rm c}$, $f_{\rm fb,out}$,
$f_{\rm fb,local}$, $f_{\rm ev}$, $\theta$.

To these we should add two parameters that are set by stellar
evolution and choice of the IMF, namely $E_{\rm SN}/M_{\star,\rm SN}$,
which is set to $10^{51}\ {\rm erg}/120\ {\rm M}_\odot$ (and is
degenerate with $f_{\rm fb,out}$ and $f_{\rm fb,local}$), and the
restoration factor $f_{\rm re}=0.2$.  Moreover, the model has two
thresholds $n_{\rm thr}$, $T_{\rm thr}$, which sets the minimum values of
density and temperature for gas particles to enter in the multi-phase
stage, and two exit conditions $n_{\rm out}$, $t_{\rm clock}$.

The diagnostic that we use to quantify the influence of parameter variations
on simulation results are the SFRs and the SK relation. The latter has good
observational constraints as far as slope, normalization and low-density
cut-off are concerned. When we vary a parameter, we keep all the others fixed
to our ``reference'' values (see Table~\ref{table:params}).

\begin{figure*}
\centerline{
\includegraphics[width=0.5\linewidth]{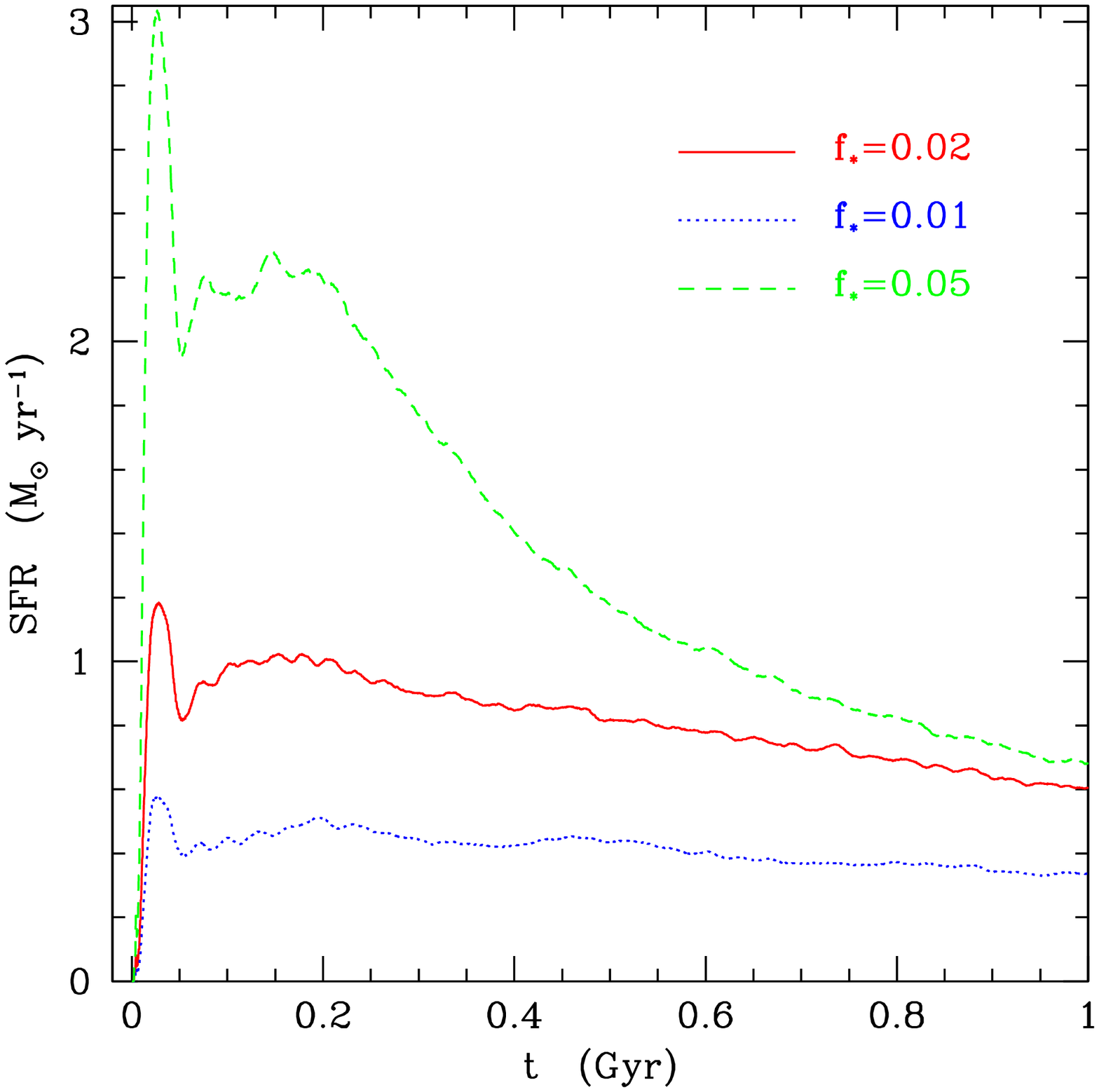}
\includegraphics[width=0.5\linewidth]{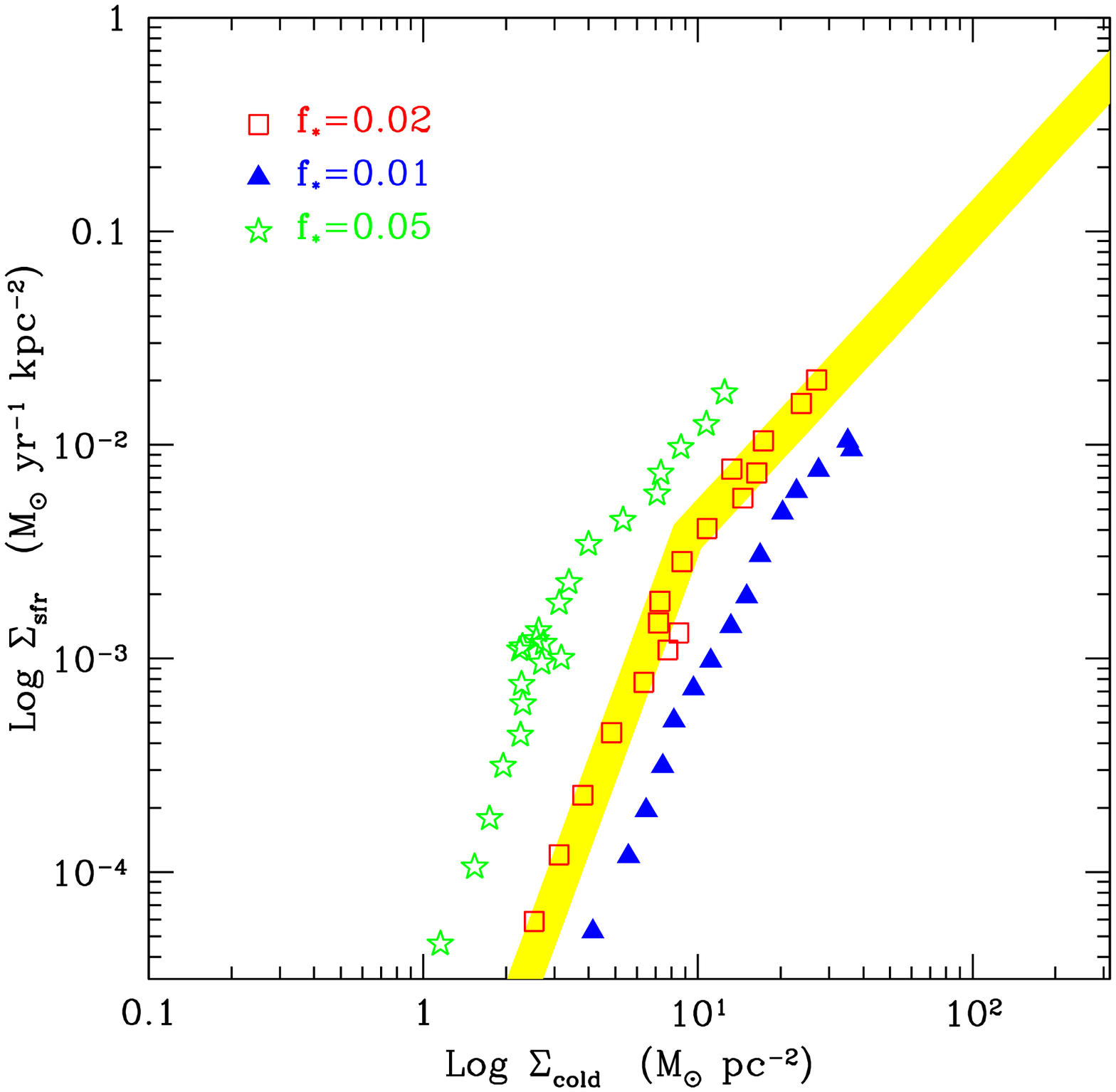}
}
\caption{Effect of varying $f_\star$ on the evolution of star
  formation rate (left panel) and SK relation after 1
  Gyr (right panel) for the MW simulation. Dashed curve and stars (in
  green), solid curve and squares (in red), dotted curve and triangles
  (in blue) correspond to $f_\star=0.05$, 0.02 and 0.01, respectively. The
  shaded area has the same meaning as in
  Fig. \protect\ref{fig:ken_mw}.}
\label{fig:fstar}
\end{figure*}

Figure \ref{fig:fstar} shows the behavior of our simulated SFR and SK
relation when we vary the star formation efficiency $f_\star$.  We used
$f_\star=0.01$,$0.02$, $0.05$.  This parameter influences the
normalization of the SK relation in a straightforward way. Indeed,
this relation fixes the consumption time-scale of cold gas into stars,
$M_{\rm c}/\dot{M}_\star$, which is regulated by $f_\star$ through
equation~\ref{eq:sfr}. We notice that the value $f_\star=0.02$,
corresponding to 2 per cent of a molecular cloud being transformed
into stars before being disrupted by stellar feedback, is well in line
with observations \citep[see, e.g.,][]{Krumholz07}.

\begin{figure*}
\centerline{
\includegraphics[width=0.5\linewidth]{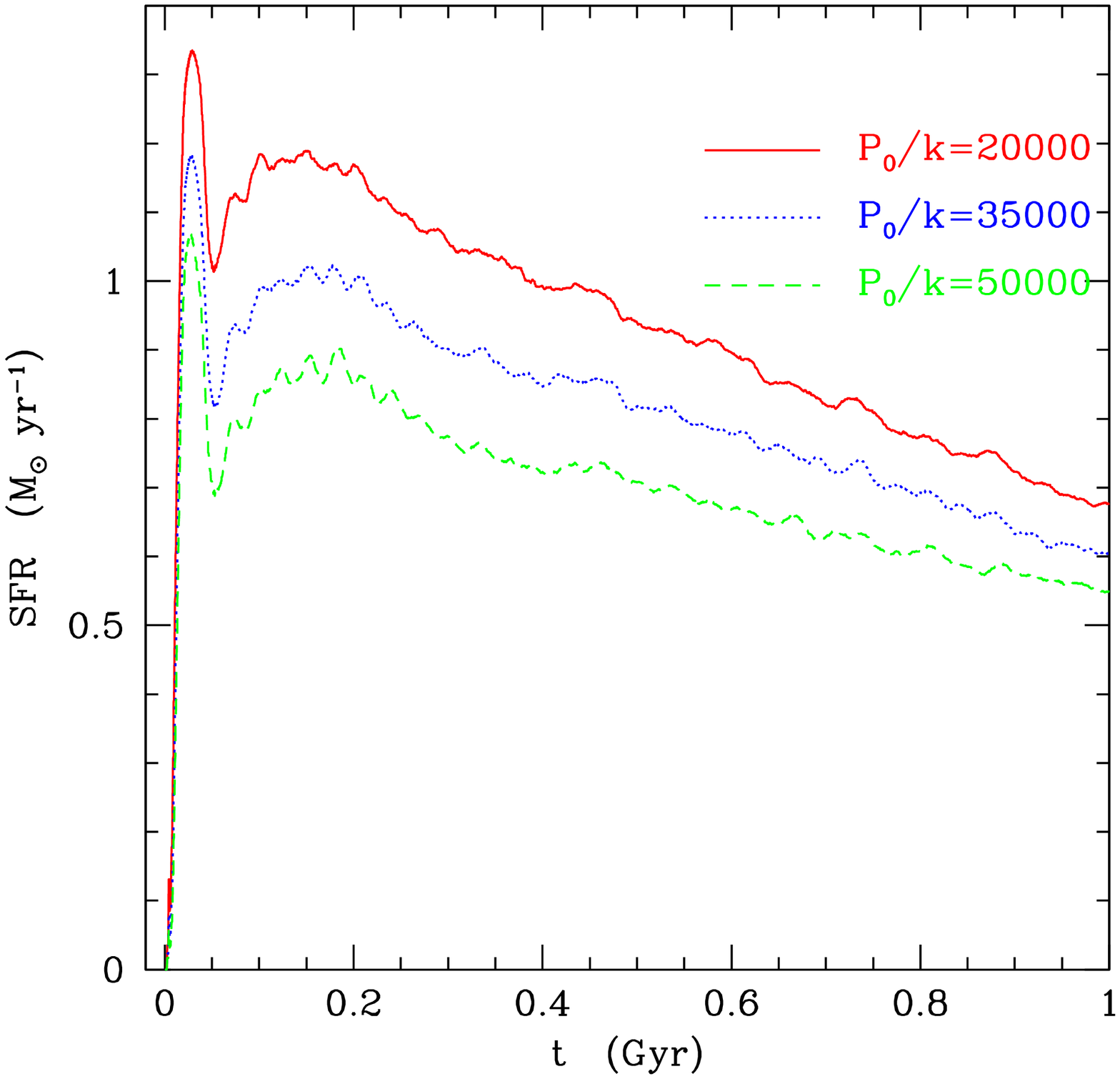}
\includegraphics[width=0.5\linewidth]{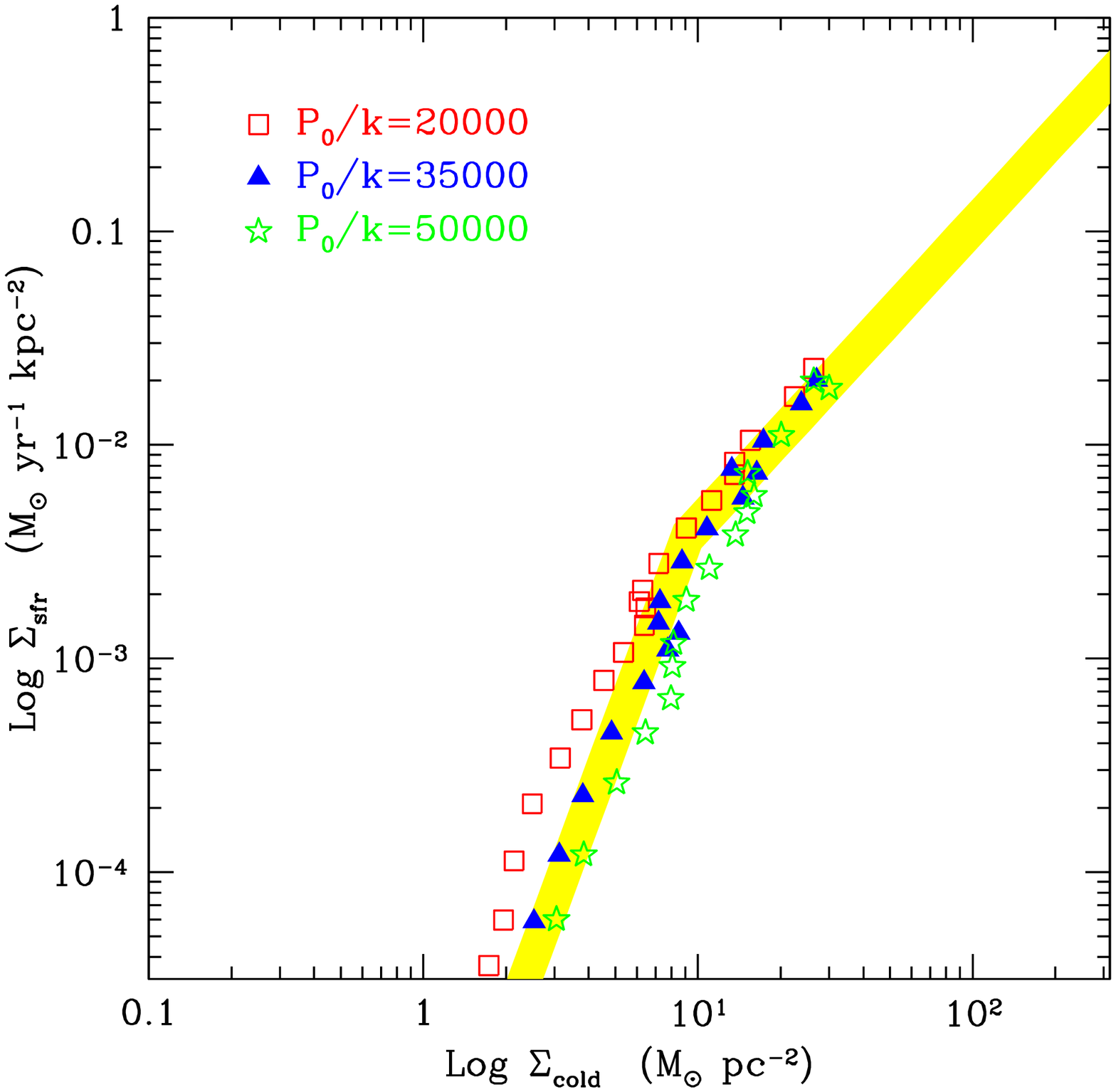}
}
\caption{The same as in Figure \protect\ref{fig:fstar}, but for
  the effect of varying the $P_0$ parameter. Short-dashed curve and
  stars (in green), dotted curve and triangles (in blue), solid curve
  and squares (in red) correspond to $P_0/k=50000$, 35000 and 20000 K
  cm$^{-3}$, respectively.}
\label{fig:presk}
\end{figure*}

As for the effect of $P_0$, it enters the model only through $f_{\rm
  mol}$, which is multiplied by $f_\star$ in equation~\ref{eq:sfr}.
As a consequence, the two parameters $P_0$ and $f_\star$ are
degenerate.  However, since the relation is non-linear, we carried out
a test obtained by varying $P_0/k$, with respect to the observational
value given by \cite{Blitz06}, to 20000 and 50000 K cm$^{-3}$. 
  The results of this test are shown in Figure~\ref{fig:presk}. We
find that the implied variation of the final results is smaller than
that caused by changing the other parameters. The SK relation is
remarkably stable against changes in $P_0$; only low values of such
parameters result in too high a $\Sigma_{\rm sfr}$ for a given
$\Sigma_{\rm cold}$.  Straightforwardly, an higher $P_0$ gives a lower
SFR.

Due to the condition of pressure equilibrium ($P/k = n_{\rm c} T_{\rm c}$), the
value of $T_{\rm c}$ fixes the proportionality between cold gas density,
directly related to dynamical time $t_{\rm dyn}$, and SPH pressure. The
lower this value is, the fastest the evolution of a MUPPI
particle. The chosen reference value, $T_{\rm c}=10^3$K, is similar to
that used in literature for other star formation and feedback schemes
\citep[e.g.,][]{SpringHern03}. We tested that lowering it to $T_{\rm c}=100$
(as, e.g., in M04) turns into very short dynamical times and very fast
evolution for our MUPPI particles, corresponding to high star
formation rates, which must then be compensated by a very low,
possibly unrealistic, value of $f_\star$.

\begin{figure*}
\centerline{
\includegraphics[width=0.5\linewidth]{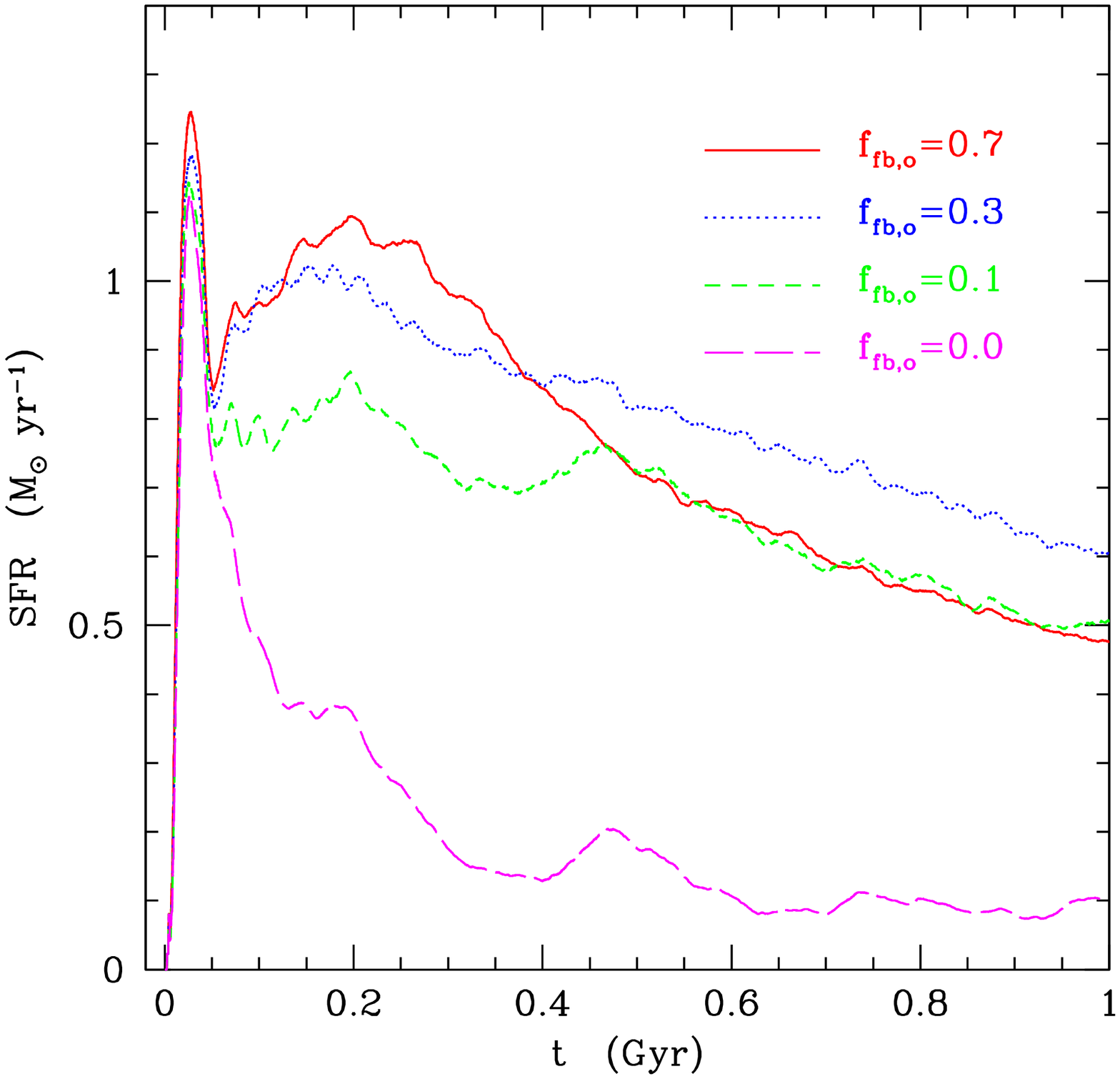}
\includegraphics[width=0.5\linewidth]{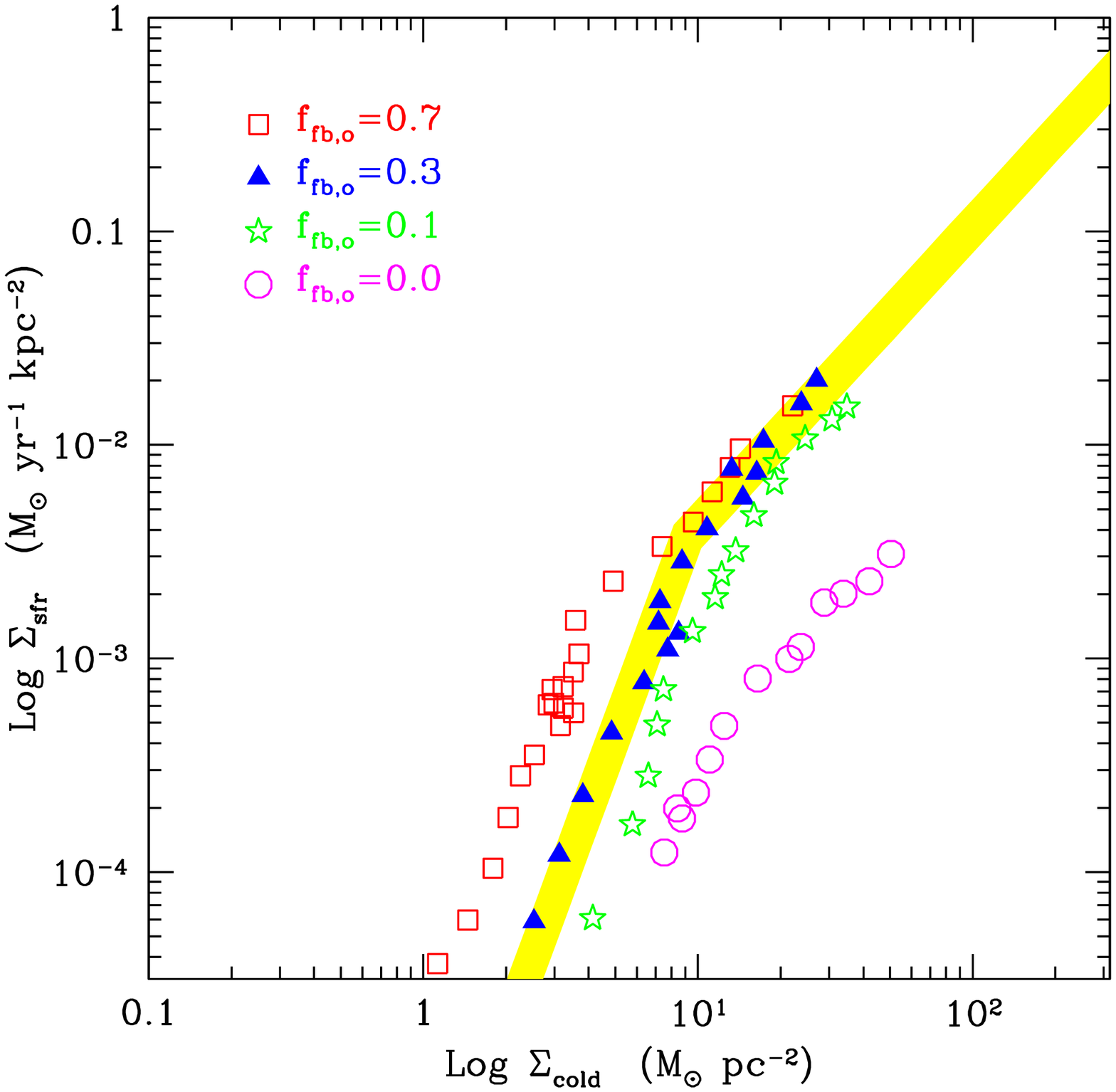}}
\caption{The same as in Figure \protect\ref{fig:fstar}, but for the
  effect of varying the $f_{\rm fb,out}$ parameter. Long-dashed curve and
  circles (in magenta), short-dashed curve and stars (in green),
  dotted curve and triangles (in blue), solid curve and squares (in
  red) correspond to $f_{\rm fb,out}=0$, 0.1, 0.3 and 0.7, respectively.}
\label{fig:fout}
\end{figure*}

Figure \ref{fig:fout} shows the effect of varying the value of the parameter
$f_{\rm fb,out}$, which regulates the amount of SN energy transferred to
neighboring gas particles. We used the values 0, 0.1, 0.3 and 0.7. The SFR
increases when this parameter ranges from 0 to 0.3. This is due to the fact
that gas particles receiving more energy are more pressurized and produce more
stars.  In particular, a relatively high value of $f_{\rm fb,out}$ is needed to
pressurize gas particles and sustain star formation, at least with the adopted
choice of $f_{\rm fb,local}$.  We checked that ISM variables (temperature,
numerical density of hot and cold gas) all increase correspondingly.  When
$f_{\rm fb,out}$ further increases, another effect takes over: a larger amount
of gas is ejected from the disk, thus becoming unavailable for star formation,
with a subsequent decrease of the cold gas mass fraction. As a consequence,
the SFR begins to decrease too. Apart from the extreme case $f_{\rm
  fb,out}=0.0$, at large values of $\Sigma_{\rm cold}$ the simulated SK relation
always stays within the observational limits.  The effect of a stronger
feedback is however to decrease the value of the surface density at which the
SK relation begins to sharply decline, thus tuning the position of the break
in the SK relation. With $f_{\rm fb,out}=0.3$ we obtain fair agreement with
observations by \cite{Bigiel08}. Therefore we choose this value as the
reference one.

Properties of the ISM, SFR and SK relation are all largely insensitive to the
exact value of $f_{\rm fb,local}$, as far as it is smaller than $f_{\rm fb,out}$.
Keeping all the other parameters fixed, a vanishing value of $f_{\rm fb,local}$
would lead to an increase in the number of particles that exit the multi-phase
regime because of catastrophic loss of hot phase thermal energy.  For this
reason we preferred the small value $f_{\rm fb,local}=0.02$, although it is
insufficient to pressurize particles in the absence of outward distributed
energy.

The amounts of cold gas evaporated by SNe, $f_{\rm ev}$, has been
fixed to 0.1 following the suggestion of M04 and \cite{Monaco04b} that
the main evaporation channel is due to the destruction of the
star-forming cloud, most of which is snow-ploughed back to the cold
phase.

\begin{figure*}
\centerline{
\includegraphics[width=0.4\linewidth]{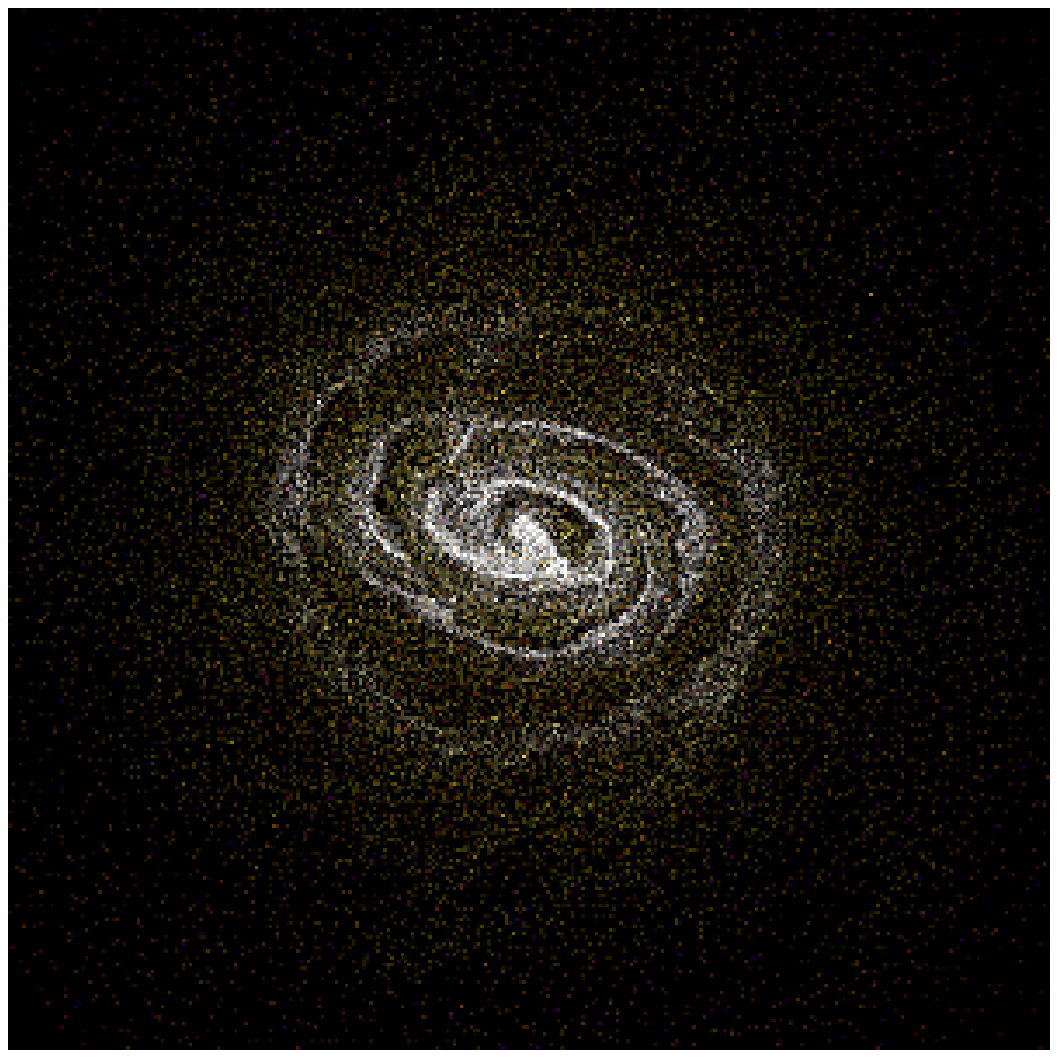}
\includegraphics[width=0.4\linewidth]{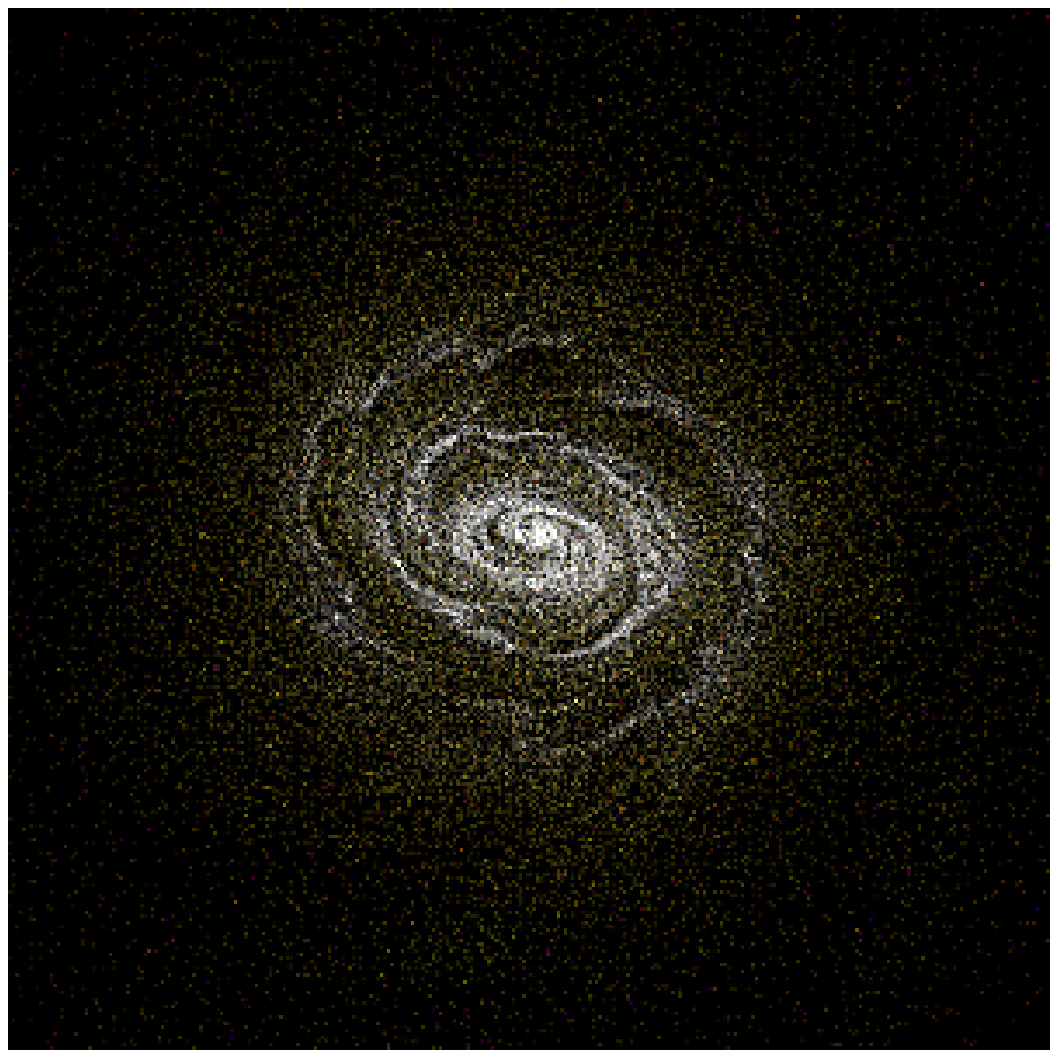}
}
\centerline{
\includegraphics[width=0.4\linewidth]{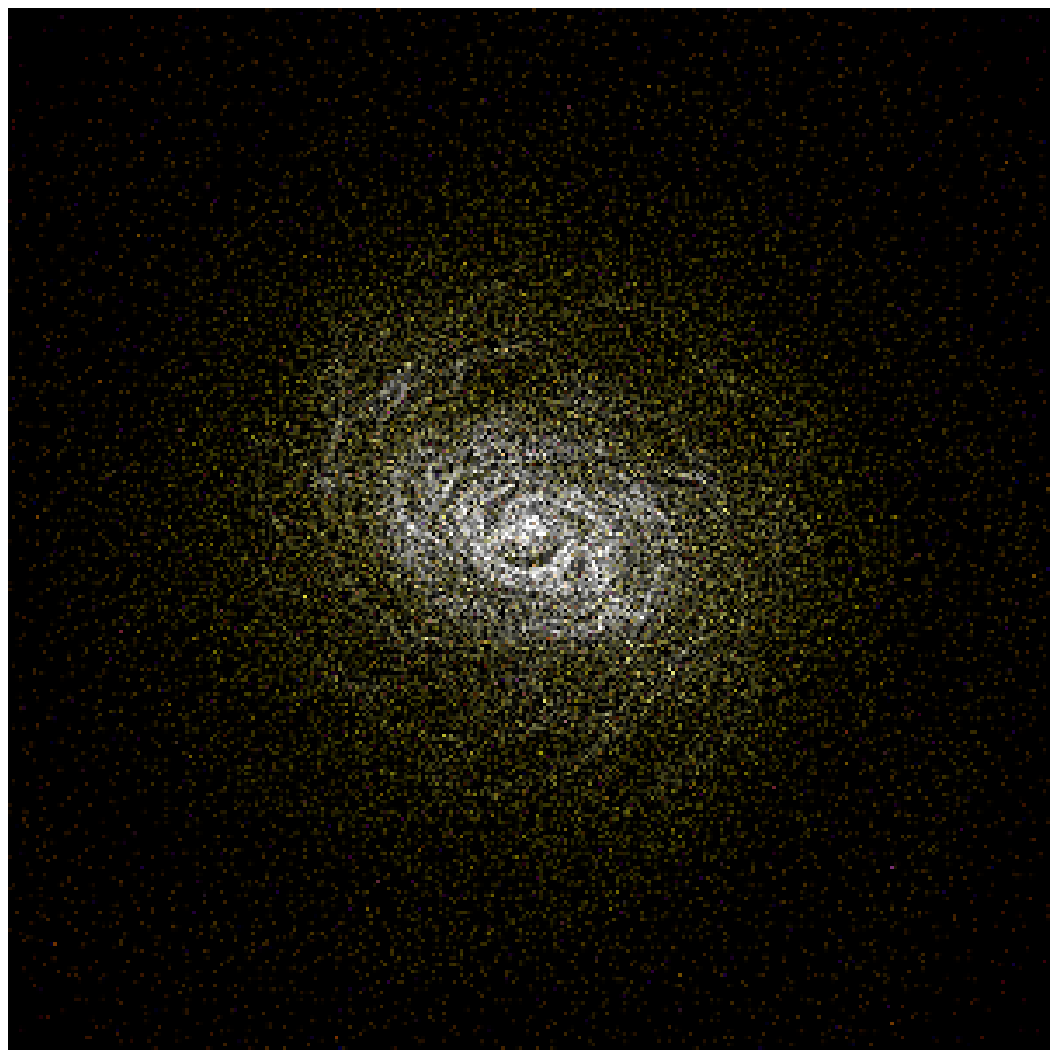}
\includegraphics[width=0.4\linewidth]{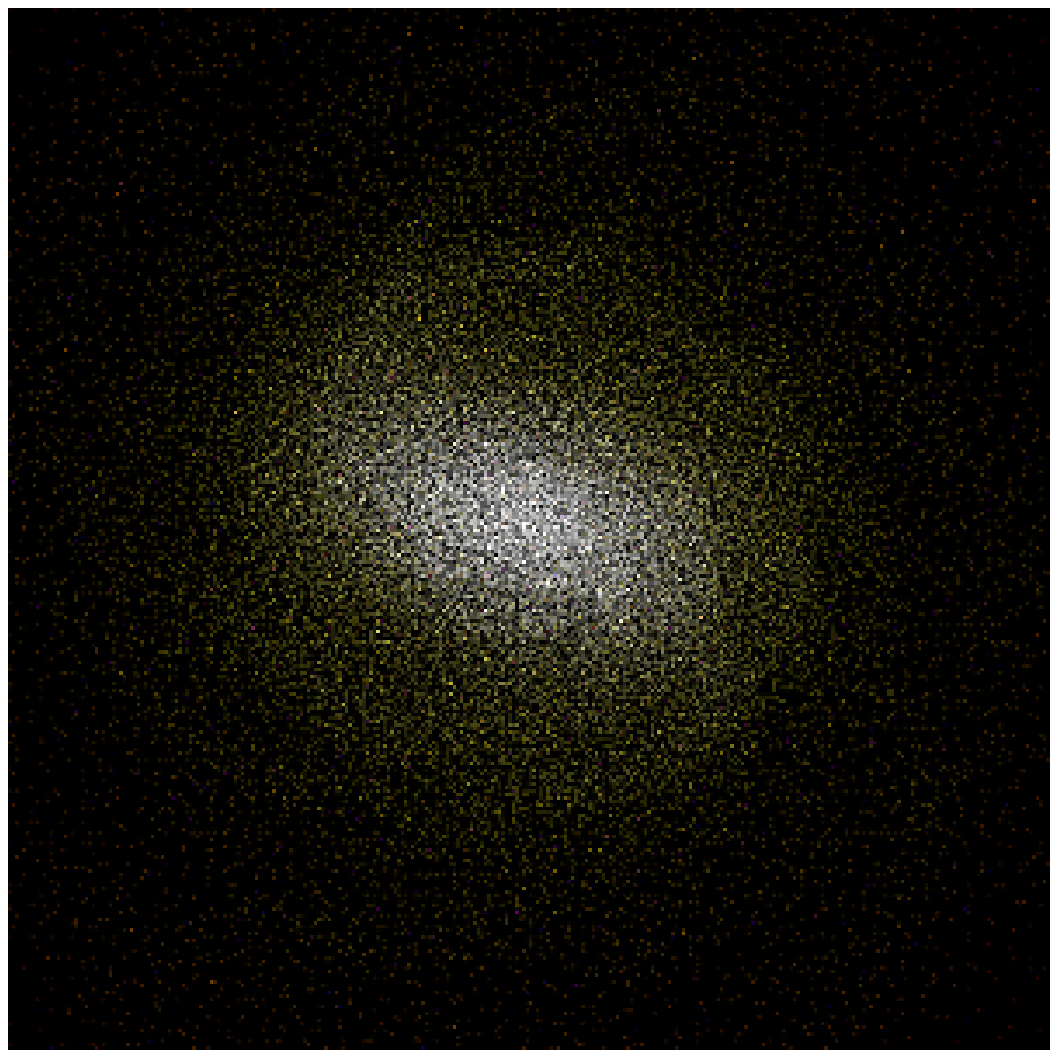}
}
\caption{Density maps of gas for the MW simulation after 1 Gyr for
  four cases 
obtained by changing the prescription to distribute feedback
  energy: 
  $\theta=30^\circ$ (upper left), the standard case $\theta=60^\circ$
  (upper right), $\theta=60^\circ$ with energy not weighted by the
  distance from the axis but only by the distance from the particle
  (lower left), isotropic distribution of energy (lower right). The
  color code is as in Fig.~\ref{fig:mw}.}
\label{fig:coni}
\end{figure*}

\begin{figure*}
\centerline{
\includegraphics[width=0.3\linewidth,angle=-90]{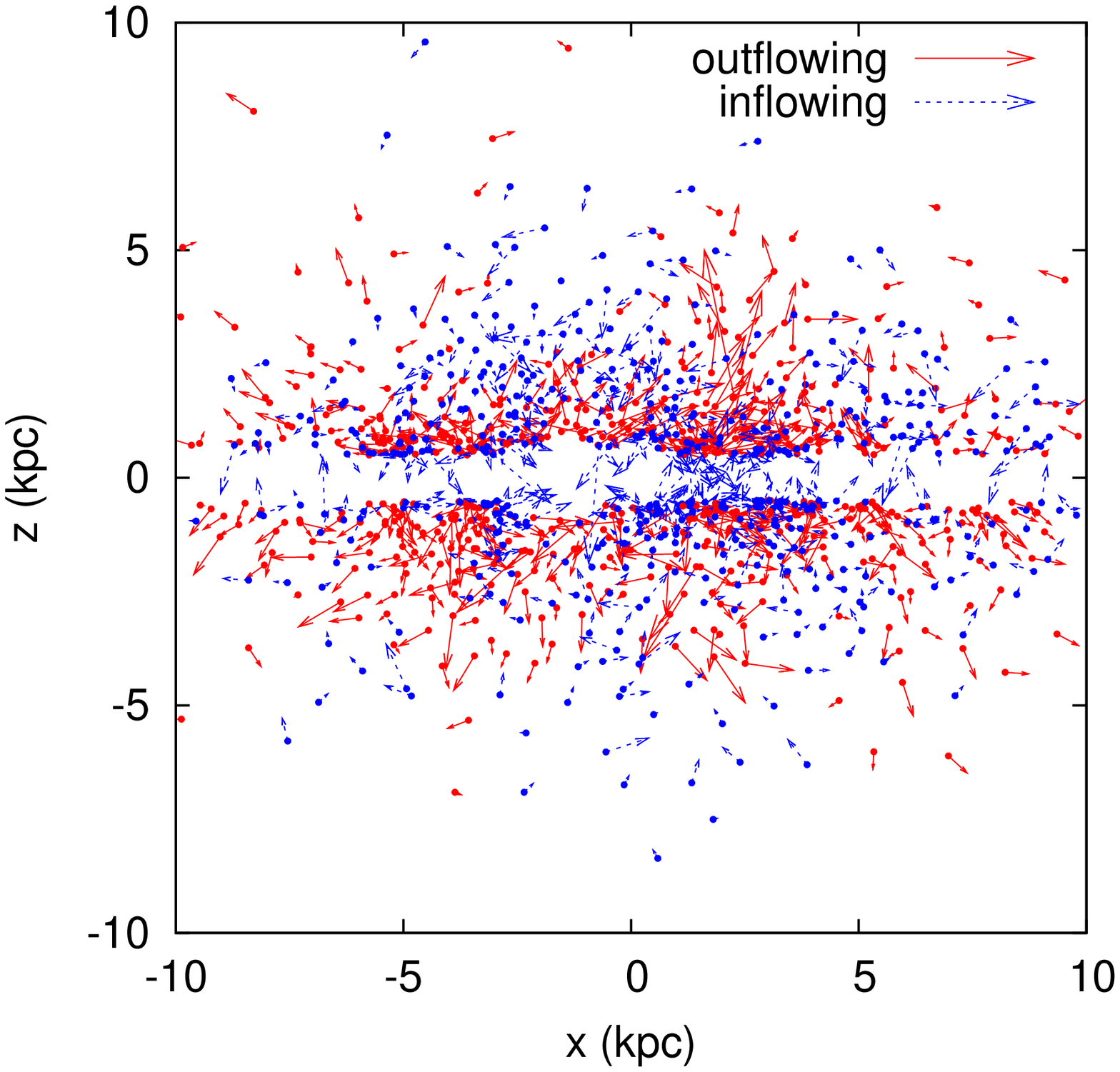}
\includegraphics[width=0.3\linewidth,angle=-90]{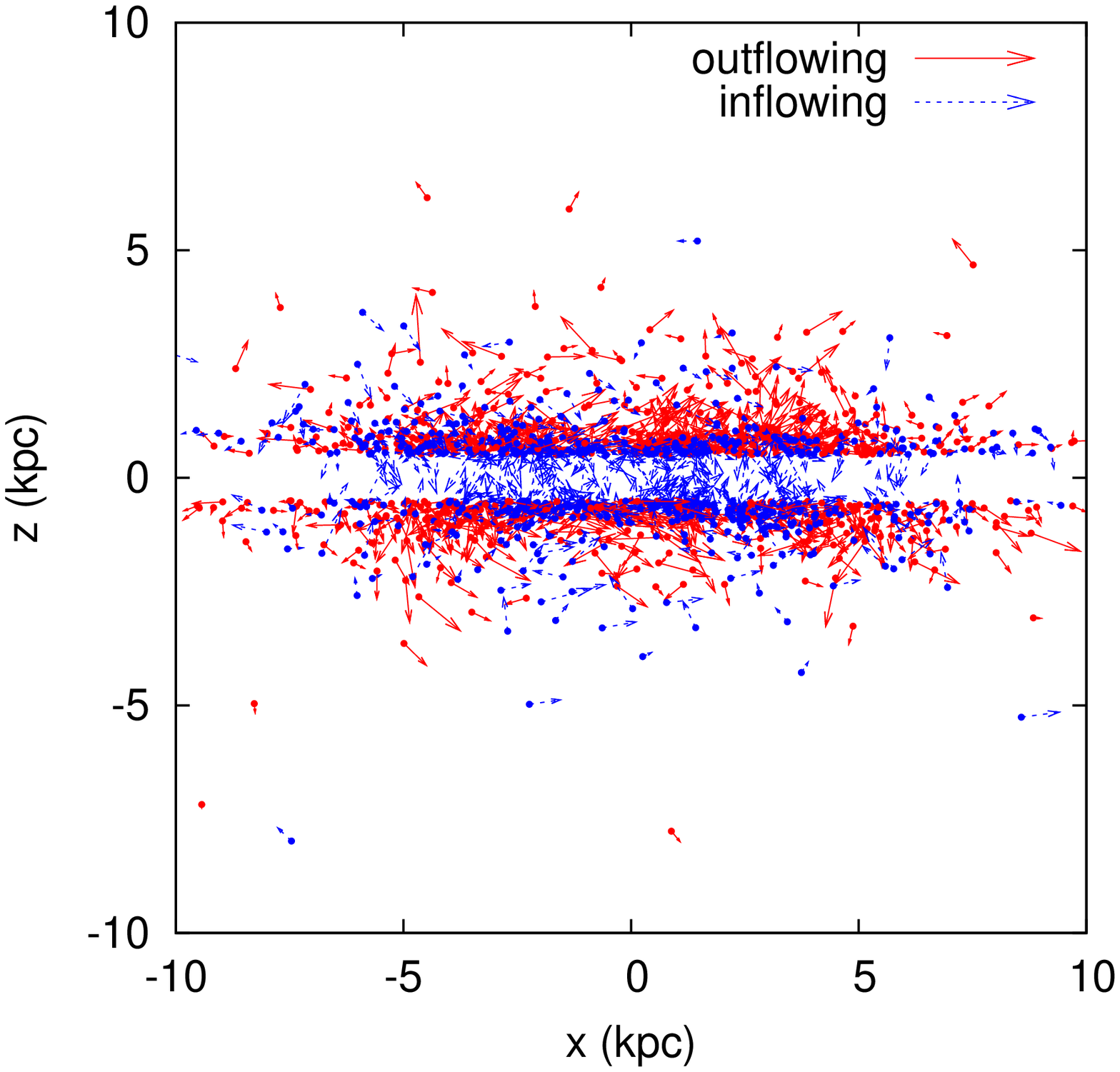}
\includegraphics[width=0.3\linewidth,angle=-90]{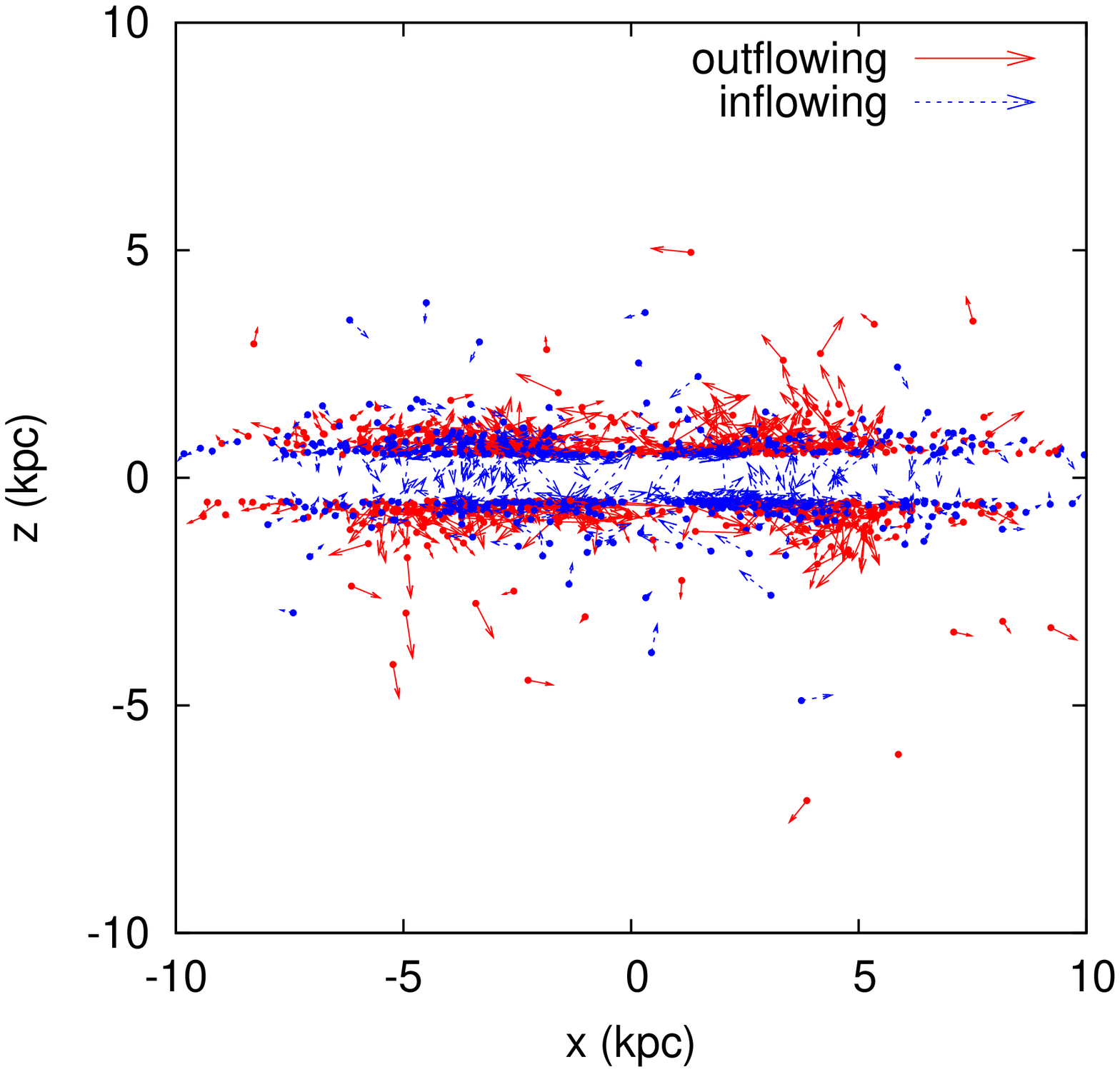}
}
\caption{Particle velocities as in
  figure~\ref{fig:vector_flows_mw}, for the MW simulation after 1 Gyr
  for three cases: $\theta=30^\circ$ (left), $\theta=60^\circ$ with
  energy weighted by the distance from the particle (middle), and for
  an isotropic distribution of energy (right).}
\label{fig:vector_coni}
\end{figure*}

Varying the angle $\theta$ or the distribution scheme of the energy to
neighboring particles does not introduce significant differences in
the global behavior of SFR and SK relation, while it changes the
morphology of the galaxy.  We show in Figure~\ref{fig:coni} the gas
distribution seen face-on after 1 Gyr in four cases,
$\theta=30^\circ$, $\theta=60^\circ$ (the standard case),
$\theta=60^\circ$ with energy {\em not} weighted by the distance from
the axis but only by the distance from the particle, and the case of
isotropic distribution of energy. Clearly the spiral pattern in the
gas distribution weakens when moving from the first to the last
case. For instance, an isotropic distribution of energy leads to the
complete disappearance of the spiral pattern. Our preference for the
wider opening angle, $\theta=60^\circ$, is motivated as follows: with
a narrower angle it frequently happens that particles have no
neighbors in the direction of decreasing density, thus leading to a
loss of the outflowing energy.  Figure \ref{fig:vector_coni}
  shows the resulting velocities of particles above or below the disc
  (see Fig.~\ref{fig:vector_flows_mw} for the standard MW case):
  distributing energy more selectively, i.e. within a smaller aperture
  angle, creates a more pronounced fountain.

As for the conditions for the onset of the multi-phase regime, we verified
that they have no strong influence on the simulations of isolated galaxies
like those presented here. This holds at least as long as the threshold density $n_{\rm
  thr}$ is kept low and the temperature threshold lies in the instability
range below $10^5$ K, that is not populated by single-phase particles (see
Figure~\ref{fig:phase}). This result may not necessarily hold in the case of
cosmological initial conditions. This specific point will be addressed in a
forthcoming work.  As far as the exit conditions are concerned, we adopted a
value of $n_{\rm out}$ slightly lower than $n_{\rm thr}$ to avoid the case of
particles rapidly bouncing between multi- and single-phase regimes.  Finally,
our chosen value for $t_{\rm clock}$ is suggested by theoretical reasons (see
Section \ref{section:entrance}). We tested the effect of taking higher values
for $t_{\rm clock}/t_{\rm dyn}$ and found that a large amount of stars is
produced outside the disk, because pressurized and blowing-out particles stay
multi-phase even when their density is rather low. This feature is not
desirable, thus giving further justification to the low value adopted for this
parameter.

\subsection{Effect of resolution}
\label{section:resolution}

\begin{figure*}
\centerline{\includegraphics[width=0.5\linewidth]{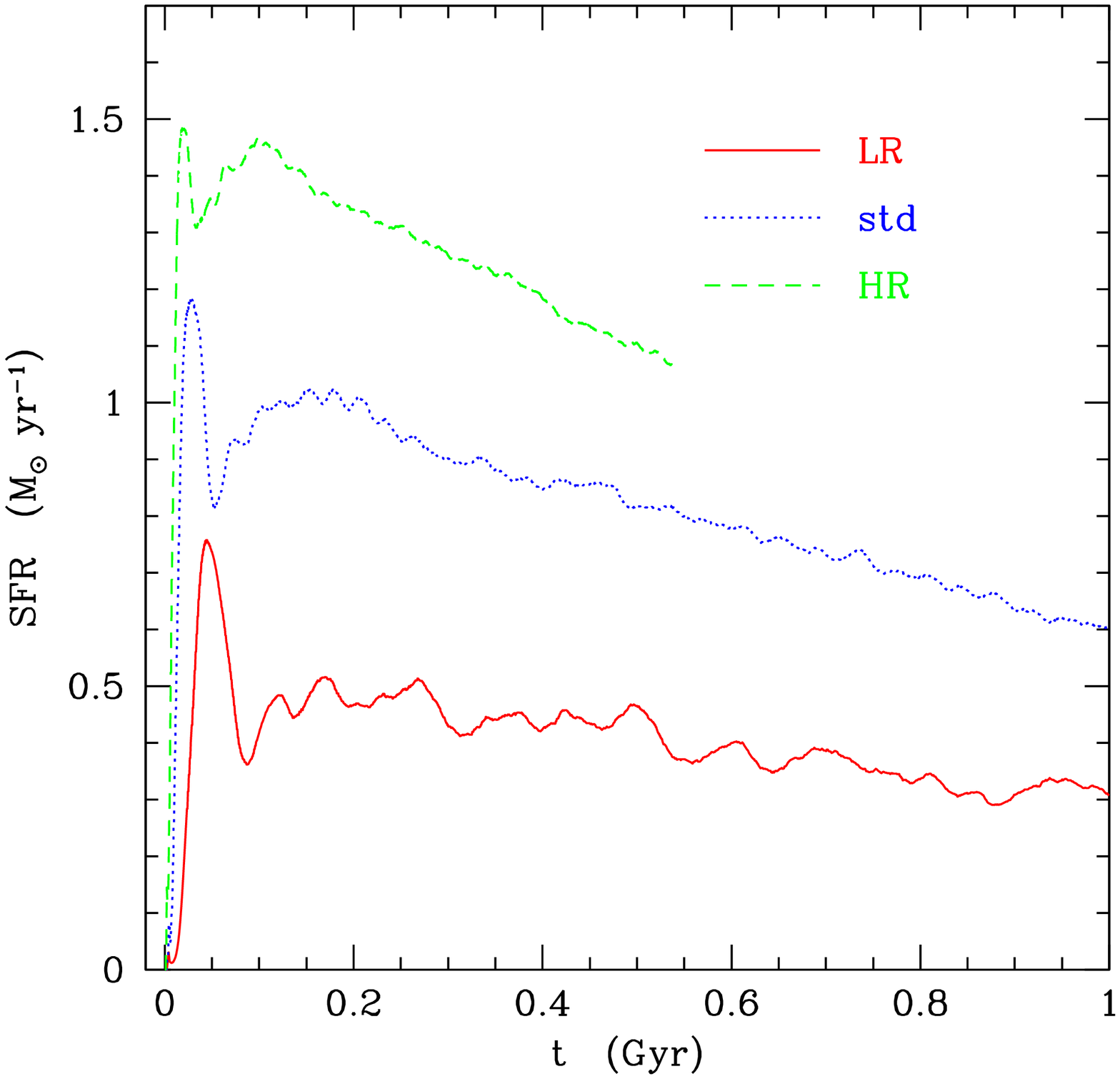}
\includegraphics[width=0.5\linewidth]{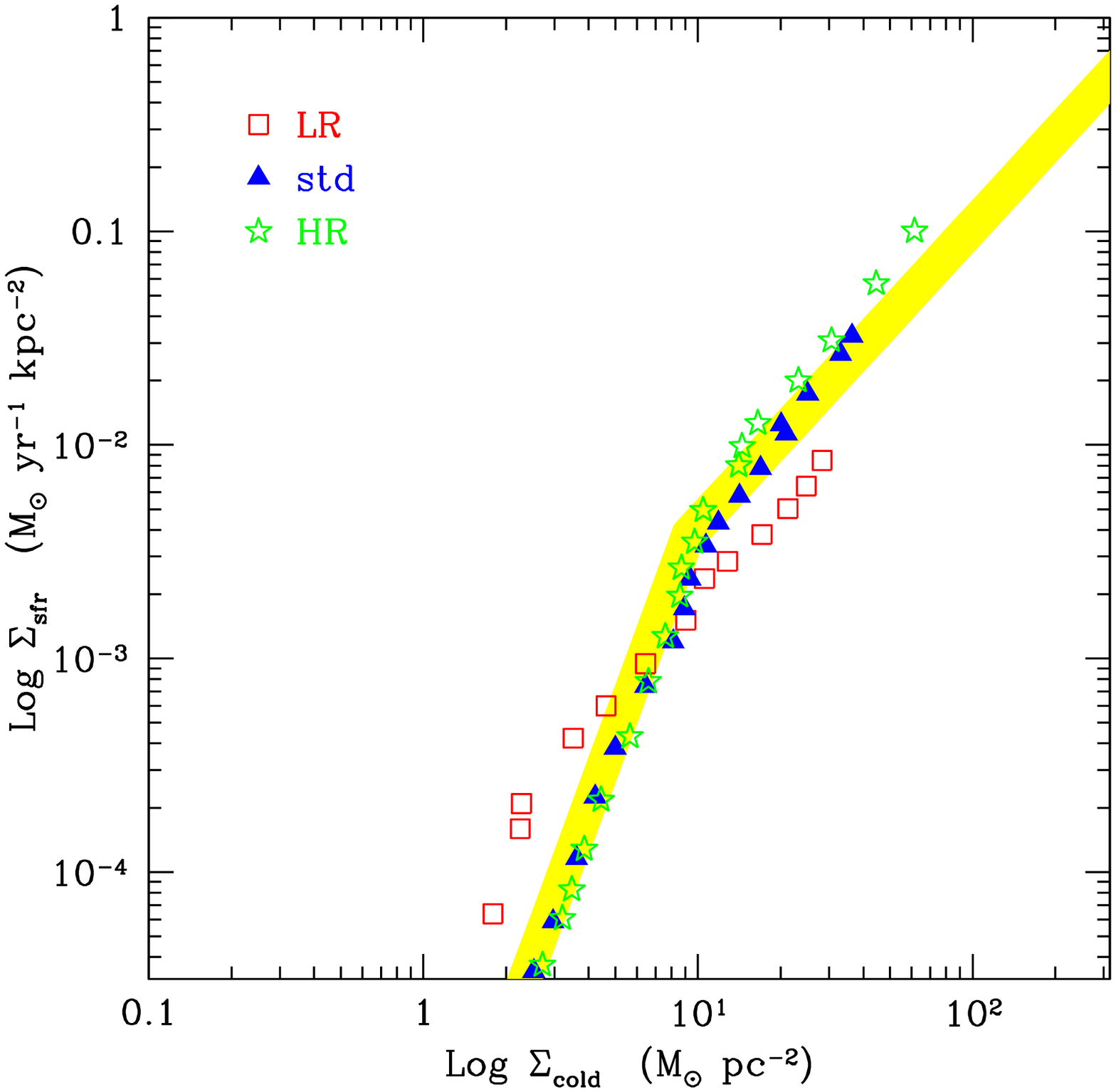}}
\centerline{\includegraphics[width=0.5\linewidth]{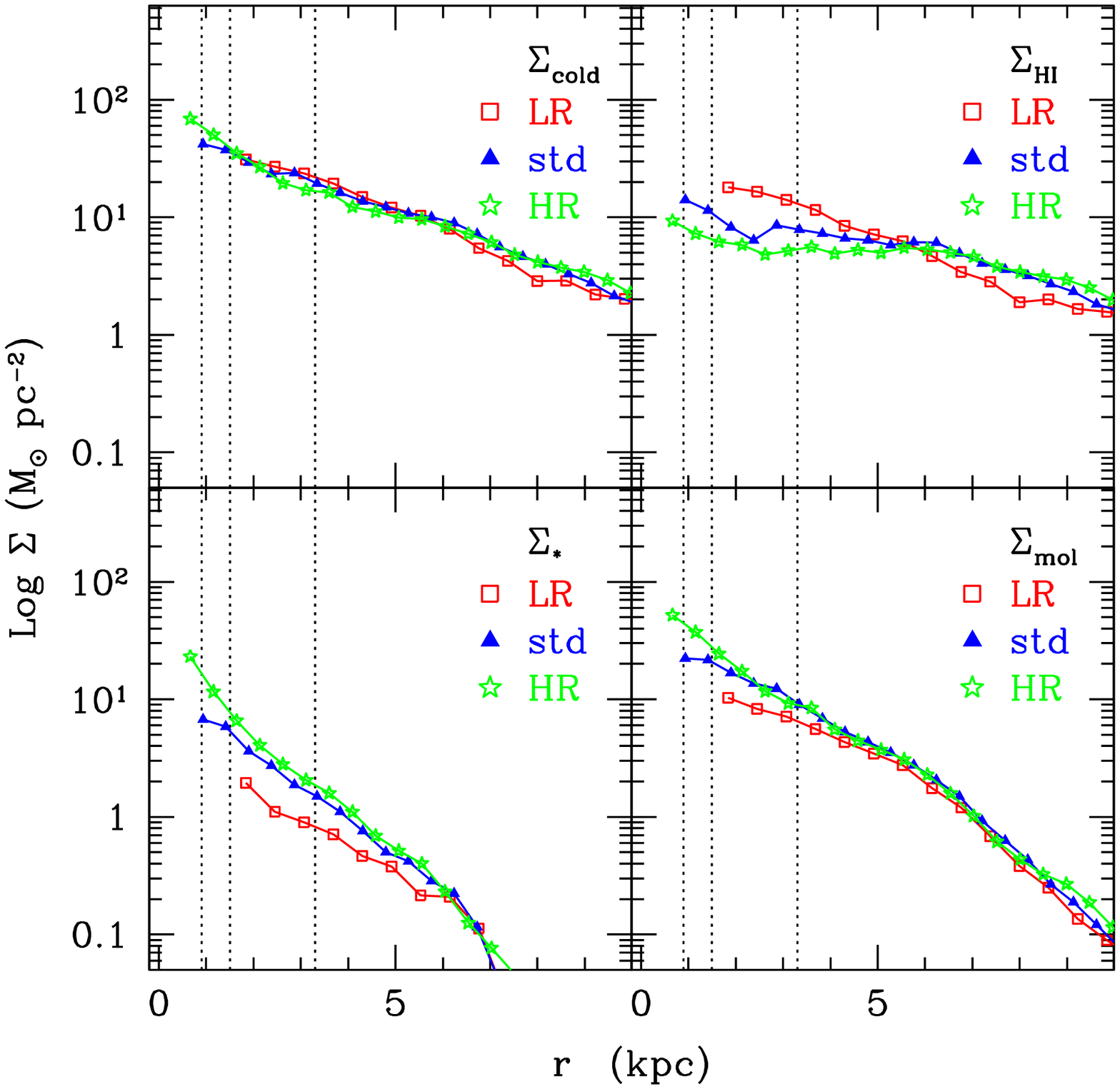}
\includegraphics[width=0.5\linewidth]{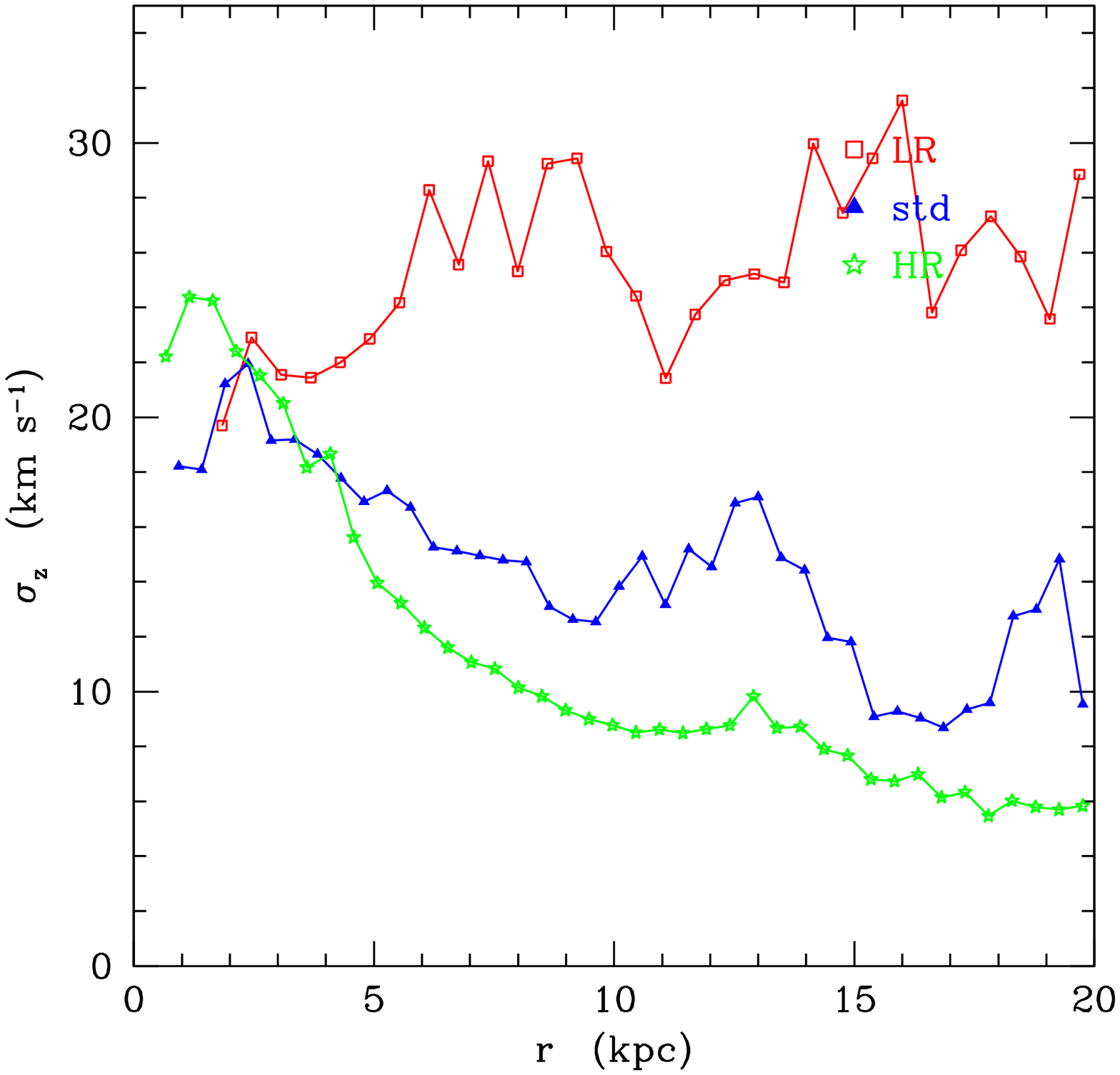}}
\caption{Stability of results of the isolated MW simulation against numerical
  resolution. Star formation rate, SK relation, surface
  density profiles of ISM properties and profiles of the r.m.s. vertical
  velocity are shown from the upper left to the bottom right panel. In the
  bottom left panel, vertical lines mark the force resolution (i.e., three
  times the plummer-equivalent gravitational softening) of our three
  simulations. All results, except those in the upper left panel, refer to 100
  Myr of evolution.}
\label{fig:resmw}
\end{figure*}

In order to test the sensitivity of MUPPI to resolution, we run the same MW
simulation with mass resolution 10 times worse (LR) and 4.62 times better
(HR), and force resolution (scaled as the cubic root of the mass resolution)
$\simeq 2.2$ times larger and $\simeq 1.7$ times smaller, respectively.  The
expensive HR simulation was stopped after 500 Myr, so we make our comparison
at this time, which is after the first (transient) peak of star formation.
Figure~\ref{fig:resmw} shows SFRs, SK relations, surface density profiles and
vertical velocity dispersions for the three runs.  The three vertical dotted
lines in the plot of the density profile mark three times the softening for
the three runs.  From these plots it is clear that the poor resolution of the
LR run does not allow to resolve the inner part of the disc, where most mass
is located.  This results in a low normalization of the SK relation and high
velocity dispersion of clouds, i.e. a thicker gas disc.  Increasing the mass
resolution by $\sim45$ times, from the LR to the HR run, improves the
description of the disc at smaller radii. In turn, this provides a reasonable
degree of convergence in density profiles, velocity dispersion and in the SK
relation, beyond three softening lengths.  In particular, we note that total,
atomic and molecular gas surface density profiles are stable when resolution
is changed.  However, this does not correspond to a convergence of the SFR,
and consequently of the stellar surface density profiles, that keeps
increasing as long as the inner regions give an important
contribution. Indeed, we notice that at $t=0.5$ Gyr the level of SFR increases
by about almost a factor 3 as we increase mass resolution by about a factor 45
from the LR to the HR run.

As a conclusion, the results of our sub-resolution model show in general a
mild dependence on numerical resolution. This dependence is more pronounced for
the SFR, owing to the fact that the bulk of star formation takes place within
the very central part of the galaxy.  We consider this as an undesired
characteristic of the model. On the other hand, it is worth pointing out that
one of the mayor advantages of our sub-resolution model lies in the fact that
ISM properties are determined by the local hydrodynamical characteristics of
the ambient gas (which are obviously resolution-dependent), respond to its
changes, and provide an effective feedback to it.  This is a focal difference
with respect to, for instance, the multi-phase effective model by
\citet{SpringHern03}.

\section{Conclusions}
\label{section:conclusions}

We presented MUPPI (MUlti-Phase Particle Integrator), a new sub-resolution
model for stellar feedback in SPH simulations of galaxy formation, developed
within the GADGET-2 code.  The code is based on a version of the model of star
formation and feedback developed by \citep{Monaco04a} (M04), which has been
modified and greatly simplified to ease the implementation in a code for
hydrodynamic simulations. In this model, each SPH gas particle is treated as a
multi-phase system with cold and hot gas in thermal pressure equilibrium, and
a stellar component. Cooling of MP particles is computed on the basis
of the density of the diluted hot phase, which carries only a small fraction
of the mass but has a very high filling factor. Energy generated by SNe is
distributed mostly to neighboring particles, preferentially along the ``least
resistance path'', as determined by the gradient of local density field. This
allows different feedback regimes to develop in different geometries: in a
thin disc energy is injected preferentially along the vertical direction, thus
creating a galactic fountain while the disc is heated to an acceptable level;
in a spheroidal configuration energy is trapped within the system and feedback
is more efficient in suppressing star formation.

One of the main ingredients of the model is the use of the phenomenological
relation of \cite{Blitz04,Blitz06}, which connects the fraction of molecular
gas with the external pressure of the star-forming disc.  Using this relation
with SPH pressure results in a system of equations that gives rise to a
runaway of star formation.  Indeed, SN energy increases pressure, and this
leads to a higher molecular fraction which turns into more star formation, up
to the saturation of the molecular fraction. This runaway does not take place
as long as the hydrodynamical response of the particle leads to a decrease of
pressure.  The consequence of this is that thermal feedback alone is efficient
in regulating star formation in an SPH simulation. This is uncommon in many
sub-resolution models of star formation and feedback currently implemented in
numerical simulations.

We tested our code with a set of initial conditions, including two isolated
disc galaxies and two spherical cooling flows.  These tests show the ability
of the code to predict the slope of the SK relation
\citep{1998ApJ...498..541K, Bigiel08} for disc galaxies, the basic properties
of the inter-stellar medium (ISM) in disc galaxies, the surface densities of
cold and molecular gas, of stars and SFR, the vertical velocity dispersion of
cold clouds and the flows connected to the galactic fountains.  
Self-regulation of star formation does not result in a bursty or
  intermittent SFR, though the DW galaxy shows damped oscillations in
  the first Gyr.  This is because the intrinsically unstable behavior
  of gas particles is stabilized by the SPH interaction with the
  surrounding particles. In the global SFR, we average over all the MP
  particles and thus smooth the intermittent
  behavior of particles and creates a more stationary
  configuration with a slowly declining SFR.

Furthermore, the cooling flow simulations show the dependence of
feedback on the geometry of the star-forming galaxies.  On the
  one hand, these simulations show that MUPPI does not trivially fix
  the SK relation, which is in fact reproduced only in the case of a
  thin rotating discs. On the other hand, the difference we obtain in
  the SK relation for our cooling flows shows that the geometry
  influences the behavior of our model, correctly distinguishing a
  thin system from a thick one.

Application to cosmological initial conditions
will be presented in a forthcoming paper.

Clearly, a number of variants of our implementation of a sub-resolution model
to describe star formation can be devised, each with its own advantages and
limitations. For instance, strictly speaking multi-phase gas should be
described by three phases, since cold gas is both in non-molecular and
molecular forms. It is possible to extend the two-phase description, adopted
in the current implementation, to three phases. However, we checked that the
increased complication is not justified by a clear advantage. The two-phase
model, though not completely consistent, is able to provide a good effective
description of an evolving ISM. Furthermore, the regulating time-scale for
star formation is the dynamical time-scale of the cold phase, computed at the
end of the first cooling transient, i.e. when most of the hot phase of a
particle that just entered the multi-phase regime has cooled.  This time-scale
has the advantage of giving plausible values for the star formation
time-scale (with $f_\star=0.02$), provided that the temperature of the cold
phase is set to the relatively high value of 1000 K. However, star formation
may be related to other time-scales, like the turbulent crossing-time of the
typical molecular cloud. Using this time-scale would be a preferable choice if
star formation is regulated by turbulence and not by gravitational
infall. Related to this point is the lack of a description in MUPPI of the
kinetic energy of the cold phase contributed by turbulent motions of cold
gas. Although we tried some simple model to include the contribution of such a
kinetic energy, we did not find any obvious advantage. Therefore, our
preference was given to the simpler formulation which neglects any kinetic
energy of the cold phase.

As a further important point, we only distribute thermal energy to particles
neighboring a star-forming one. This is deliberately done to demonstrate the
ability of MUPPI to provide efficient stellar feedback without any injection
of kinetic energy. Clearly SN explosions inject both thermal and kinetic
kinetic. It is straightforward to extend MUPPI so as to include also the
distribution of kinetic energy, at the very modest cost of adding another
parameter which specifies the fraction of the total energy at disposition to
be distributed as thermal and kinetic.  We will test and use a version of
MUPPI with kinetic feedback in forthcoming papers.

Finally, the version of the code presented here does not include any explicit
description of chemical enrichment, so that cooling is always computed for gas
with zero metalicity. On the other hand, a number of detailed implementations
of chemo-dynamical models in SPH simulation codes have been recently presented
\citep[e.g.,][ and references therein]{Tornatore07,Wiersma09}, which describe
the production of metals from different stellar populations, also accounting
for the life-times of such populations and including the possibility of
treating different initial mass-functions.  We are currently working on the
implementation of MUPPI within the version of the GADGET code which includes
chemical evolution as implemented by \cite{Tornatore07}. This will ultimately
provide a more complete and realistic description of star formation in
simulations of galaxies.

\section*{Acknowledgements}
We are highly indebted with Volker Springel, for providing us with the
non-public version of the GADGET-2 code. We also wish to thank Simone
Callegari and Lucio Mayer for providing us with the initial conditions for the
MW and DW simulations.  We acknowledge useful discussions with Luca Tornatore,
Klaus Dolag and Bruce Elmegreen.  We acknowledge the anonymous referee
  for useful comments who helped to improve this work.  The simulations were
carried out at the ``Centro Interuniversitario del Nord-Est per il Calcolo
Elettronico'' (CINECA, Bologna), with CPU time assigned under INAF/CINECA and
University-of-Trieste/CINECA grants. This work is supported by the ASI-COFIS
and ASI-AAE (I/088/06/0) grants and by the PRIN-MIUR grant ``The Cosmic Cycle
of Baryons''. Partial support by INFN-PD51 grant is also gratefully
acknowledged.

\bibliographystyle{mn2e}
\bibliography{master}

\label{lastpage}

\end{document}